\documentclass[a4,fleqn,usenatbib]{mnras}
\usepackage{etoolbox}
\makeatletter
\patchcmd\@combinedblfloats{\box\@outputbox}{\unvbox\@outputbox}{}{%
   \errmessage{\noexpand\@combinedblfloats could not be patched}%
}%
 \makeatother

\usepackage[T1]{fontenc}
\usepackage{ae,aecompl}

\usepackage{graphicx}	
\usepackage{amsmath}	
\usepackage{amssymb}	
\usepackage{stmaryrd}
\usepackage{bm}
\usepackage{times}
\usepackage{amsmath}
\usepackage{amssymb}
\usepackage{textcomp}
\usepackage{verbatim}

\newcommand {\hi} {\ifmmode \ion{H}{I} \else $\ion{H}{I}$ \fi}
\newcommand {\hii} {\ifmmode \ion{H}{II} \else $\ion{H}{II}$ \fi}
\newcommand {\hei} {\ifmmode \ion{He}{I} \else $\ion{He}{I}$ \fi}
\newcommand {\heii} {\ifmmode \ion{He}{II} \else $\ion{He}{II}$ \fi}
\newcommand {\heiii} {\ifmmode \ion{He}{III} \else $\ion{He}{III}$ \fi}
\newcommand {\ts}{\textsubscript}
\newcommand*\diff{\mathop{}\!\mathrm{d}}

\title[Radiation hydrodynamics in {\sc Arepo}]{{\sc Arepo-RT}: Radiation hydrodynamics on a moving mesh}

\author[R. Kannan et al.]{Rahul Kannan$^{1}$\thanks{E-mail: rahul.kannan@cfa.harvard.edu}\thanks{Einstein Fellow},
Mark Vogelsberger$^{2}$\thanks{Alfred P. Sloan Fellow},
Federico Marinacci$^{2}$,
Ryan McKinnon$^2$,
\newauthor
R\"udiger Pakmor$^{3}$
and Volker Springel$^{3,4,5}$
\\ 
\\
$^{1}$Harvard-Smithsonian Center for Astrophysics, 60 Garden Street, Cambridge 02138, MA, USA\\
$^{2}$Kavli Institute for Astrophysics \& Space Research, Massachusetts Institute of Technology, 77 Massachusetts Ave, Cambridge 02139, MA, USA\\
$^{3}$Heidelberg Institute for Theoretical Studies, Schloss- Wolfsbrunnenweg 35, D-69118 Heidelberg, Germany\\
$^{4}$Zentrum f\"ur Astronomie der Universit\"at Heidelberg, ARI, Mönchhof-str. 12-14, D-69120 Heidelberg, Germany\\
$^{5}$Max-Planck-Institut f\"ur Astrophysik, Karl-Schwarzschild-Str. 1, 85741 Garching, Germany\\
}

\date{Accepted XXX. Received YYY; in original form ZZZ}

\pubyear{2018}

\begin{document}
\label{firstpage}
\pagerange{\pageref{firstpage}--\pageref{lastpage}}
\maketitle

\begin{abstract}
We introduce {\sc Arepo-RT}, a novel radiation hydrodynamic (RHD) solver for
the unstructured moving-mesh code {\sc Arepo}. Our method solves the
moment-based radiative transfer equations using the M1 closure relation. We
achieve second order convergence by using a slope
limited linear spatial extrapolation and a first order time prediction step to obtain the values of the primitive variables on both sides of the cell interface. 
A Harten-Lax-Van Leer flux function, suitably modified for moving meshes, is then
used to solve the Riemann problem at the interface. The implementation is
fully conservative and compatible with the individual timestepping scheme of {\sc Arepo}.
It incorporates atomic Hydrogen (H) and Helium (He)
thermochemistry, which is used to couple the ultra-violet (UV) radiation field
to the gas. Additionally, infrared radiation is coupled to the gas
under the assumption of local thermodynamic equilibrium between the gas
and the dust. We successfully apply our code to a large number of test problems,
including applications such as the expansion of \ion{H}{II} regions, radiation
pressure driven outflows and the levitation of optically thick layer of gas
by trapped IR radiation.     
The new implementation is suitable for studying various important
astrophysical phenomena, such as the effect of radiative feedback in driving
galactic scale outflows, radiation driven dusty winds in high redshift quasars,  
or simulating the reionization history of the Universe in a  self consistent manner. 
\end{abstract}


\section{Introduction}
Radiation fields are ubiquitous in Nature.  In the earliest epochs the Universe
was a hot dense soup of matter and radiation. The mean free path of photons
was very small due to Thomson scattering off free electrons. As the Universe
cooled to ${\sim 4000 \ \text{K}}$ about $300,000$ years (${z\simeq1100}$) after
the Big Bang, protons began to capture the free electrons and form atomic
hydrogen. These recombinations diminished the number density of free electrons
allowing the matter and radiation to decouple, making the Universe transparent to light.

The radiation from this epoch is observed today as the nearly uniform
${\sim2.72\,\text{K}}$ Cosmic Microwave background (CMB; \citealt{Alpher1948,
Penzias1965}).  The CMB is currently our best probe to explore the early Universe. The
slight temperature inhomogeneities present in the CMB (${\Delta T/T \simeq
10^{-5}}$, \citealt{Smoot1992}) correspond to the primordial density
fluctuations, which grow through gravitational collapse and form the structures
that we see today \citep{White1978}.  

After recombination, the Universe went through a period of darkness with
no sources of visible light. Once the gas in the high density regions became
dense enough, it  started forming stars and eventually proto-galaxies. These
early stars and galaxies have an important effect on the surrounding
environment. They emit copious amounts of ionizing Hydrogen radiation (${\geq
13.6\,\text{eV}}$), which is believed to re-ionize the neutral inter-galactic
medium \citep[IGM]{Shapiro1987, Madau1996, Gnedin1997, Madau1999, Gnedin2000}.
Reionization is believed to be initially patchy with pockets of ionized plasma
surrounding the most energetic sources. As the pockets grow larger and become
more numerous, they overlap and eventually reionize the whole Universe. In
addition, the photons heat the IGM, altering its thermal state which in turn
affects the observed Lyman-$\alpha$ distribution \citep{Gnedin1998, Gnedin1998b,
Schaye2000}.  

The epoch of reionization (EoR), carries plenty of information about the process
of structure formation in the Universe and provides evolutionary links between
the smooth matter distribution at early times revealed by CMB, and the
large-scale structure observed at low redshifts.  Gaining insights into this
epoch is challenging, but recent observations of the steep faint end slope of the
ultra-violet (UV) luminosity function indicates that the cosmic ionizing photon
budget is dominated by faint galaxies \citep{Bouwens2011, Bouwens2015}.
Full reionization within ${z\sim6}$ can only be achieved if the
observed luminosity function is extrapolated by two orders of magnitude below
the observational limit of the {\it Hubble Space Telecscope} \citep[][HST]{Finkelstein2015}. However,
photoheating can suppress star formation in low mass haloes, and it is
therefore unclear if such an extrapolation in valid.  Furthermore, the escape
fraction of ionizing photons from the interstellar (ISM) and the circumgalactic
(CGM) media is impossible to measure directly above ${z\sim4}$.  While some
indirect measurements of the escape fraction have been made
\citep{Zackrisson2013, Dijkstra2016, Reddy2016, MasRibas2017}, they are not
very constraining and allow the escape fraction to be anywhere between ${1\%}$
and ${30\%}$.

With the imminent launch of the {\it James Webb Space Telescope} (JWST;
\citealt{Gardner2006}), the study of EoR enters a new era. It is expected to
increase the quality of high redshift data and extend it beyond ${z=10}$.  It is
therefore important for theoretical models to achieve enough accuracy to
interpret the observational results. Numerical radiation
hydrodynamic simulations offer the most accurate and realistic theoretical
models of reionization. Hence, an implementation of an accurate and efficient
radiative transfer algorithm becomes imperative to study the high redshift
Universe.

Radiation fields also play an important role in many physical processes that
occur inside dark matter (DM) haloes. For example, a long standing puzzle in
galaxy formation theory has been the low star formation efficiency in DM haloes
\citep{Silk2012}. Star formation efficiency peaks at about
${M_\ts{halo} \sim 10^{12} M_\odot}$ and decreases in both higher and lower mass
haloes \citep{Moster2010, Moster2012, Behroozi2013}.  Feedback from stars
\citep{Navarro1996, Springel2003, Stinson2006, DV2008,  Agertz2013,
Vogelsberger2013, Hopkins2014, Hopkins2017} and the central active galactic
nuclei (AGN; \citealt{Springel2005, Sijacki2006, Booth2009, Choi2012,
Kannan2017, Weinberger2017}) are invoked to explain this low star formation
efficiency in low and high mass galaxies respectively. 

Early galaxy formation simulations showed that the coupling between supernovae
(SNe) feedback energy and the ISM is very inefficient \citep{Katz1996,
Navarro1996}, as most of the injected energy is radiated away very
efficiently. Various sub-grid models have been proposed in order to avoid
cooling loses such as delaying the cooling of gas particles around a star
\citep{Thacker2001, Stinson2006, Agertz2013}, stochastically injecting energy
into the surrounding gas such that it is heated up to the temperatures where
gas cooling becomes inefficient \citep{DV2008, Schaye2015} and injecting
kinetic energy that adds velocity kicks to gas particles to remove them from
the inner regions of galactic discs \citep{Springel2003, Oppenheimer2006,
Vogelsberger2013}. These ad-hoc methods have been quite successful in
reproducing the properties of galaxies in a broad sense
\citep{Vogelsberger2014, Vogelsberger2014N, Schaye2015}. However, they require
fine tuning of free parameters in the model in order to reproduce the low mass
end of the luminosity function \citep{Vogelsberger2014, Schaye2015,
Pillepich2018} and in some cases require unrealistic values of SNe feedback
energy (${>10^{51}\,\text{erg}}$; \citealt{Guedes2011, Schaye2015}) or
excessively large gas outflow velocities \citep{Pillepich2018}. 
 
Many recent works have pointed out that young massive stars deposit large amounts
of energy in the form of  photons and stellar winds before they go SNe, which can have a significant dynamical impact on the ISM \citep{Murray2010,
Walch2012}. \citet{Stinson2013} showed that the high energy photons emitted by
OB stars can ionize and photoheat the surrounding regions helping to regulate
star formation especially at high redshifts \citep{Kannan2014a}. However, this
requires efficient thermalisation in the injected radiation energy close
to the source, which is not guaranteed in the high density regions where stars
form. Radiation pressure, both direct UV and multiscattered infrared (IR), is
another mechanism hypothesised to drive significant outflows (${\sim
100\,\text{km s}^{-1}}$) \citep{Hopkins2011, Agertz2013, Hopkins2014}.
However, it seems that unphysically large optical depths of IR
radiation (${\tau_\ts{IR}\sim 50}$) are required to effectively trap the photons
and boost the momentum injection to the levels required to efficiently suppress
star formation \citep{Roskar2014}. Alternatively, if enough radiation escapes
the star forming regions and the ISM of galaxies, it can in principle reduce
the gas cooling rates of the circum-galactic medium (CGM) thereby reducing gas inflows into the centers
of galaxies \citep{Cantalupo2010, Gnedin2012, Kannan2014b, Kannan2016a}.  While
these works hint towards the importance of radiation fields, the crude nature of
these sub-grid models makes it difficult to gauge the exact
mechanisms and significance of radiation fields in regulating the star
formation rates of low mass galaxies. Therefore, full radiation hydrodynamic
simulations are necessary in order to gain a fundamental understanding of
stellar feedback \citep{Rosdahl2015b, Kim2017, Peters2017}. 
 
The impact of stellar feedback decreases as the mass and potential depth of DM haloes
increases. A more energetic source of feedback is needed, which is conveniently
found in the form of the central AGN. The inability of galaxy formation
simulations to resolve the region around the central super massive black hole
(SMBH), necessitates sub-grid prescriptions to account for this feedback
channel. These models generally discriminate between a high accretion rate
quasar mode \citep{Springel2005} feedback and a low accretion rate mechanical
radio mode feedback \citep{Weinberger2017}. However, the details of how the AGN
energy couples to the gas in and around galaxies is still uncertain, so
modeling efforts have so far been necessarily crude. 

In general the AGN deposits energy and momentum into the surrounding gas
driving winds. Observations show compelling evidence for galaxy scale AGN
driven outflows \citep{Heckman1981, Nelson1995, Greene2005, Karouzos2016}.
These are generally high mass-loaded (${M_\ts{out} > 10^8\,\text{M}_\odot}$),
fast (${v_\ts{out} \geq 1000\,\text{km s}^{-1}}$)  and multiphase winds that can
extended over up to several tens of kpc \citep{Cicone2015, Tombesi2015,
Zakamska2016}.  The efficiency with which these winds suppress star formation
is still  unknown mainly because the exact mechanism of the coupling between
the AGN feedback energy and the gas is not understood. Ultra-fast outflows
(${\geq 10,000\,\text{km s}^{-1}}$) can develop a two-temperature structure,
where most of the thermal pressure support is provided by the protons while the
cooling processes operate directly only on the electrons. This significantly
slows down inverse Compton cooling, maintaining the thermal structure of the
wind and generating large momentum boosts (${\sim 20L/c}$) and high kinetic
luminosities \citep{FG2012, Costa2014}.  The AGN luminosity can also couple
directly to the surrounding gas through radiation pressure on dust
\citep{Fabian1999, Murray2010, Thompson2016}. Observations indicate a large
fraction of the the optical and UV radiation is absorbed and re-emitted at IR
wavelengths by a surrounding envelope of dusty gas before escaping the galactic
nucleus \citep{Fabian1999b}.  If the column density of the gas is high enough,
the optical depth to IR radiation can be rather large, trapping this
radiation within the AGN nucleus. The multiscattered IR photons boost the
momentum injection by a factor proportional to the optical depth (${\dot{p}\sim
\tau_\ts{IR}L/c}$). 

Many recent studies have tried to model this mechanism using simple analytic
models \citep{Ishibashi2015, Thompson2016} and idealised high-resolution
radiative transfer simulations of the dusty AGN torus \citep{Roth2012,
Ishibashi2017, Costa2018}. Recent advances in numerical techniques have allowed
 for the estimation  of the impact of radiation pressure on galactic scale using isolated
disc simulations \citep{Bieri2017, Cielo2017} and more recently in cosmological
simulations \citep{Costa2017}. These simulations suggest that IR radiation
pressure can drive fast (${\geq 1000\,\text{km s}^{-1}}$), high mass loaded and
short lived winds during the obscured phase of the AGN. In fact, these winds
are able to remove enough material from the center to completely shut down star
formation, indicating the need to employ accurate radiation hydrodynamic (RHD)
simulations to understand radiative feedback from AGNs.

On smaller scales the radiation from massive protostars can unbind gas in
its surroundings and create cavities. These cavities can prevent any further
accretion onto the star from the direction of the bubble, cutting off fuel
supply to the star and stalling the mass growth \citep{Kuiper2012}, although
the development of Rayleigh-Taylor instabilities can destroy the cavity and
allow gas to fall back onto the star \citep{Rosen2016}.  The stellar radiation
fields also have a large impact on the structure of the protoplanetary disks
\citep{Flock2016} and on the climate of exoplanets \citep{Heng2011}. 

To summarise, radiation plays a crucial role in a large variety of
astrophysical systems, and its impact ranges all the way from small planetary
systems to the large-scale thermal and ionization history of the Universe. The
complexity of radiative transfer requires accurate radiation hydrodynamic
simulations to precisely capture and model its impact. Consequently, an
accurate and efficient radiative transfer implementation is needed to improve
current astrophysical simulations by taking into account the effects of radiation.
In this work, we present a moment-based radiative transfer implementation for
the moving mesh hydrodynamics code {\sc Arepo}.  

The paper is structured as follows. In Section~\ref{sec:RT} we briefly outline
the various RT schemes used in literature and discuss the advantages and
shortcomings of the scheme used in this paper.  Section~\ref{sec:methods}
describes the spatial discretization and time integration techniques used to
solve the radiative transfer equations for our scheme. In
Section~\ref{sec:tests} we present several test problems to quantify the
accuracy of our implementation. Finally, we present our conclusions in
Section~\ref{sec:conclusions}.

\section{The Radiative Transfer Equations}
\label{sec:RT}
Here we first discuss the relevant radiative transfer equations that we are
going to solve.  We start by defining the specific intensity, ${I_\nu({\bf x},
t, {\bf n}, \nu)}$, at position ${\bf x}$ and time $t$,  as the rate of
radiation energy ($E_\nu$) flowing per unit area (d${\bf A}$), in the direction
(${\bf n}$), per unit time (d$t$), per unit frequency interval (d$\nu$)
centered on frequency $\nu$ and per unit solid angle (d$\Omega$)
\begin{equation}
\text{d}E_\nu = I_\nu({\bf x}, t, {\bf n}, \nu)  \ ({\bf n} \cdot \text{d}{\bf A})\, \text{d}t \, \text{d}\nu \, \text{d}\Omega \, .
\end{equation}  
The propagation of the radiation field and interactions with the surrounding
medium such as absorption and emission of radiation leads to a change in the
radiation energy at that spatial position. Taking these processes into account
we can write down the continuity equation for the specific intensity as
\citep{Mihalas1984}
\begin{equation}
\frac{1}{c} \frac{\partial {I_\nu}}{\partial t} + {\bf n} \cdot  {\bf \nabla} I_\nu =  j_\nu - \kappa_\nu \,  \rho \, I_\nu \, ,
\label{eq:RT}
\end{equation}
where $j_\nu$ is the emission term and $\kappa_\nu$ is the absorption
coefficient. 

Radiative transfer is a complex process due its high dimensionality.  An
accurate numerical solution requires discretising Eq.~\ref{eq:RT} in angular
and frequency variables in addition to spatial and time discretization. A
variety of different numerical algorithms have been proposed to solve the
radiative transfer problem. The most common method is the long characteristic
ray-tracing scheme \citep{Mihalas1984, Abel1999, Abel2002, Greif2014,
Jaura2017} that cast rays from each source through the simulation domain and
solve Eq.~\ref{eq:RT} along each ray. This method, although very accurate, is
computationally expensive (${{\mathcal O}(N^2)}$) and requires high angular
resolution to capture the correct transport of radiation.  Furthermore,
parallelising this algorithm requires significant data exchanges between
different processors. In order to reduce the complexity of the problem, some
works have resorted to short characteristics methods \citep{Ciardi2001,
Whalen2006, Trac2007, Pawlik2008, Petkova2011} which integrates the radiative
transfer (RT) equation only along lines that connect nearby cells making it
easier to parallelise.

The radiative transfer equations can also be solved with Monte Carlo methods.
These schemes \citep{Oxley2003, Semelin2007,
Dullemond2012}, emit individual photon packets to sample the interaction
lengths and scafttering angles of the photons from the underlying probability
density functions. While they perform remarkably well, they are computationally
very demanding and the Poisson noise inherent to the statistical description of
the radiation field leads to a signal-to-noise ratio that grows only with the
square root of the number of photon packets emitted.  

Solving the moments of the radiative transfer equation has gained popularity in
recent years \citep{Levermore1984, Gonzalez2007, Rosdahl2013, Rosdahl2015}.
A fluid description of the radiation field is obtained by taking the
zeroth and first moments of Eq.~\ref{eq:RT} 
\begin{equation}
\begin{split}
&\frac{\partial E_r}{\partial t} + {\bf \nabla} \cdot {\bf F_r} = S - \kappa_\ts{E} \, \rho \, {\tilde c} \, E_r \, ,
\end{split}
\label{eq:E}
\end{equation}
\begin{equation}
\begin{split}
&\frac{\partial {\bf F_r}}{\partial t} + {\tilde c}^2 {\bf \nabla} \cdot {\mathbb P_r} =  - \kappa_\ts{F} \, \rho \, {\tilde c} \, {\bf F_r}\, ,
\end{split}
\label{eq:F}
\end{equation}
where the radiation energy density ($E_r$), flux (${\bf F_r}$) and pressure
(${\mathbb P_r}$) are defined as 
\begin{equation}
\{{\tilde c}E_r,  \ {\bf F_r},  \  {\mathbb P_r}\} = \int_{\nu_1}^{\nu_2}\int_{4\pi} \{1, {\bf n}, \  ({\bf n}\otimes{\bf n})\}I_\nu  \, \text{d}\Omega \, \text{d}\nu \, .
\end{equation}
Here (In Eqs.~\ref{eq:E} and \ref{eq:F}) $S$ denotes the source term which
quantifies the amount of radiation energy emitted, $\kappa_\ts{E}$ and $\kappa_\ts{F}$ are
the radiation energy density and radiation flux weighted mean opacities within
the frequency range defined by ${[\nu_1, \nu_2]}$ and $\rho$ is the density of
gas in the cell. We note that we have re-formulated the equations in terms of
signal speed (${\tilde c}$) of radiation transport, which can be different from
the actual speed of light ($c$) when the reduced speed of light approximation
(RSLA) is used (see Section~\ref{sec:transport} for more details).

This set of hyperbolic conservation equations defines the rate of change of
$E_r$ and ${\bf F}_r$ as a function of time and position.  However, in order to
solve these equations an estimate of ${\mathbb P}_r$ is required. We therefore
need to obtain the pressure tensor by invoking a closure relation. One popular
choice, which works remarkably well in optically thick media, is to recast
Eqs.~\ref{eq:E}~and~\ref{eq:F} into a diffusion equation \citep{Lucy1977,
Krumholz2012, Gonzalez2015} by assuming that the photon flux is proportional to
the gradient of the photon energy density (${{\bf F}_r = -{\tilde c} {\bf
\nabla}E_r/{3\kappa \rho}}$). While this approximation performs well in highly
optically thick media, its accuracy in optically thin cases is not well
understood. Furthermore, maintaining the directionality of photon propagation
is quite difficult as the photon flux is assumed to be directed along the
gradient of the photon energy density, forcing the photons to diffuse
isotropically. This makes it difficult to form sharp shadows behind optically
thick barriers \citep{Zhang2017}. A better approximation can be obtained by
ignoring the term of the order $c^{-1}$ in Eq.~\ref{eq:F} and set
\begin{equation}
{\bf F}_r = -\frac{{\tilde c}}{\kappa_\ts{F} \, \rho} {\bf \nabla} \cdot {\mathbb P}_r \, ,
\end{equation}
which yields
\begin{equation}
\frac{\partial E_r}{\partial t} + {\bf \nabla} \cdot \left( -\frac{{\tilde c}}{\kappa_\ts{F} \rho} {\bf \nabla} \cdot {\mathbb P}_r \right) = S - \kappa_\ts{E} \, \rho \, {\tilde c} \, E_r \, .
\end{equation}
The Eddington tensor formalism can then be used to equate the
radiation energy density ($E_r$) and the radiation pressure tensor (${\mathbb
P}$) by defining a proportionality tensor called the Eddington tensor
(${\mathbb D}$)
\begin{equation}
{\mathbb P_r} = E_r \, {\mathbb D} \, .
\label{eq:edd}
\end{equation} 
This Eddington tensor essentially encodes the direction of photon transport at
each point in the domain.  In this form, the RT equations transform into an
anisotropic diffusion equation. The discretization of this equation is
surprisingly non-trivial and widely used methods give rise to unphysical
oscillations \citep{Parrish2005}.  Some works get around this problem by adding
an isotropic component to the anisotropic diffusion tensor, however, this
reduces the accuracy with which the algorithm preserves the directionality of
the underlying photon field \citep{Petkova2009}. Therefore this method suffers
from the same problems as the isotropic diffusion approximation. Furthermore,
the timestep limitations imposed by the parabolic diffusion equation requires
the implementation of implicit or semi-implicit schemes in order to make the
algorithm fast enough \citep{Kannan2016b, Kannan2017}. These time
integration techniques are difficult to implement and parallelise efficiently. 

For these reasons, we chose to discard the approximations above and instead
solve the coupled hyperbolic conservation laws for the photon energy density
(Eq.~\ref{eq:E}) and the photons flux (Eq.~\ref{eq:F}), coupled with the
Eddington closure relation (Eq.~\ref{eq:edd}).  It is straightforward to
compute ${\mathbb D}$ in the case of a single or few sources. However, the
computation becomes quite arduous when considering galaxy scale simulations
which can have millions of sources within the simulation domain.  Many works
have tried to derive approximate estimates of ${\mathbb D}$, such as, for
instance, the optically thin variable Eddington tensor (OTVET) formalism
\citep{Gnedin2001, Finlator2009, Petkova2009} or the M$1$ \citep{Levermore1984,
Dubroca1999, Ripoll2001} method. The OTVET formalism computes ${\mathbb D}$ by
assuming that the intervening material between the radiation source and sink is
optically thin. The obvious drawback of this method is that direction of
radiation field behind any optically thick material will not be correctly
captured. Additionally the computational cost associated with estimating the
positions of every source relative to every volume element can be quite
significant. 

On the other hand, the M$1$ closure requires only local quantities of a
given cell to compute the Eddington Tensor ${\mathbb D}$ as
\begin{equation}
{\mathbb D} = \frac{1-\chi}{2} {\mathbb I} + \frac{3\chi-1}{2} {\bf n} \otimes {\bf n} \, ,
\label{eq:M1}
\end{equation} 
where
\begin{equation}
{\bf n} = \frac{\bf F_r}{|{\bf F_r}|}, \,\,\, \chi = \frac{3+4f^2}{5+2\sqrt{4-3f^2}}, \,\,\, {\text{and}} \ f = \frac{|{\bf F_r}|}{{\tilde c}E_r} \, .
\end{equation}
Since the radiation flux cannot be larger than the signal speed times the
radiation energy density, the reduced flux ($f$) will always be limited by
${0\le f \le1}$. The local nature of the M$1$ closure implies that the
computational cost is independent of the number of sources and only depends on
the number of resolution elements within the domain. This has allowed recent
works to perform galaxy scale simulations with relatively low computational
cost \citep{Rosdahl2015b, Costa2017, Bieri2017}.  Since we are mostly
interested in performing simulations containing a large number of sources we
adopt the M$1$ closure formalism in our work. 

Having established our numerical scheme to evolve the radiation field, we now have
to couple the radiation to the gas. The radiation field couples to the gas
hydrodynamics via photon absorption and scattering (RHS of Eqs.~\ref{eq:E} and
\ref{eq:F}). These physical processes are quantified using the average
opacities ($\kappa_\ts{E}$ and $\kappa_\ts{F}$) of the gas. Energy and momentum
conservation then dictates that photon absorption introduces source terms into
hydrodynamic momentum and energy conservation equations \citep{Gonzalez2007}:
\begin{equation}
\frac{\partial (\rho {\bf v})}{\partial t} + {\bf \nabla} \cdot (\rho \, {\bf vv}^T + P{\mathbb I}) = \frac{\kappa_F \, \rho \, {\bf F_r}}{c} \, ,
\label{eq:momentum}
\end{equation}
\begin{equation}
\frac{\partial (\rho E)}{\partial t} + {\bf \nabla} \cdot [(\rho E + P){\bf v}] = -\Lambda + \kappa_E \, \rho \, {\tilde c} \, E_r + \frac{\kappa_F \, \rho}{c} {\bf F_r} \cdot {\bf v} \, ,
\label{eq:energy}
\end{equation}
where $\Lambda$ is the gas cooling rate which is function of the abundance of the ionic species present in the gas, which is in turn dependent on the incident radiation field, $P$ its thermal pressure, $E$ its
total energy per unit mass and {\bf v} the gas velocity field.  
\section{Methods}
\label{sec:methods}
In this section, we describe in detail the spatial discretization and time integration
techniques used to solve the radiation hydrodynamic equations introduced in
Section~\ref{sec:RT}.  The
transport equations (setting the R.H.S of Eq. \ref{eq:E} and Eq. \ref{eq:F}) and the source terms are solved separately using an operator split approach. This is achieved using a Strang split scheme \citep{Strang1968}, which involves a half step update of the primitive RT variables ($E_r, {\bf F}_r$) due to the source terms, a full step update due to the transport operations, and finally another half step update with the source operations. This makes the solution formally converge at second order.

\subsection{The transport equations}
\label{sec:transport}
Let us first consider the free transport of photons, which is obtained by
setting the RHS of Eqs.~\ref{eq:E} and \ref{eq:F} to zero 
\begin{equation}
\frac{\partial E_r}{\partial t} + {\bf \nabla} \cdot {\bf F_r} = 0 \, ,
\label{eq:consE}
\end{equation}
\begin{equation}
\frac{\partial {\bf F_r}}{\partial t} + {\tilde c}^2 {\bf \nabla} \cdot {\mathbb P_r} = 0 \, .
\label{eq:consF}
\end{equation}

Equations \ref{eq:consE} and \ref{eq:consF} are conservation laws for the
photon energy density and photon flux that take the form of a system of
hyperbolic partial differential equations. They can be written in compact
conservative form as 
\begin{equation}
\frac{\partial \boldsymbol{\mathcal U}}{\partial t} +  {\bf \nabla} \cdot {\bf {\mathcal F}(\boldsymbol{\mathcal U})} = 0 \, ,
\end{equation}
where 
\begin{equation}
\boldsymbol{\mathcal U} = \left(\begin{array}{cc}  E_r \\ {\bf F_r} \end{array}\right) \, ,
\end{equation}
and
\begin{equation}
{\mathcal F}(\boldsymbol{\mathcal U}) = \left(\begin{array}{cc} {\bf F_r} \\ {\tilde c}^2 {\mathbb P_r}\end{array}\right) \, .
\end{equation}

A finite volume simulation code like {\sc Arepo} divides the computational domain into a
set of control volumes. The fluid's state is described by the cell averages of
the conserved quantities, which are obtained by integrating the primitive
quantities over the volume of the cell
\begin{equation}
\boldsymbol{\mathcal Q} = \int_V \boldsymbol{\mathcal U} \text{d}V \, .
\end{equation}
Using the Gauss' theorem we can estimate the change in these quantities with
time as
\begin{equation}
\frac{\partial \boldsymbol{\mathcal Q}}{\partial t} = - \int_{\partial V} [{\mathcal F}(\boldsymbol{\mathcal U}) - \boldsymbol{\mathcal U}{\bf w}^T]\text{d}{\bf A} \, ,
\label{eq:move}
\end{equation}
where ${\bf w}$ is the velocity of each point of the cell boundary. For
Eulerian schemes, the mesh is static (${{\bf w} = 0}$), while in a fully
Lagrangian approach, the surface would be allowed to move at every point with
the local flow velocity (${{\bf w} = {\bf v}}$). For moving-mesh codes like
\textsc{arepo}, it is not possible to follow the distortions of the shapes of
fluid volumes exactly in multi-dimensional flows and therefore the general
formula of Eq.~\ref{eq:move} is used.  Practically, this requires then to
solve the total flux as a combination of the flux over a static interface
(${{\mathcal F}(\boldsymbol{\mathcal U})}$) and an advection step owing to the
movement of the interface (${-\boldsymbol{\mathcal U}{\bf w}^T}$).

The hydro and magneto-hydrodynamic (MHD) schemes solve the respective equations in the reference frame of the
moving face. This ensures full Galilean invariance, a property that is of
significant importance for cosmological simulations where highly supersonic
bulk flows are common. Unfortunately, this approach is not possible for a
photon fluid, where Galilean invariance does not have any meaning. We instead choose to
modify the Riemann solution such that it takes the motion of the mesh into
account. 

We therefore write down the total flux on a moving mesh (${\mathcal F}^m$) as 
\begin{equation}
 \mathcal{F}^m =  \left(\begin{array}{cc} {\bf F_r} - E_r{\bf w}^{T}  \\ {\tilde c}^2 {\mathbb P_r} - {\bf F}_r {\bf w}^T \end{array}\right) \, ,
 \label{eq:fluxmove}
\end{equation}
where the velocity of the cell interface (${\bf w}$) is calculated using the method
outlined in \citealt{Springel2010} (Eqs. 32 and 33).  Godunov's approach
\citep{Godunov1959} is used to compute ${\mathcal F}^m$ and solve the
approximate Riemann problem normal to the interface. Since \textsc{arepo} uses
unstructured Voronoi meshes, the dimensionally operator split framework cannot
be applied. Rather the unsplit approach is used as described in
\citet{Springel2010}. In a nutshell, this involves defining the Riemann problem
normal to the cell face by rotating the relevant primitive variables into the
coordinate system defined by setting the $x-$axis normal to the cell face. Once
the flux has been calculated, it is transformed back to the lab frame. 

We employ the Harten-Lax-van Leer \citep{Harten1983} framework, that splits the
solution of the Riemann problem at each interface into three possible flux
estimates
\begin{equation}
 {\mathcal F}^m =  
\begin{cases}
 {\mathcal F}_L^m &\quad\text{if} \ \lambda^- \geqslant 0 \, , \\
  {\mathcal F}_{hll}^m &\quad\text{if} \ \lambda^- \leqslant 0 \leqslant \lambda^+ \, , \\
  {\mathcal F}_R^m &\quad\text{if} \ \lambda^+ \leqslant 0 \, ,
\end{cases}
\end{equation}
where
\begin{equation}
{\mathcal F}_{hll}^m = \frac{\lambda^+ {\mathcal F}_L^m - \lambda^- {\mathcal F}_R^m + \lambda^+ \lambda^- (\boldsymbol{\mathcal U}_R - \boldsymbol{\mathcal U}_L)}{\lambda^+ - \lambda^-} \, .
\label{eq:riemann}
\end{equation}
The subscripts `$L$' and `$R$' refer to the value of the variables
(${\boldsymbol{\mathcal U}, {\mathcal F}(\boldsymbol{\mathcal U})}$) on the left
and right states of the cell interface.  $\lambda^+$ and $\lambda^-$ are the
maximum or minimum eigenvalues of the of the Jacobian ${\partial
{\mathcal F}^m/ \partial \boldsymbol{\mathcal U}}$ defined as
\begin{equation}
\begin{split}
 \lambda^+ &= \text{max}(\lambda_{L}^{max}, \lambda_{R}^{max}) \, ,  \\
\lambda^- &= \text{min}(\lambda_{L}^{min}, \lambda_{R}^{min}) \, .
 \end{split}
\end{equation}
The eigenvalues represent the wave speeds of system of equations which in our
formulation are estimates for the lower and upper bounds of the signal
velocities. The eigenvalues of the system are obtained by solving for $\lambda$
in
\begin{equation}
 \left|  \frac{\partial {\mathcal F}^m}{\partial \boldsymbol{\mathcal U}} - \lambda_m \mathcal{I} \right| = 0 \, .
 \label{eq:eigen1}
\end{equation}
However, from  Eq.~\ref{eq:fluxmove}
\begin{equation}
 \frac{\partial {\mathcal F}^m}{\partial \boldsymbol{\mathcal U}} =  \frac{\partial {\mathcal F}^s}{\partial \boldsymbol{\mathcal U}} - w {\mathcal I} \, ,
 \label{eq:eigensplit}
\end{equation}
where ${w = {\bf w}\cdot\hat{\bf n}}$ is the component of the velocity of the
face along the face normal ($\hat{\bf n}$) and ${\mathcal F}^s$ is the total
flux on a static mesh. Substituting the value of ${\partial{\mathcal{F}^m}/
\partial\boldsymbol{\mathcal{U}}}$ from Eq.~\ref{eq:eigensplit} into
Eq.~\ref{eq:eigen1} we get
\begin{equation}
 \left|  \frac{\partial {\mathcal F}^s}{\partial \boldsymbol{\mathcal U}} - (\lambda_m+w) \mathcal{I} \right| = 0 \, .
 \label{eq:eigen2}
\end{equation}
Comparing Eqs.~\ref{eq:eigen1} and \ref{eq:eigen2}, gives ${\lambda_s =
\lambda_m+w}$, meaning that the eigenvalues of the system on a moving mesh
($\lambda_m$) are just a linear combination of the eigenvalues of the system on
static mesh ($\lambda_s$) and the face velocity ($w$),
\begin{equation}
\begin{split}
 \lambda^+ &= \lambda_m^+ = -w + \lambda_s^+ \, , \\ 
  \lambda^- &= \lambda_m^- = -w + \lambda_s^-  \, .
 \end{split}
\end{equation}
This is equivalent to rotating the eigenvectors by an angle ${x/t=-w}$. We note that if the face velocity is superluminal or greater than the largest signal speed of the static system, then the Riemann solver chooses to purely advect the fluxes in an upwind manner, analogous to supersonic fluid flow in non-relativistic hydrodynamics. 

The eigenvalues of the Jacobian matrix are determined by interpolating the
tabulated values obtained by \citet{Gonzalez2007}.\footnote{We obtain the table
from the public version of the {\sc Ramses-rt} code}. By setting ${\lambda_s^+ =
-\lambda_s^- = {\tilde c}}$ we obtain the Rusanov \citep{Rusanov1961} flux
function, which is also implemented (also known as the Global-Lax-Friedrichs or
GLF flux function). 

A conservative time integration of Eq.~\ref{eq:move} is obtained by using the method outlined in  \citet[Eqs. 19-22]{Pakmor2016}, which
employs Heun's method, a variant of the second order Runge-Kutta scheme. The
fluxes are computed as an average of fluxes at the beginning and end of the
timestep. The mesh geometry of the second half of the current timestep is used
for the first half of the next timestep, essentially requiring only one mesh
construction per timestep. Therefore the update of the conservative variables are given by
\begin{equation}
\boldsymbol{\mathcal Q}_i^{n+1} = \boldsymbol{\mathcal Q}_i^n - \frac{\Delta t}{2} \left(\sum_j A_{ij}^n \, {\bf \hat{F}}_{ij}^{n}(\boldsymbol{\mathcal U}^n) +   A_{ij}^{'} \, {\bf \hat{F}}_{ij}^{'}(\boldsymbol{\mathcal U}^{'})\, \right) .
\end{equation}
This makes the scheme easily compatible
with the hierarchical timestepping scheme used in \textsc{arepo} (see
\citealt{Pakmor2016} for more details).

A higher order accuracy is obtained  by replacing the
piecewise constant approximation of Godunov's scheme with a slope-limited piece-wise
linear spatial extrapolation and a first order prediction forward in
time to obtain the states of the primitive variables on both sides of the
interface \citep{vanLeer1979}
\begin{equation}
\boldsymbol{\mathcal U}_{L,R}' = \boldsymbol{\mathcal U}_{L,R} + \left. \frac{\partial \boldsymbol{\mathcal U}}{\partial \boldsymbol{r}} \right \vert_{L,R}  ({\bf f} - {\bf s}_{L,R}) + \left. \frac{\partial \boldsymbol{\mathcal U}}{\partial t} \right \vert_{L,R} \Delta t,
\label{eq:extpr}
\end{equation}
where ${\bf s}$ is the center of mass of the cell and  ${\bf f}$ is the center
of the cell face. The time derivatives of the primitive variables (${\partial
\boldsymbol{\mathcal U}/\partial t}$ in Eq.~\ref{eq:extpr}) are expressed in
terms of their spatial derivatives using Eqs.~\ref{eq:consE} and\ref{eq:consF}
\begin{equation}
\frac{\partial E_r}{\partial t}  = - {\bf \nabla} \cdot {\bf F_r}\, ,
\label{eq:timeE}
\end{equation}
\begin{equation}
\frac{\partial {\bf F_r}}{\partial t} = - {\tilde c}^2 {\bf \nabla} \cdot {\mathbb P_r}\, .
\label{eq:timeF}
\end{equation}

We use the local least square fit (LSF) method described in \citet{Pakmor2016}
to obtain the gradient estimates (${\partial \boldsymbol{\mathcal U}/\partial
\boldsymbol{r}}$ in Eq.~\ref{eq:extpr}). They are constructed such that they
reproduce the cell centered values of the neighboring cells as well as possible. If
$\phi_i$ is the primitive variable defined at the center of cell `$i$', and
$\phi_j$'s are the values of the primitive variable for all neighboring cells
`$j$', then the best linear approximation of ${\left <{\bf \nabla}\phi \right
>_i}$ is obtained by minimising the sum of the deviations for all neighbors
>($S_{\text {tot}}$)
\begin{equation}
S_{\text {tot}} = \sum_j \frac{A_{ij}}{|{\bf s_j} - {\bf s_i}|^2} (\phi_j - \phi_i - \left <{\bf \nabla}\phi \right >_i ({\bf s_j} - {\bf s_i}))^2 \, ,
\end{equation}
where $A_{ij}$ is the area of the interface between cells `i' and `j'.  The
monotonicity of the gradients is imposed by requiring that the linearly
reconstructed quantities on the face centroids are bounded by the maxima and
minima of all the cell centered values of the neighbouring cells \citep[Eqs.
28-30]{Springel2010}. This general gradient estimate retains the necessary
accuracy even for large offsets between the mesh-generating points and the
center of mass of cells.

Ordinarily the spatial and time extrapolations of the primitive variables are
carried out independent of each other. Unfortunately, these independent
estimates cannot ensure that the reduced flux at the interface (${f_{L,R} =
(|{\bf F}_{r_{L,R}}|/{\tilde c}E_{r_{L,R}})}$) remains bounded between $[0, 1]$.
We can enforce this condition by limiting the value of the extrapolated photon
flux as ${{\tilde {\bf F}}_{r_{L,R}} = \alpha {\bf F}_{r_{L,R}}}$ where ${\alpha =
\text{min}(1, {\tilde c}E_{r_{L,R}}/|{\bf F}_{r_{L,R}}|)}$. However, this form
of a limiter introduces too much noise in the solution and degrades
the convergence of the code in addition to increasing the diffusivity.

Instead, the extrapolation of ${\bf F}_r$ is made dependent on value of $E_r$ and
the reduced flux.  Accordingly, the gradients of $E_r$, ${{\bf F}_\text{N} =
{\bf F}_r/E_r}$  and ${|{\bf F}_\text{N}|}$ are calculated instead of just $E_r$
and ${\bf F}_r$. First, the time prediction step of the photon
energy density (Eq.~\ref{eq:timeE}) is performed\footnote{For the calculation of the fluxes between two
cells on different time steps, the time extrapolation is done for each cell always from the last time the
cell was active.}
\begin{equation}
E_r^{'} = E_r - {\bf \nabla} \cdot {\bf F}_r  \, \Delta t \, ,
\end{equation}
where 
\begin{equation}
\begin{split}
{\bf \nabla} \cdot {\bf F}_r &= \text{Tr}({\bf \nabla}{\bf F}_r)\\
&= \text{Tr}\left(E_r \, {\bf \nabla} {\bf F}_\text{N} + \frac{1}{E_r} {\bf F}_r \otimes {\bf \nabla}E_r\right) \, .
\end{split}
\end{equation}
The spatial extrapolation is then carried out by extrapolating the photon
energy density using the estimate of the gradient of $E_r$
\begin{equation}
 E_{r_{L,R}}' = E_r^{'} + \nabla E_r \cdot ({\bf f} - {\bf s}_{L,R}) \, ,
\end{equation}
and then we extrapolate
\begin{equation}
|{\bf F}_{\text{N}_{L,R}}| = |{\bf F}_\text{N}| + {\bf \nabla} |{\bf F}_\text{N}| \cdot ({\bf f} - {\bf s}_{L,R})\, .
\end{equation}
Finally, we impose 
\begin{equation}
{\bf F}_{r_{L,R}}' = \psi {\bf F}_r\, ,
\end{equation}
where
\begin{equation}
\psi = \frac{|{\bf F}_{\text{N}_{L,R}}|}{|{\bf F}_r |} E_{r_{L,R}}' \, .
\end{equation}

We take advantage of the fact that the gradients are limited such that they are
bounded by the maximum and minimum cell centered values of the neighbouring
cells. This method of extrapolating the photon flux ensures that the reduced
flux at the interface ($f_{L,R}$) is always within $[0,1]$ without affecting
the directionality of the underlying photon field. Additionally, the photon
energy density and the reduced flux at the interface is not slope limited in
any way except for the condition that it does not create local maxima or
minima. This ensures stability without adding too much diffusion into the
scheme.  We note that the prediction step is not performed
for the photon flux (${\bf F}_r$) because, it introduces too much noise into
the solution, in addition to making it difficult to preserve
the directionality of photon field.

Since the transport equations are solved using an explicit scheme, the timestep
of each cell is constrained by the von Neumann stability condition
\begin{equation}
 \Delta t_\ts{RT} \leq \eta \frac{\Delta x}{\tilde{c} + \left|{\bf v}_{c}\right|} \, ,
 \label{eq:timestep}
\end{equation}
where ${\Delta x}$ is the cell width, ${\bf v}_c$ is the velocity of the cell in the lab frame and ${\eta \sim 0.3}$. Since the RT scheme is coupled to hydrodynamics and gravity the final timestep will be
\begin{equation}
 \Delta t = \text{min}(\Delta t_\ts{RT}, \Delta t_\ts{hydro}, \Delta t_\ts{grav}) \, .
\end{equation}
The high speed of light demands very small ${\Delta t_\text{RT}}$ forcing other
computationally expensive parts of the code, such as, mesh construction and
gravity force calculations, to be called more often than actually required. In
many physical problems this can be overcome by using the reduced speed of light
approximation (RSLA), which is applicable for systems where the characteristic
velocity is much smaller than the speed of light. For large scale cosmological
simulations, this is no longer true. For example, to track the ionization fronts (I-fronts)
properly in the IGM one must use the full speed of light \citep{Rosdahl2013, Bauer2015}.  Implicit/semi-implicit time integration schemes can be used to
overcome this problem. Unfortunately, scalable implementation of these
schemes is quite involved as it requires inverting large sparse matrices
\citep[e.g.][]{Kannan2016b}. The other option is to sub-cycle the RT steps
\citep{Commercon2014}, i.e., perform N${_{sub} \geq 1}$ number of RT steps per
hydrodynamical step. This effectively reduces the frequency with which the time
consuming routines are called. We chose to implement the latter and the details
of this method are described in Appendix~\ref{sec:subcycle}. 

\subsection{The thermochemistry and cooling equations}
In this section we describe the numerical methods to evaluate the source terms
of the radiative transfer equations (R.H.S. of Eqs.~\ref{eq:E} and \ref{eq:F}). 

\subsubsection{Ultra-Violet (UV) thermochemistry, photoheating and radiation pressure}
\label{sec:chemistry}
We first look at the single scattering regime, where a particular photon
interacts with the surrounding medium only once i.e, it is absorbed only once,
which destroys the photon, and there is little scattering. This occurs mainly
in the UV regime, where the photons have enough energy to ionize atoms such
as Hydrogen (H) and Helium (He). Since the ionization potential varies quite
considerably, an accurate treatment of UV radiative transfer requires multiple frequency bins centered around the energy of each
ionization state of the gas. 

In the single scattering regime it is easier to work with photon number
densities instead of the radiation energy densities ($E_r$). Therefore, we
define the photon number density $N_\gamma^i$, photon number flux (${\bf
F}_\gamma^i$) and the associated pressure tensor (${\mathbb P}_\gamma^i$)  in
each frequency bin `i' as
\begin{equation}
\{{\tilde c}N_\gamma^i, {\bf F}_\gamma^i, {\mathbb P}_\gamma^i\} = \int_{\nu_{i1}}^{\nu_{i2}} \frac{1}{h\nu} \ d\nu \ \int_{4\pi}\{1, {\bf n}, \  ({\bf n}\otimes{\bf n})\}I_\nu  \ \text{d}\Omega \, ,
\end{equation}
where ${\nu_{i1} \le \nu_i < \nu_{i2}}$.

Accordingly, Eqs.~\ref{eq:E} and \ref{eq:F} can be reformulated as 
\begin{equation}
\begin{split}
\frac{\partial N_\gamma^i}{\partial t} + {\bf \nabla} \cdot {\bf F}_\gamma^i &= - \sum_j {\tilde c} \, n_j \, N_\gamma^i \, {\bar \sigma_{ij}} - \kappa_i \, \rho \, \tilde{c} \, N_\gamma^i \\ &+ \sum_j s_{ij} \, ,
\end{split}
\label{eq:UVE}
\end{equation}
\begin{equation}
\begin{split}
&\frac{\partial {\bf F}_\gamma^i}{\partial t} + {\tilde c}^2 {\bf \nabla} \cdot {\mathbb P_\gamma^i} =  - \sum_j {\tilde c} \, n_j \, {\bf F}_\gamma^i \, {\bar \sigma_{ij}} - \kappa_i \, \rho \, \tilde{c} \, {\bf F}_\gamma^i \, ,
\end{split}
\label{eq:UVF}
\end{equation}
where $n_j$ is the number density of a particular ionic species, which in our
case consists of ${j \in \{\hi, \hei, \heii \}}$. $\kappa_i$ is the dust opacity
of photon group `i' and ${\bar {\sigma}}_{ij}$ is the mean ionization cross
section of species `j' in the frequency bin `i'
\begin{equation}
{\bar \sigma_{ij}} = \frac{\displaystyle\int_{\nu_{i1}}^{\nu_{i2}} \frac{4\pi J_\nu}{h\nu} \sigma_{j_\nu} \, \text{d}\nu}{\displaystyle\int_{\nu_{i1}}^{\nu_{i2}} \displaystyle\frac{4\pi J_\nu}{h\nu} \, \text{d}\nu}\, ,
\end{equation}
where
\begin{equation}
J_\nu = \frac{1}{4\pi}\int_{4\pi} I_\nu \, \text{d}\Omega\, .
\end{equation}
Finally, $s_{ij}$ is the source term that accounts for recombination radiation.
Most of the previous RT implementations employ the On-The-Spot-Approximation
(OTSA), which assumes that any radiation emitted by recombinations is
immediately absorbed in the surroundings also called case B recombinations.
This is a good approximation in optically thick media but is not valid in
optically thin environments where the case A recombination rates are more
relevant. Here we include the option of not applying OTSA and define
 \begin{equation}
s_{ij} =  
\begin{cases}
 0 &\quad\text{if} \ \text{OTSA} \\
 \sum_j \delta_{ij} \, (\alpha_j^A - \alpha_j^B ) \, n_j \, n_e &\quad \text{Otherwise} \, ,
\end{cases}
\end{equation}
where $\delta_{ij}$ is unity if the recombination radiation from species `j'
emits into the frequency bin `i', else it is set to zero and $\alpha_j^A$ and
$\alpha_j^B$ are the case A and case B recombination rates . 

As commonly done we use the operator split approach to solve these equations. First
the transport equations (setting R.H.S. of Eqs.~\ref{eq:UVE} and \ref{eq:UVF} to zero)
are solved as described in \ref{sec:transport}. Then the thermochemistry
equations are solved, which involves solving the equation for the change in the
photon number density and photon number flux
\begin{equation}
\frac{\partial N_\gamma^i}{\partial t} = -{\tilde c} \, N_\gamma^i \left(\sum_j n_j \, {\bar \sigma}_{ij} + \kappa_i \, \rho \right) + \sum_j s_{ij} \, ,
\label{eq:uvchemN}
\end{equation}
\begin{equation}
\frac{\partial {\bf F}_\gamma^i}{\partial t} = -{\tilde c} \, {\bf F}_\gamma^i \left( \sum_j n_j \, {\bar \sigma}_{ij} + \kappa_i \, \rho \right) \, ,
\label{eq:uvchemF}
\end{equation}
which are coupled with the equations which govern the number density evolution
of the ionic species
\begin{equation}
\frac{\text{d} n_\hii}{\text{d}t} = -\alpha_\hii \, n_\hii \, n_e + \sigma_{e\hi} \, n_e \, n_\hi + {\tilde c} \, n_\hi \sum_i {\bar \sigma}_{i\hi} \, N_\gamma^i \, ,
\label{eq:uvchemnhii}
\end{equation}
\begin{equation}
\begin{split}
\frac{\text{d} n_\heii}{\text{d}t} &= \alpha_\heiii \, n_\heiii \, n_e +  \sigma_{e\hei} \, n_e \, n_\hei \\ &+ {\tilde c} \, n_\hei \sum_i {\bar \sigma}_{i\hei} \, N_\gamma^i -\alpha_\heii \, n_\heii \, n_e \\ & - \sigma_{e\heii} \, n_e \, n_\heii - {\tilde c} \, n_\heii \sum_i {\bar \sigma}_{i\heii} \, N_\gamma^i   ,
\end{split}
\label{eq:uvchemnheii}
\end{equation}
\begin{equation}
\begin{split}
\frac{\text{d} n_\heiii}{\text{d}t} &= -\alpha_\ion{He}{III} \, n_\heiii \, n_e + \sigma_{e\heii} \, n_e \, n_\heii \\  &+ {\tilde c} \, n_\heii \sum_i {\bar \sigma}_{i\heii} \, N_\gamma^i ,
\end{split}
\label{eq:uvchemnheiii}
\end{equation}
where $\sigma_{ej}$ and $\alpha_j$ are the collisional ionization and
recombination rates of the ionic species `j'.  These equations are supplemented
with the following closure relations:
\begin{equation}
  n_\text{H} = n_\hi + n_\hii \, ,
 \label{eq:uvchemh}
\end{equation}
\begin{equation}
  n_\text{He} = n_\hei + n_\heii + n_\heiii \, ,
 \label{eq:uvchemhe}
\end{equation}
\begin{equation}
  n_e = n_\hii +  n_\heii + 2n_\heiii \, .
 \label{eq:uvchemend}
\end{equation}
We note that case B recombination rates are used if OTSA is applied otherwise,
case A recombination rates are used. Finally, the dust reprocessed optical/UV
radiation is added onto the IR energy density
\begin{equation}
 \frac{\partial E_\ts{IR}}{\partial t} = \sum_{i}^{\notin IR} \kappa_i \, \rho \, {\tilde c} \, e_i \, N_\gamma^i \, ,
\end{equation}
where, $e_i$ is the mean energy per photon of the frequency group `i' defined
as
\begin{equation}
 e_i = \frac{\displaystyle\int_{\nu_{i1}}^{\nu_{i2}} 4 \pi J_\nu \text{d} \nu}{\displaystyle\int_{\nu_{i1}}^{\nu_{i2}} \frac{4 \pi J_\nu}{h\nu} \text{d} \nu} \, ,
\end{equation}
and $E_\ts{IR}$ is the energy density of the Infra-red (IR) radiation field.

In addition to changing the ionization state of the gas, photons can also
deposit energy through photoheating. Quantitatively, the photoheating rate
($\Gamma$) of a given ionic species is given as 
\begin{equation}
 \Gamma_j = \int_{\nu_{tj}}^{\infty} \frac{4\pi J_\nu}{h\nu} \, \sigma_{j\nu} \, (h\nu - h\nu_{tj}) \, \text{d}\nu \  ,
 \end{equation}
where $h$ is the Planck constant and ${{\mathcal E}_j = h\nu_{tj}}$ is the
ionization potential of the ionic species `j'.  In order to be compatible with
a multi-frequency approach, we need to discretize this equation into finite
frequency bins. We do so by splitting the above integral as follows
\begin{equation}
 \Gamma_j = \sum_i \int_{\nu_{i1}}^{\nu_{i2}} \frac{4\pi J_\nu}{h\nu} \, \sigma_{j_\nu} \, (h\nu - h\nu_{tj}) \, \text{d}\nu \,  ,
 \end{equation}
which can in turn be written down as
\begin{equation}
 \Gamma_j =  {\tilde c} \sum_i  N_\gamma^i \, \bar{\sigma}_{ij} \, \epsilon_{ij} \, ,
 \end{equation}
where the photoheating rate in the frequency bin `$i$' due to the ionization of
species `$j$' is defined as
 \begin{equation}
\epsilon_{ij} = \frac{\displaystyle\int_{\nu_{i1}}^{\nu_{i2}} \frac{4\pi J_\nu}{h\nu} \, \sigma_{j_\nu} \, (h\nu-h\nu_{tj}) \, \text{d}\nu}{\displaystyle\int_{\nu_{i1}}^{\nu_{i2}} \frac{4\pi J_\nu}{h\nu} \, \sigma_{j_\nu} \, \text{d}\nu} \, .
\end{equation}
Therefore the total amount of energy deposited into the gas through
photoheating  (${\mathcal H}$) is then 
\begin{equation}
{\mathcal H} = \sum_j n_j \Gamma_j.
\label{eq:UVheating}
\end{equation}

Finally, the momentum injection rate by photon absorption is discretised in the
same way and is given by
\begin{equation}
\frac{\partial \rho v}{\partial t} =  \frac{1}{c} \sum_i {\bf F}_\gamma^i \, \left(\sum_j \, n_j \, \bar{\sigma}_{ij} \, p_{ij}  + \kappa_i \, \rho \, e_i\right)\, ,
\label{eq:uvmomentum}
\end{equation}
where
\begin{equation}
 p_{ij} = \frac{\displaystyle\int_{\nu_{i1}}^{\nu_{i2}} 4\pi J_\nu \, \sigma_{j_\nu} \, \text{d}\nu}{\displaystyle\int_{\nu_{i1}}^{\nu_{i2}} \frac{4\pi J_\nu}{h\nu} \, \sigma_{j_\nu} \, \text{d}\nu} \, .
\end{equation}

 The numerical integration of this thermo-chemical network is quite challenging as
small changes in the photon density can lead to rapid changes in the ionization
state and temperature of the gas. Therefore, an explicit time integration of these equations
would require very small time steps making the thermochemistry step
computationally expensive. In order to overcome this problem, we use a variant of the  method
outlined in \citet{Petkova2009}, which employs a semi-implicit approach, that
first solves for the number density evolution of the ionic species implicitly
using the values of the $n_e$ and $N_\gamma^i$ from the previous timestep.
$n_e$ is then updated with the revised values of the number density of the
ionic species according to Eq.~\ref{eq:uvchemend}. The change in temperature of the gas is then
calculated explicitly using the updated abundances of the ionic species (a detailed description of this scheme is discussed in
Appendix~\ref{sec:uvchem}).

If the temperature or one of the abundances changes by more than $10\%$ during a timestep, then we resort
to using the publicly available ordinary differential equation solver {\sc SUNDIALS} {\sc CVODE} \citep{hindmarsh2005} 
which employs a variable order, variable step, multistep backward differencing scheme to compute the temperature and chemical abundances. This  
scheme is stable enough to allow us to set the thermochemistry timestep to the transport timestep given by
Eq.~\ref{eq:timestep}.

\subsubsection{Infrared(IR) dust-gas coupling}
\label{sec:IR}
Dust grains are solid, macroscopic particles composed of dielectric and
refractory materials.  Many of the physical details are empirical as we do not
yet know the precise composition of dust grains, nor do we know their precise
physical properties. They do however, play a major role in the physics of the
interstellar medium (ISM). Although, they only make up about $1\%$ of the ISM
\citep{Gilmore1989, Zubko2004}, they absorb and reprocess almost $50\%$ of the
starlight in the galaxy \citep{Battisti2016}. The surface of dust grains host a
variety of chemical reactions that changes the chemical composition of the ISM
significantly, which in turn leads to changes in the star formation rate. They
are also thought be in radiative equilibrium with the local IR radiation field
\citep{Krumholz2012}. 
 
Modelling the coupling between IR radiation, gas and dust can be quite
challenging. Ideally, this requires treating the dust as a separate fluid and
accurately accounting for the energy and momentum exchange between gas, dust
and the IR radiation field, which can get rather complicated. We
instead choose to couple the IR RT scheme to the semi-empirical dust model
of \citet{McKinnon2016}. This model accounts for the stellar production of
dust, accretion of gas-phase metals onto existing grains, destruction of dust
through  local  supernova  activity, thermal sputtering of dust,  and  dust
driven  by  winds  from  star-forming regions. It reproduces the dust content
in low redshift galaxies \citep{McKinnon2017}. 

In this model, the dust is treated as a passive scalar, whose motion is coupled
to the gas, which is a good approximation for short stopping timescales found
in the ISM. We further assume that the system is close to local thermodynamic
equilibrium, which is good approximation for cold high density regions of the
ISM \citep{Goldsmith2001, Krumholz2013}. Under these conditions the gas emits
as a black body. If the IR photon frequency bin covers a sufficiently large
range, then the source function can be approximated by the frequency integral
of a Planck spectrum
 \begin{equation}
  S_\text{IR} = \kappa_\ts{P} \, \rho \, c \, a \, T^4 \ ,
 \end{equation}
where $a$ is the radiation constant, $\kappa_\ts{P}$ is the Planck mean opacity and
$T$ is the temperature of the gas. If the IR SED is dominated by reprocessed
radiation, then it can be approximated to a Planckian, implying that $\kappa_\ts{E}
= \kappa_\ts{P}$. Additionally, for a system with a large IR optical depth, we can
assume that the flux weighted mean opacity is similar to the Rosseland mean
opacity ${\kappa_\ts{F} \sim \kappa_\ts{R}}$. Using these assumptions, we can then write
Eqs.~\ref{eq:E} and \ref{eq:F} as
\begin{equation}
\begin{split}
&\frac{\partial E_\ts{IR}}{\partial t} + {\bf \nabla} \cdot {\bf F_\ts{IR}} = \kappa_\ts{P} \, \rho \, \left(c \, a  \, T^4 - {\tilde c} \, E_\ts{IR}\right) \ ,
\end{split}
\label{eq:eir}
\end{equation}
\begin{equation}
\begin{split}
&\frac{\partial {\bf F_\ts{IR}}}{\partial t} + {\tilde c}^2 {\bf \nabla} \cdot {\mathbb P_\ts{IR}} =  - \kappa_\ts{R} \, \rho \, {\tilde c} \, {\bf F_\ts{IR}} \ .
\end{split}
\label{eq:fir}
\end{equation}
The dust opacities (for both IR and UV radiation bins) are calculated
self-consistently based on the empirical dust model (see
Appendix~\ref{sec:dust} for more details). The destruction of dust by the UV
radiation field is not included as it is sub dominant compared to other dust
destruction mechanisms such as SNe and thermal sputtering and is only important
in highly luminous systems with an extremely hard spectra such as Gamma ray
bursts \citep{Waxman2000, Draine2002}. 

Using the usual operator split approach, we first solve the pure transport
equations (setting the R.H.S. of Eqs.~\ref{eq:eir}~and~\ref{eq:fir} to zero) using the
algorithm described in Section~\ref{sec:transport}. The source terms are then 
\begin{equation}
\frac{\partial E_\ts{IR}}{\partial t} = \kappa_\ts{P} \, \rho \, \left(c \, a \, T^4 - {\tilde c} \, E_\ts{IR}\right) \, ,
\label{eq:eirsource}
\end{equation}
\begin{equation}
\frac{\partial {\bf F_\ts{IR}}}{\partial t} = - \kappa_\ts{R} \,  \rho  \, {\tilde c} \, {\bf F_\ts{IR}} \, ,
\end{equation}
and the change in the internal energy ($u$) and momentum of the gas are given
by
\begin{equation}
 \frac{\partial u}{\partial t} = - \kappa_\ts{P} \, \rho \, \left(c \, a \, T^4 - {\tilde c} \, E_\ts{IR}\right) \, ,
 \label{eq:uir}
\end{equation}
\begin{equation}
 \frac{\partial \rho v}{\partial t} =  \frac{\kappa_R \, \rho}{c} {\bf F}_\text{IR}\, .
 \label{eq:pir}
\end{equation}

Eqs.~\ref{eq:eirsource}~and~\ref{eq:uir} form a set of coupled equations. The
fourth power dependence on the temperature makes the equations stiff requiring
very small timesteps to maintain stability of the solution. We use a variant of
the approach presented in \citet[][R15]{Rosdahl2015} to solve these equations. First
we get a semi-implicit estimate of the change in ${{\bf U}_\ts{E} \in
(E_\ts{IR}, u)}$ 
\begin{equation}
 \Delta {\bf U}_\ts{E} = {\bf \dot{U}}_\ts{E} \Delta t ({\bf I} - {\bf J} \Delta t)^{-1} \ ,
\end{equation}
where $\Delta t$ is the timestep and ${{\bf J} = \frac{\partial {\bf
{\dot{U}}}_\ts{E}}{\partial {\bf {U}}_\ts{E}}}$ is the Jacobian matrix.  Using
the symmetry of the problem (${\Delta E_\ts{IR} = -\Delta u}$) the change over
timestep $\Delta t$ can be written as
\begin{equation}
 \Delta E_\ts{IR} = -\Delta u = \frac{c\,a\,T^4 - {\tilde c} \,E_\ts{IR}}{(\kappa_\ts{P} \,\rho \,\Delta t)^{-1} + {\tilde c} + 4\,c\,a\,T^3 C_v^{-1}} \ ,
\end{equation}
where ${C_v = k_B/((\gamma -1)\mu)}$ is the specific heat at constant volume,
$\gamma$ is the adiabatic index and $\mu$ is the mean molecular weight. If the
relative change in both $E_\ts{IR}$ and $u$ is less than $10\%$ we keep this
estimate for the solution.  Otherwise, we discard it and switch to solving these equations using the {\sc SUNDIALS} {\sc CVODE} library as mentioned in the previous section. This method is
quite stable and avoids large iterative loops even when sudden changes in the
photon energy density occur.

\section{Test Problems}
\label{sec:tests}
This section presents various test problems of our radiation hydrodynamic
implementation. The relevant hydrodynamic equations are evolved using the
moving mesh finite volume scheme outlined in \citet{Springel2010}, with the
improved time integration and gradient estimation techniques described in
\citet{Pakmor2016}. 

In Section~\ref{sec:adv} we test the accuracy and the convergence order of our scheme.  Sections~\ref{sec:stromgren}~and~\ref{sec:multstrom} discuss a test for the ionizing
chemistry scheme (Section~\ref{sec:chemistry}).  Section~\ref{sec:shadow} tests
the UV chemistry and cooling implementation and the ability of the scheme to
capture and maintain the directionality of the  underlying photon field.  The
simulations presented in Sections~\ref{sec:h2exp}~and~\ref{sec:radpress} are
evolved using the full RHD equations and test the coupling between the photons
and the gas. Section~\ref{sec:disc} investigates how the M$1$ closure relation
performs in a multisource setup.  Sections~\ref{sec:dustabs}~and~\ref{sec:diff}
analyse the performance of the multiscattered IR+coupled dust implementation
(Section~\ref{sec:IR}), specifically, the ability of our scheme to capture
accurate results in optically thick media. Finally,
Section~\ref{sec:lev} tests the temperature coupling multi-scattering, coupling
to the hydrodynamics module and explores the competition between gravity and
radiation pressure. The simulations in Sections~\ref{sec:adv},
\ref{sec:stromgren}, \ref{sec:multstrom}, \ref{sec:shadow}, \ref{sec:disc},
\ref{sec:dustabs} and \ref{sec:diff} are run without hydrodynamics and are
intended to test the accuracy and the stability of the RT implementation only.
In Sections \ref{sec:h2exp}, \ref{sec:radpress} and \ref{sec:lev} we
additionally test the accuracy of the our scheme when hydrodynamics andionizing
radiative transfer are coupled. In full RHD tests, we initially start with a
regular Cartesian grid, which is then allowed to move and adapt to the local
fluid motion \citep{Springel2010}. In addition, the mesh is regularised where
needed using the scheme outlined in \citet{Vogelsberger2012}.

In many of these tests we also vary the numerical scheme of the basic radiation
transport equations. We explore the differences between the flux functions used
to solve the Riemann problem at the cell interfaces, specifically the
differences between the Global-Lax-Friedrichs (GLF) and the Harten Lax-van Leer
(HLL) flux functions.  The HLL flux function uses accurately calculated
eigenvalues that represent the wave speeds of system of transport equations.
These depend highly on the reduced flux and the angle between the photon flux
and the cell interface. The GLF flux function on the other hand uses a single
wave speed set equal to the reduced speed of light in our calculations. So
theoretically, the HLL flux function is less diffusive and better maintains the
directionality of the photon field. Additionally, we also compare the accuracy
and stability  of the piecewise constant (PC) and the linear
gradient extrapolation methods. The PC scheme assumes that the primitive
variables do not vary within a volume element and therefore the input values
for the Riemann solver are the cell centered values. On the other hand, the
gradient extrapolation scheme linearly extrapolates the cell centered values
onto the face of the cell and then solve the Riemann problem with these
extrapolated values (see Section~\ref{sec:transport} for more information). In
summary, we test the following four numerical transport schemes:
\begin{itemize}
\item {\bf PC-GLF:} piecewise constant approximation with GLF flux function.
\item {\bf PC-HLL:} piecewise constant approximation with HLL flux function.
\item {\bf GLF:} linear extrapolation with GLF flux function.
\item{\bf HLL:} linear extrapolation scheme with HLL flux function.
\end{itemize}
We note that our fiducial scheme is the HLL scheme. 

Finally, we also explore the performace of our code on different mesh configurations. We mainly use three different mesh geometries, namely a
regular Cartesian mesh, a regular staggered mesh constructed by two
Cartesian meshes that are displaced from each other by $0.45$ ${\Delta x}$ in
every direction, where ${\Delta x}$ is the cell size for the given resolution and an irregular mesh, which is obtained from the Cartesian mesh by adding a random offset of up to $0.2$
${\Delta x}$, mimicking the typical maximum deviation between
mesh-generating points and cell centers in real problems
\citep{Vogelsberger2012}. Finally, please note that we use the labels `regular
mesh' and `staggered mesh' interchangeably.

\subsection{Radiation wave propagation}
\label{sec:adv}

\begin{figure}
\includegraphics[width=\columnwidth]{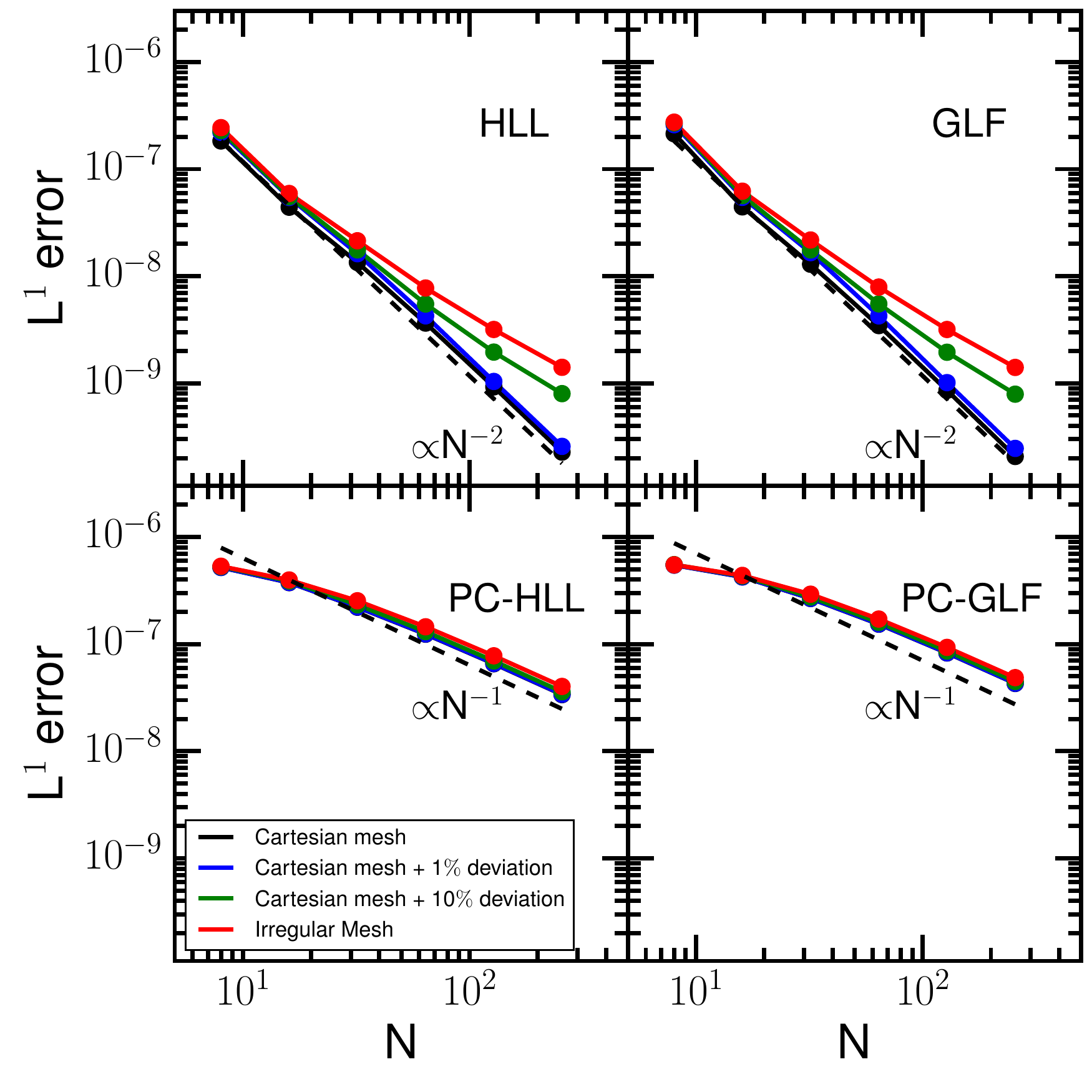}
\caption{ {\bf Radiation wave propagation}: L$^1$ error, at ${t= 1}$, as a function of the number of resolution elements ($N$) for different numerical schemes. While the piecewise constant approximation converges very slowly with a slope of $ \lesssim 1.0$, our fiducial scheme converges much faster with a slope of ${\sim 2.0}$, irrespective of the flux function used. }
\label{fig:convergence}
\end{figure}

 We start with a test to gauge the accuracy of our RT scheme. We investigate  the propagation of small-amplitude,
free-streaming  radiation  wave  in  a  purely  absorbing,  homogeneous medium with low optical depth, like the test described in \citet{Gardiner2005} and \citet{Skinner2013}. A
2D box of sidelength ${L_\ts{box} \{x,y\} = \{2,1\} }$ is initialised with a photon number
density as follows
\begin{equation}
 E(r) = E_{bg}  + \epsilon \sin\left(\frac{2 \pi (x+2y)/\sqrt{5}}{\lambda}\right),
\end{equation}
where $E_{bg}$ is a uniform background photon density field, $\epsilon=10^{-6}$ and $\lambda = 2/\sqrt{5}$. The radiation flux points in the direction of the wave and has a value ${|{\bf F}_r| = c E_r}$,
with ${c=1}$. The optical depth
per wavelength is set to $\tau_\lambda = \rho \kappa \lambda = 0.1$
and periodic boundary
conditions are used everywhere.

We employ four different background grids a regular Cartesian grid, a cartseian grid with $1\%$ deviation between the mesh generating point and the mesh centroid, Cartesian grid with $10\%$ deviation and finally a Cartesian grid with $20\%$ deviation, which we denote as a irregular mesh as it mimics the typical maximum deviation between
mesh-generating points and cell centers in real problems. We start the simulation at time ${t=0}$ and evolve the system until ${t=0.89}$ so that there is
one complete wave period in each of the $x-$ and $y-$ directions. The solution at $t=1$ is then given by
\begin{equation}
 E(r) = E_{bg}  + \epsilon \ e^{-\rho \kappa t} \sin\left(\frac{2 \pi (x+2y)/\sqrt{5}}{\lambda}\right) \ .
\end{equation}

The $L^1$ norm measured for different resolutions for the PC and fiducial
schemes is shown in Fig.~\ref{fig:convergence}. The PC-GLF and PC-HLL
schemes converge rather slowly with a convergence order of $\lesssim 1.0$. The fiducial
schemes on the other hand start out with much smaller errors and converge
significantly faster towards the analytic solution with an order of ${\sim 2.0}$.
There is essentially no dependence of the accuracy of the solution on the type of the flux function used. The convergence order decreases as we move to more distorted meshes,  especially at higher resolutions, which shows the importance of using mesh regularization schemes in {\sc Arepo}.

These results illustrate the advantages of our algorithm. It is less diffusive and highly accurate even at
relatively low resolutions.  A gradient extrapolated predictor-corrector scheme coupled to a Strang split approach for sources terms is necessary to achieve second order
accuracy.

\begin{figure}
\includegraphics[width=\columnwidth]{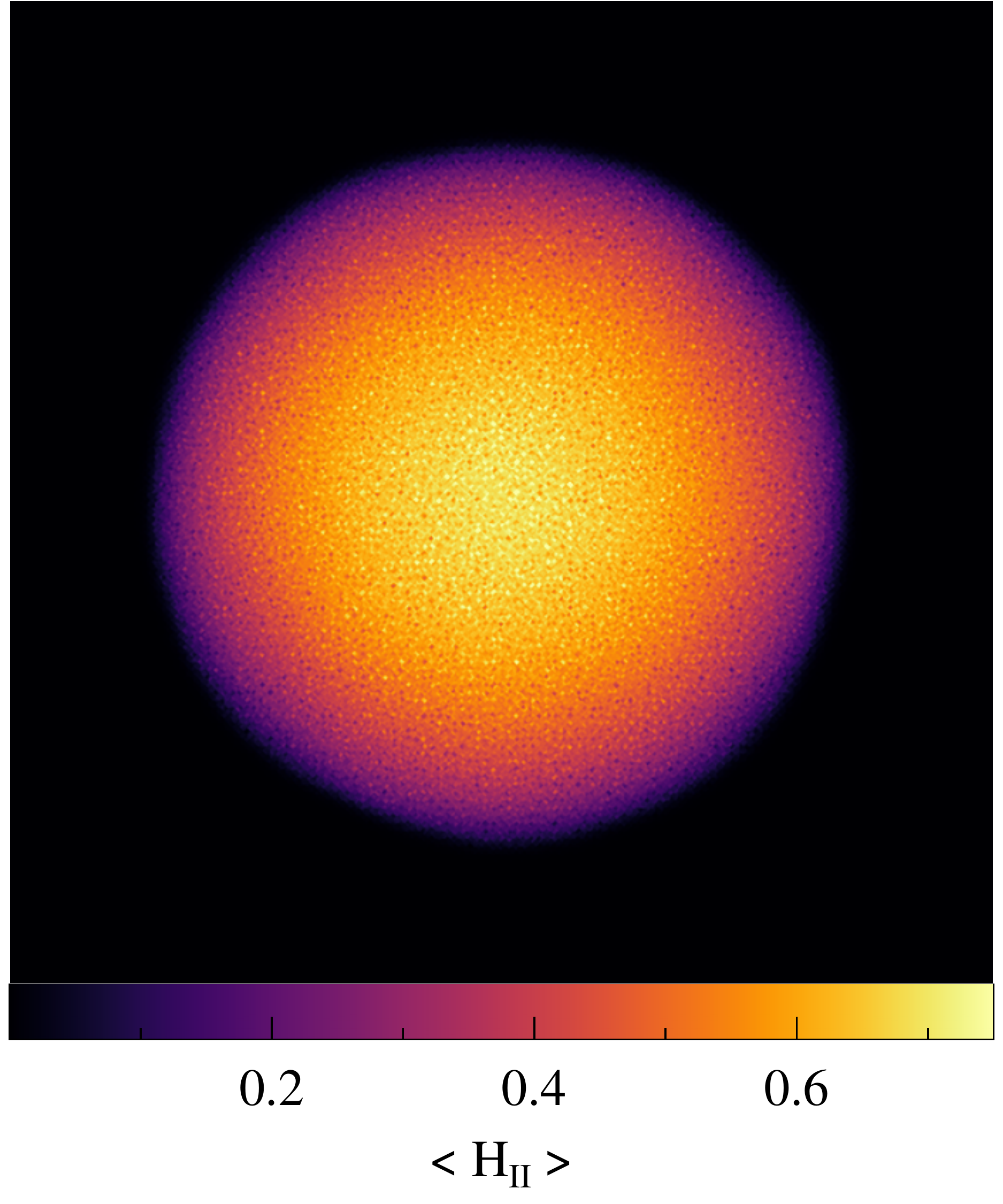}
\caption{ {\bf Str\"omgren sphere}: A projected map of the ionized hydrogen fraction in the Str\"omgren sphere test for the highest resolution (${2\times128^3}$) simulations performed using HLL (top panels) and GLF (bottom panels) functions with underlying regular (left panels) and irregular (right panels) meshes. We obtain quite spherical Str\"omgren regions, because the photon injection region is well resolved. }
\label{fig:stromgenmap}
\end{figure}

\subsection{Str\"omgren sphere}
\label{sec:stromgren}

We now test our scheme for the classical problem of a \ion{H}{II} region
expanding in a constant density and temperature medium \citep{Stromgren1939,
Spitzer1978}. This tests the radiation transport scheme and the \ion{H}{}
chemistry implementation in parallel. A single monochromatic radiation source is
placed at the center of a domain of sidelength ${16\,{\rm kpc}}$, which is
emitting hydrogen ionizing photons with an energy ${13.6\,{\rm eV}}$. The source
outputs a constant stream of photons at a rate of ${\dot{N}_\gamma= 5 \times
10^{48}\,{\rm photons/s}}$.  The density of the surrounding gas is set to
${n_{\rm H} = 10^{-3}\,{\rm cm}^{-3}}$ and the gas has temperature ${T =
10^4\,{\rm K}}$. The Case B recombination rate of \ion{H}{II} under these
conditions is ${\alpha_{\rm B} = 2.69 \times 10^{-13}\,{\rm cm}^{-3}{\rm
s}^{-1}}$. Assuming all the emitted photons are used to ionize the surrounding
hydrogen the maximal extent of the Str\"omgren radius ($r_{s,0}$) is given by
\begin{equation}
 r_s = \left(\frac{3\dot{N}_\gamma}{4\pi \, \alpha_{\rm B} \, n_{\rm H}^2}\right)^{1/3} = 5.38 \ \text{kpc} \, .
 \label{eq:str_rad}
\end{equation}
The evolution of the radius of the I-front can be obtained by assuming
that it is infinitely thin
\begin{equation}
 r_{I,0}(t) = r_s[1 - \exp(-t/t_{\rm rec})]^{1/3}\, ,
 \label{eq:strom_evolve}
\end{equation}
where
\begin{equation}
 t_{\rm rec} = \frac{1}{n_{\rm H} \alpha_{\rm B}} = 125.13 \ \text{Myr}\, ,
 \label{eq:rec}
\end{equation}
is the recombination time for our choice of the gas density.

The radial profiles of the neutral (${{\tilde n}_{\ion{H}{I}} =
n_{\ion{H}{I}}/n_{\ion{H}{}}}$) and ionized hydrogen (${{\tilde
n}_{\ion{H}{II}}}$) fractions can analytically be computed from
\citep{Osterbrock2006}
\begin{equation}
 \frac{{\tilde n}_{\ion{H}{I}}(r)}{4\pi r^2} \int d\nu \, \dot{N}_\gamma (\nu) \, e^{-\tau _{\nu} (r)} \sigma_\nu  = {\tilde n}_{\ion{H}{II}}^2(r) \, n_{\ion{H}{}} \, \alpha_{\rm B}\, ,
 \label{eq:strom1}
\end{equation}
where
\begin{equation}
 \tau_\nu (r) = n_{\ion{H}{}} \, \sigma_\nu \int_0^r dr' \, {\tilde n}_{\ion{H}{I}}(r') \, .
 \label{eq:strom2}
\end{equation}

\begin{figure}
\includegraphics[width=\columnwidth]{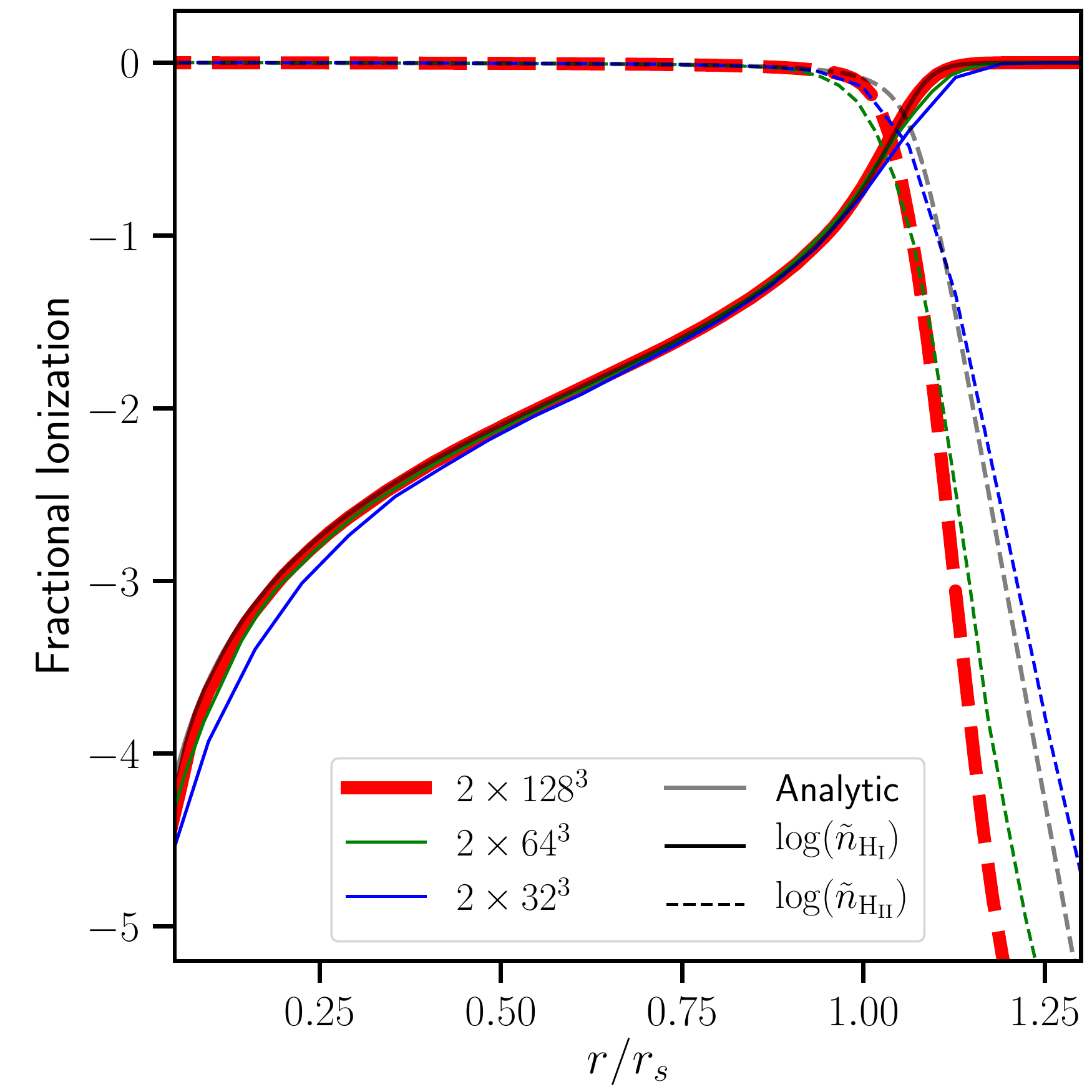}
\caption{{\bf Str\"omgren sphere}: Profiles of the neutral (solid curves) and ionized (dashed curves) hydrogen fractions as a function of radius at the end of the Str\"omgren sphere test for the simulations with $2 \times 128^3$ (red curves), ${2\times64^3}$ (green curves) and ${2\times32^3}$ (blue curves) runs. The simulations match the analytic results (black curves) quite well.}

\label{fig:cstromgenprof}
\end{figure}

 We perform the simulation with ${2\times128^3}$,  ${2\times64^3}$ and ${2\times32^3}$ resolution elements using our fiducial scheme on a regular mesh.  We use the reduced speed of light approximation in these runs, with
${f_{\rm r} = {\tilde c}/c = 0.01}$. In order to understand the effect of reducing the speed of light we perform additional simulations with ${f_{\rm r} = 10^{-4}}$,  $2 \times 10^{-4}$ and $10^{-3}$.  Fig.~\ref{fig:stromgenmap} shows the projected maps of ionized hydrogen, at
$0.5~\text{Gyr}$ for the run with $2 \times 128^3$ resolution elements. There are
small departures from a perfect spherical symmetry due to the geometry of the
injection region and to a lesser extent due to the geometry of the underlying
mesh. Resolving the injection radius with multiple cells improves the
sphericity of the solution. 	

\begin{figure}
\includegraphics[width=\columnwidth]{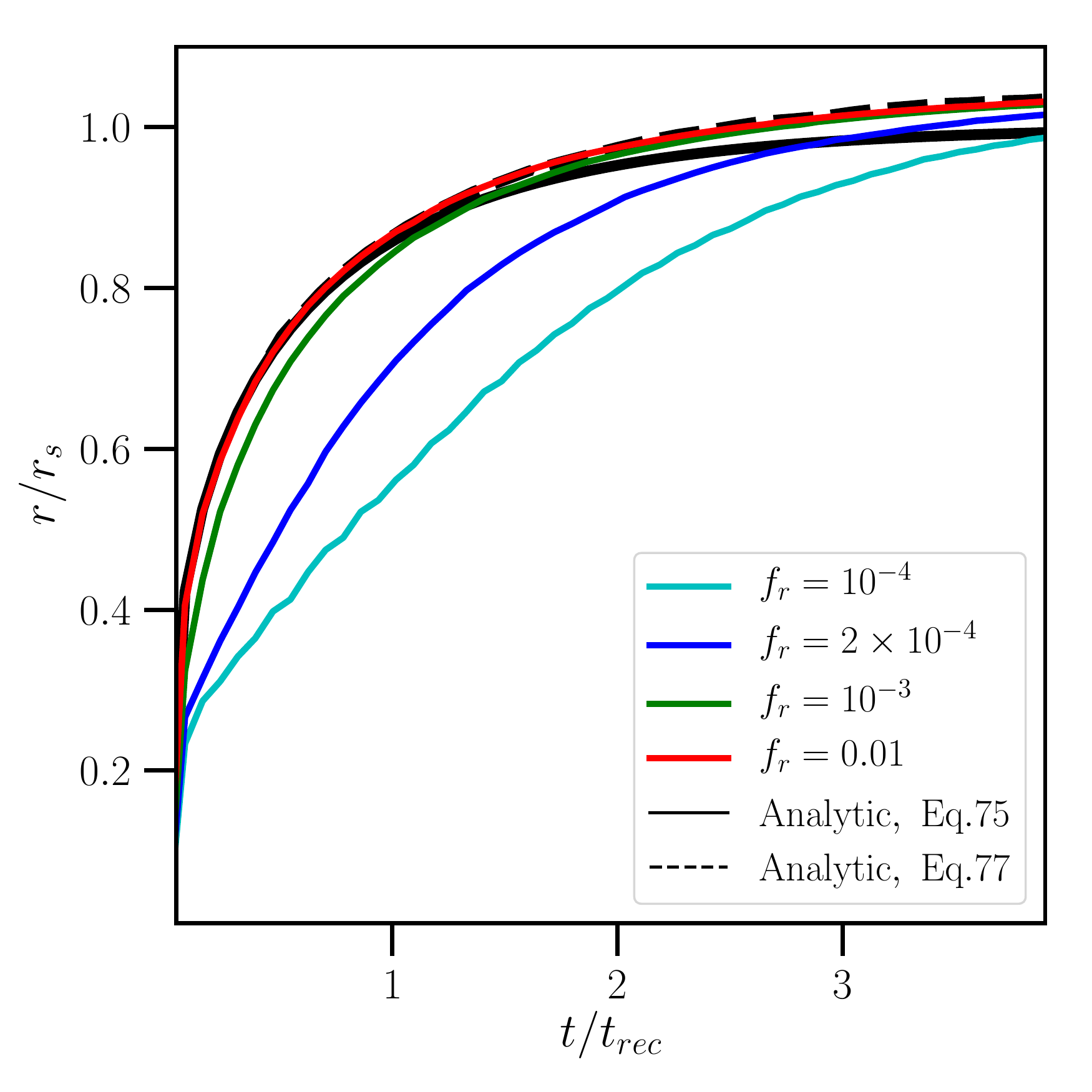}
\caption{{\bf Str\"omgren sphere}: Radius of the ionization front as a function of time for simulations with ${2\times32^3}$ - resolution elements performed with $f_r = 10^{-4}$ (cyan curve), $f_r = 2 \times 10^{-4}$ (blue curve),  $f_r = 10^{-3}$ (green curve) and $f_r = 0.01$ (red curve). The analytic solutions obtained from Eq.~\ref{eq:strom_evolve} and Eq.~\ref{eq:strom1} are plotted as solid black and dashed black curves respectively.  The run with the highest speed of light matches the analytic expectations well, while the runs with lower light speeds take longer to reach the expected analytic $r_s$.}
\label{fig:stromgenexp}
\end{figure}

The profiles of neutral (solid curve) and ionized (dashed curve) hydrogen at
$0.5$ Gyrs are shown in Fig.~\ref{fig:cstromgenprof}. The analytic solution obtained
from Eqs.~\ref{eq:strom1} and \ref{eq:strom2} is plotted in black. We can see
that our method reproduces the analytic solution and the accuracy of the solution increases with
resolution as expected.

In Fig.~\ref{fig:stromgenexp}, we show the time evolution of the ionizing
front, for the runs with different light speeds. The ionization front is
defined as the radius at which the ionization fraction equals $0.5$. We
note that at late times the analytic expectations derived from
Eq.~\ref{eq:strom_evolve} (solid black curve) and Eq.~\ref{eq:strom1} (dashed
black curve) diverge because Eq.~\ref{eq:strom_evolve} assumes an infinitely
thin transition region and fully ionized gas within $r_s$ which is not an accurate description of the Str\"omgren
sphere. The simulation results match the more accurate analytic expectation
described by Eq.~\ref{eq:strom1}.

The position of the I-front after $0.5 \ \rm{ Gyr}$ is quite similar in all the runs, implying that, given time, the simulations will converge to the right solution irrespective of the speed of light used. However, this is only true if we are interested in the final state of the Str\"omgren sphere and not in its evolution, which is clearly different for the different speeds of light used. As shown in R13, the behaviour of the solution can be quantified by comparing the light crossing time ($t_{\rm cross} = r_s/{\tilde c}$) to the recombination time ($t_{\rm rec}$, Eq. \ref{eq:rec}); $q = t_{\rm cross}/t_{\rm rec}$. When, $q\ge1$, the ionization front expands at the reduced speed of light and reaches $r_s$ at $t=t_{\rm cross}$ and this is insensitive to the recombination timescale.  For $q<1$, the reduced speed of light is larger than the analytic expectation of the speed of the ionization front and hence the simulated ionization front travels at the reduced speed of light till it reaches the analytic solution and starts following the analytic results. Of course, these considerations are only valid if we assume full ionization within $r_s$.  In most realistic situations, there is a complex ionization structure given by Eq. \ref{eq:strom1} and shown in Fig. \ref{fig:cstromgenprof}. In these realistic situations, the time to reach the analytic solution is longer. For example, in our simulation with $f_{\rm r} = 10^{-4}$ ($q=1.4$), the ionization front should start to follow the analytic solution by $t/t_{\rm rec}  =1.4$, but it takes $t/t_{\rm rec} > 3.5$ to achieve this because of the complicated internal ionization structure of the Str\"omgren sphere. 

If $\tau_{\rm min}$ is the shortest relevant timescale of a simulation, then ensuring that {$t_{\rm cross}<<\tau_{\rm min}$} will ensure that the right solution for the evolution of the I-front will be recovered without affecting other timescales of the system. Therefore, we can typically set 
\begin{equation}
 f_{r} = {\rm min}\{1, \ \eta \ t_{\rm cross}/\tau_{\rm min}\} \ ,
 \label{eq:rsla}
\end{equation}
where $\eta \simeq 10$. The value of $\tau_{\rm min}$ is of course problem dependent. For re-ionization simulations, where the speed of the ionization front can reach speeds of $10^{4} \ {\rm km \ s^{-1}}$ \citep{Bauer2015}, there is very little room to reduce the speed of light. However, if we want to simulate the ISM, then the high densities and low gas velocities allows for large reductions in the speed of light $f_r \lesssim 10^{-3}$. Since most of the tests in the paper are performed for relatively high density initial conditions, a value of $f_r=0.01$ is good enough to capture accurate results. \footnote{See R13 and \citep{Skinner2013} for a more detailed discussion on the reduced speed of light approximation and its applicability.}

\subsection{Multi-Frequency H-He Str\"omgren sphere}
\label{sec:multstrom}

\begin{figure}
\includegraphics[width=\columnwidth]{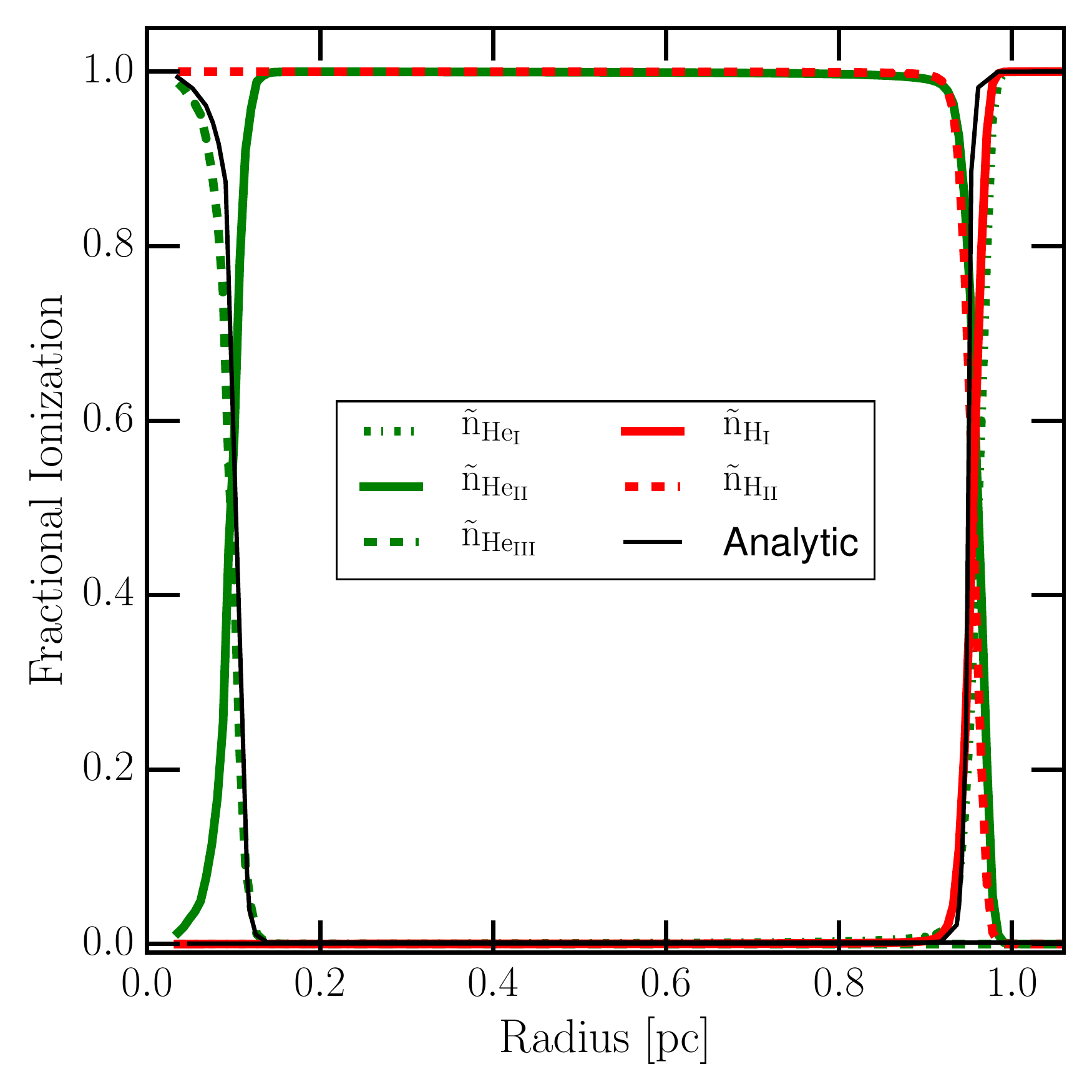}
\caption{{\bf Multi-Frequency H-He Str\"omgren sphere}: Profiles of the neutral Hydrogen (solid red curve), ionized hydrogen (dashed red curve), neutral helium (dot dashed green curve), singly ionized helium (solid green curve) and doubly ionized helium (dashed green curve) fractions as a function of radius around a O4V star at the end of the multi-frequency H-He Str\"omgren sphere test. The analytic expectations of doubly ionized helium and neutral helium are over plotted as black curves.}
\label{fig:multprof}
\end{figure}

As an extension of the previous test, we simulate the ionization structure
around an O4V star. For this test, we include both atomic Hydrogen and Helium
chemistry. In order to appropriately model the absorption of photons at
different frequencies by the different ionic species, a multi frequency
approach is required. Qualitatively, the ionization structure in a nebula with
hydrogen and helium depends on the helium abundance as well as the spectrum of
the ionizing star.  The first ionization potential of Helium is
${24.6\,\text{eV}}$ and it can be doubly ionized when photons with energies
greater than  ${54.4\,\text{eV}}$ are present.  The photons with energies between
$13.6$ and ${24.6\, \text{eV}}$ can ionize only neutral hydrogen while photons
with energies above ${24.6\, \text{eV}}$ can ionize both neutral hydrogen and
neutral helium. In order to accurately capture this behaviour, we split the
spectrum into three bins with frequency ranges corresponding to the ionization
potentials of \ion{H}{I} [$13.6\,{\rm eV}, \, 24.6\,{\rm eV}$), \ion{He}{I}
[$24.6\,{\rm eV}, \, 54.4\,{\rm eV}$) and \ion{He}{II} [$54.4\,{\rm eV}, \,
100.0\,{\rm eV}$).

The star is assumed to emit a black body spectrum with ${T_{\text{eff}}~=~4.87~\times~10^4\,~{\text K}}$ and a luminosity of
${7.6~\times~10^5\,~{\text L}_\odot}$, which translates to a Lyman continuum
(Lyc) photon rate of ${5~\times~10^{49}\, {\text{photons s}}^{-1}}$.  We simulate
a domain of size ${3 \ \text{pc}}$ with ${2\times128^3}$ resolution elements and a
regular staggered grid. The surrounding gas has a constant density of
${10^3\,~{\text{cm}}^{-3}}$ and a temperature of ${10^4\,~{\text K}}$.  

The approximate size of the $\ion{He}{II}$ zone is given by \citep{Tielens2005}
\begin{equation}
  r_s (\text{He}) = \left(\frac{3\dot{N}_{\gamma,\text{Lyc}}(\text{He})}{4\,\pi \,y \,(1+y) \,\alpha_{\rm B} (\text{He}) \, n_{\rm H}^2}\right)^{1/3}\, ,
\end{equation}
where `y' is the Helium number fraction and for this test ${r_s (\text{He}) \sim
1\, \text{pc}}$. This estimate is only correct if there is no hydrogen ionization.
For realistic conditions we need to solve the radiative transfer equation and
the coupled chemistry equations as described in Section~\ref{sec:chemistry}.

The fractional ionization profiles \ion{H}{I} (solid red curve), \ion{H}{II}
(dashed red curve), \ion{He}{I} (dot-dashed green curve), \ion{He}{II} (solid
green curve) and \ion{He}{III} (dashed green curve) are plotted in
Fig.~\ref{fig:multprof}. Helium is doubly ionized in the central
${0.1\,\text{pc}}$ while both Hydrogen and Helium are singly ionized within the
central ${1\,\text{pc}}$.  Even a very hot O4V star is only  able to produce a
small doubly ionized He region as the amount of photons above
${54.4\,\text{eV}}$ is rather low. 

As a comparison we also plot the analytic fractional ionization profiles of $\ion{He}{III}$ and $\ion{He}{I}$ in black. While the position of
the transition region matches the analytic solution well, the width is
larger than the expected value. This is because the width of the
transition region is of the order of the mean free path of ionizing photons, 
\begin{equation}
l = 1/(n_{\text H}\,\sigma_e) \sim 10^{-3}\,~\text{pc} \, ,
\end{equation}
which is much smaller than the cell size in our simulation. As the ionization
cross section is highly peaked towards the ionization potential, the stellar
radiation field is least attenuated at the highest frequencies. Since the H
cross section for highly energetic photons is smaller than that of He, helium
can stay ionized slightly further out than H as seen in the figure. We have
only plotted the analytic expectation for two of the five ionic species in the
simulation, in order to increase the readability of the plot. We note that we
match the profiles of the other ionic species equally well (See Fig.~7.2 of
\citealt{Tielens2005} for more details). This test confirms the accuracy of our
multi-frequency radiative transfer scheme coupled to the H-He chemistry.

\begin{figure*}
\includegraphics[scale=0.6]{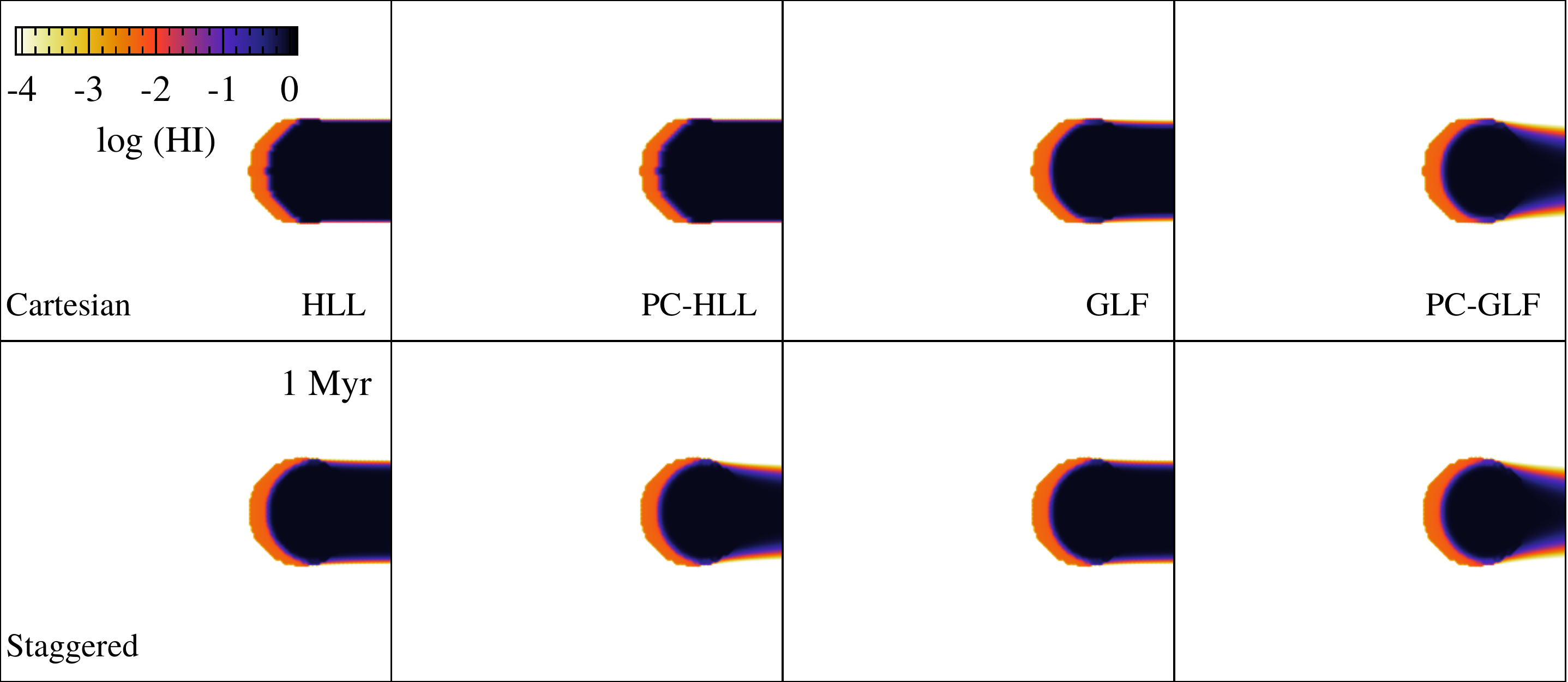}
\caption{{\bf I-front trapping in a dense clump and the formation of a shadow}: The \ion{H}{I} maps showing slices at ${z=0.5L_\text{box}}$ after ${1\,\text{Myr}}$ of evolution for HLL (first column), PC-HLL (second column), GLF (third column) and PC-GLF (fourth column) schemes. The top panels depict the results for simulations using a underlying Cartesian mesh, while the bottom panels depict the results for a staggered mesh. }
\label{fig:h11}
\end{figure*}

\begin{figure*}
\includegraphics[scale=0.6]{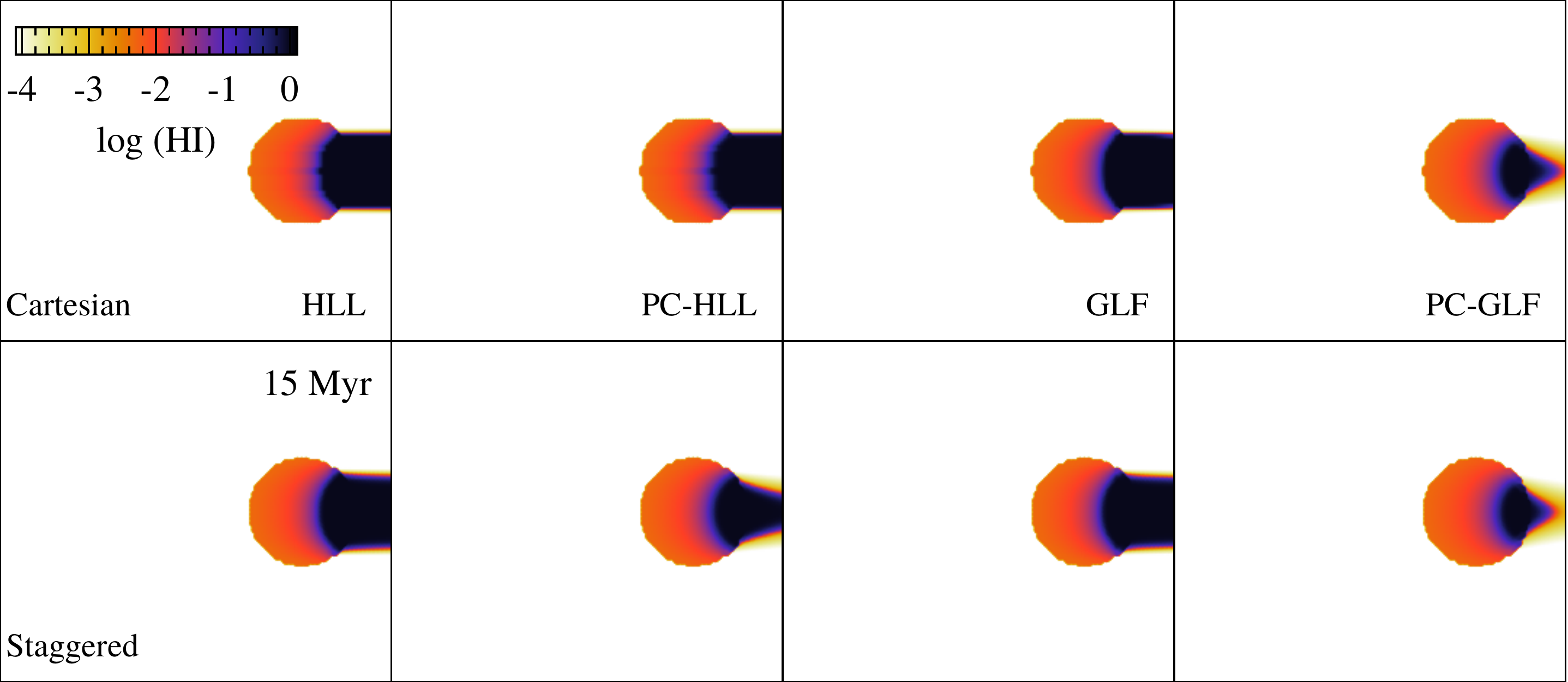}
\caption{{\bf I-front trapping in a dense clump and the formation of a shadow}: Same as Fig.~\ref{fig:h11} but at simulation time of ${t=15\,\text{Myr}}$. The fiducial HLL and GLF schemes are able to form sharp shadows irrespective of the mesh geometry used. The PC-HLL  scheme is only able to obtain sharp shadows if the mesh interfaces are exactly parallel/perpendicular to the photon propagation direction as is the case in a Cartesian mesh, but fails to do so in a staggered mesh which has cross mesh transport. The PC-GLF scheme is unable to form sharp shadows irrespective of the mesh geometry.}
\label{fig:h115}
\end{figure*}

\subsection{I-front trapping in a dense clump and the formation of a shadow}
\label{sec:shadow}

\begin{figure*}
\includegraphics[scale=0.6]{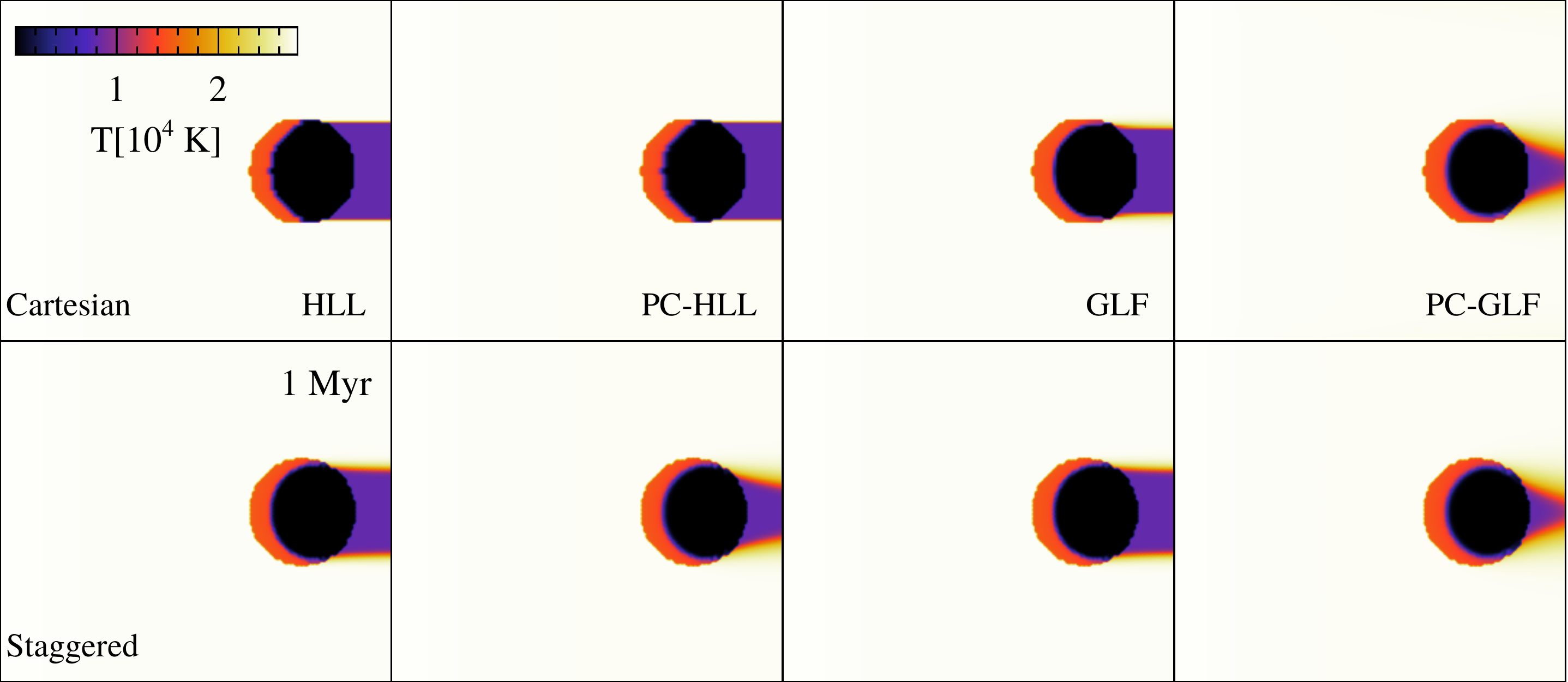}
\caption{{\bf I-front trapping in a dense clump and the formation of a shadow}: The temperature maps showing slices at ${z=0.5L_\text{box}}$ after ${1\,\text{Myr}}$ of evolution.}
\label{fig:t1}
\end{figure*}

\begin{figure*}
\includegraphics[scale=0.6]{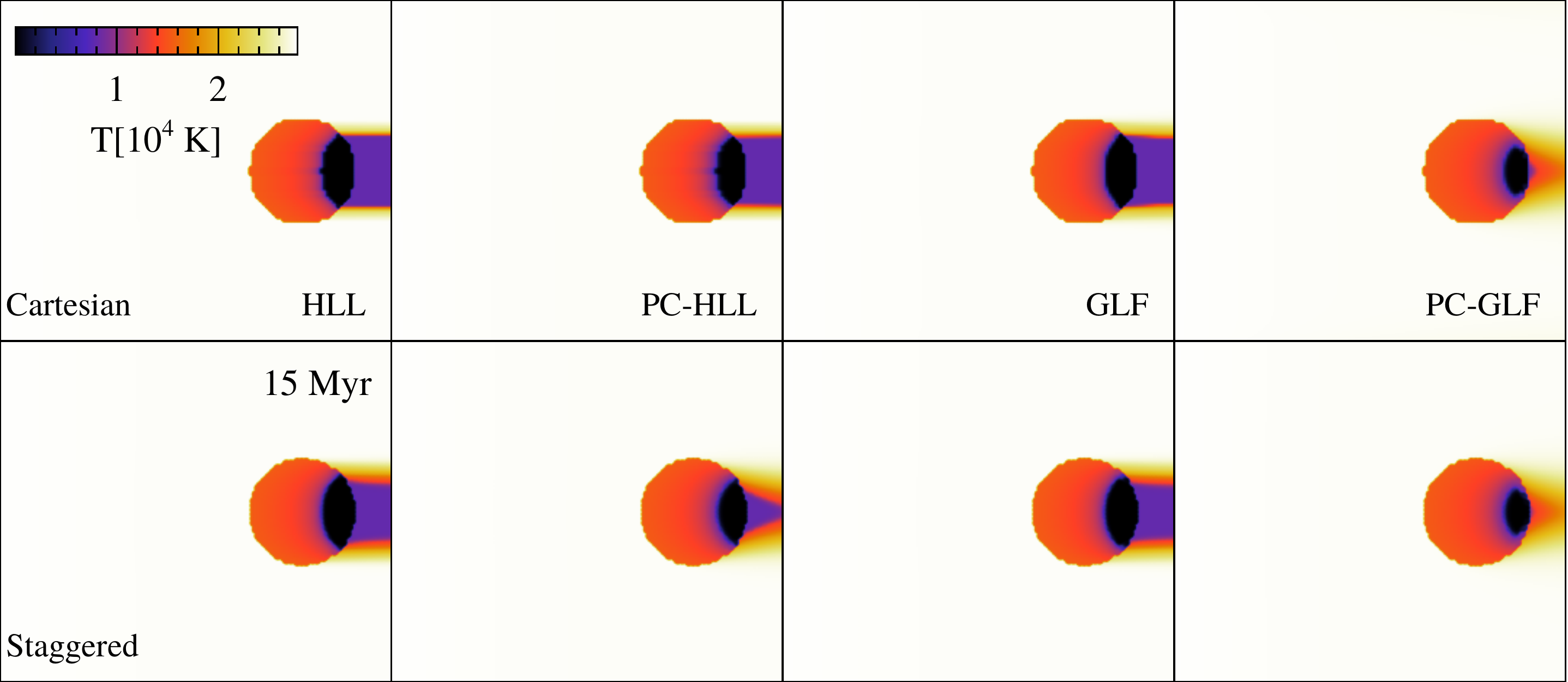}
\caption{{\bf I-front trapping in a dense clump and the formation of a shadow}: Same as Fig.~\ref{fig:t1} but at simulation time of ${t=15\,\text{Myr}}$. The temperature maps replicate the trends seen in the \ion{H}{I} maps.}
\label{fig:t15}
\end{figure*}

Next we simulate the trapping of a plane parallel I-front by
a dense, uniform, spherical clump. This test mimics self-shielding within a
high density gas cloud illuminated by ionizing UV photons. Ideally, the other
side of the clump should be shielded from the ionizing radiation, producing
sharp shadows. The ability of a clump to trap an ionization front depends on
the strength of the ionizing flux ($F$), the clump density ($n_\ion{H}{}$), the
radius of the clump ($r_\text{clump}$) and the case B recombination rate
($\alpha_\text{B}$). We can define the `Str\"omgren Number' for a clump as $L_s
= 2r_\text{clump}/l_s(0)$, where $l_s(0)$ is the Str\"omgeren length at zero
impact parameter
\begin{equation}
l_s(0) = \frac{F}{\alpha_\text{B}\, n_\ion{H}{}^2} \, ,
\end{equation} 
%

\begin{equation}
L_s = \frac{2r_\text{clump} \, \alpha_B \, n_\ion{H}{}^2}{F}\, .
\end{equation}
If ${L_s>1}$ the clump is able to trap the I-front, while if ${L_s<1}$, the clump
would be unable to trap the I-front and would instead be flash ionized by its
passage.

In our setup, the UV ionizing radiation has a black body spectrum with an
effective temperature ${T_\text{eff} = 10^5\,\text{K}}$ and an ionizing flux of ${F
= 10^6\,\text{photons s}^{-1}\text{cm}^{-2}}$. This flux is a plane parallel wave
travelling in the $+x$ direction and is incident on the $x=0$ boundary. The
domain size is ${6.6\,\text{kpc}}$ on a side with $128^3$ resolution elements. A
spherical high density clump of radius ${r_\text{clump}=0.8\,\text{kpc}}$ is
placed within the domain, with the center of the clump at ${(x,y,z)=(5, 3.3,
3.3)\,\text{kpc}}$. The density and temperature of the clump are
${n_\ion{H}{}^{\text{clump}} = 4 \times 10^{-2}\,\text{cm}^{-3}}$ and
${T^{\text{clump}}=40\,\text{K}}$ respectively. The remaining domain is filled
with a hot, low density gas with a temperature of
${T^{\text{out}}=8000\,\text{K}}$ and density of ${n_\ion{H}{}^{\text{out}} =
n_\ion{H}{}^{\text{clump}}/200 = 2 \times 10^{-4}\,\text{cm}^{-3}}$.  Initially,
the domain is neutral with an ionization fraction of ${{\tilde
n}_\ion{H}{II}=10^{-6}}$.  For these parameters and assuming a Case B
recombination rate of ${\alpha_\text{B}(T) = 2.59 \times 10^{-13}
(T/10^4\,\text{K})^{-3/4}\,\text{cm}^3\text{s}^{-1}}$, we obtain ${l_s
\simeq 0.78(T/10^4\,\text{K})^{3/4}\,\text{kpc}}$ and ${L_s \simeq
2.05(T/10^4\,\text{K})^{-3/4}}$.  Therefore, along the axis of symmetry
the I-front should be trapped approximately at the centre of the clump for
${T = 10^4\,\text{K}}$. In reality, the temperature could be expected to
be somewhat different and spatially varying, but to a rough first approximation
this estimate should hold. We run the simulation for ${15\,\text{Myr}}$ with
$f_r=0.1$. A single frequency approximation is used, so that there is no
leakage of photons beyond the I-front. 

We perform this simulation with four different types of numerical schemes, HLL,
PC-HLL, GLF and PC-GLF, which are described in Section~\ref{sec:tests}. We
note that `PC' denotes a piecewise-constant approximation, which is used in
many recent works \citep{Rosdahl2015, Bieri2017, Costa2017, Cielo2017,
Lupi2017}. However, we have shown in Section~\ref{sec:adv} that this
approximation is exceedingly diffusive. Just switching to a less diffusive HLL
flux function from GLF will only slightly improve the numerical diffusivity of
the scheme. We have shown that it is important to have gradient extrapolated
values at the interface, in order to reduce the numerical diffusivity and
increase the convergence order of the scheme.  We use this test to further show
the limitations of the `PC' approximation and demonstrate the importance of using
higher order schemes for radiative transfer.  We use a reduced light speed
fraction of $f_r= 0.1$. This high speed is needed for the light to have reached the cloud
in the first snapshot under consideration, at $1$ Myr.

\begin{figure}
\includegraphics[width=\columnwidth]{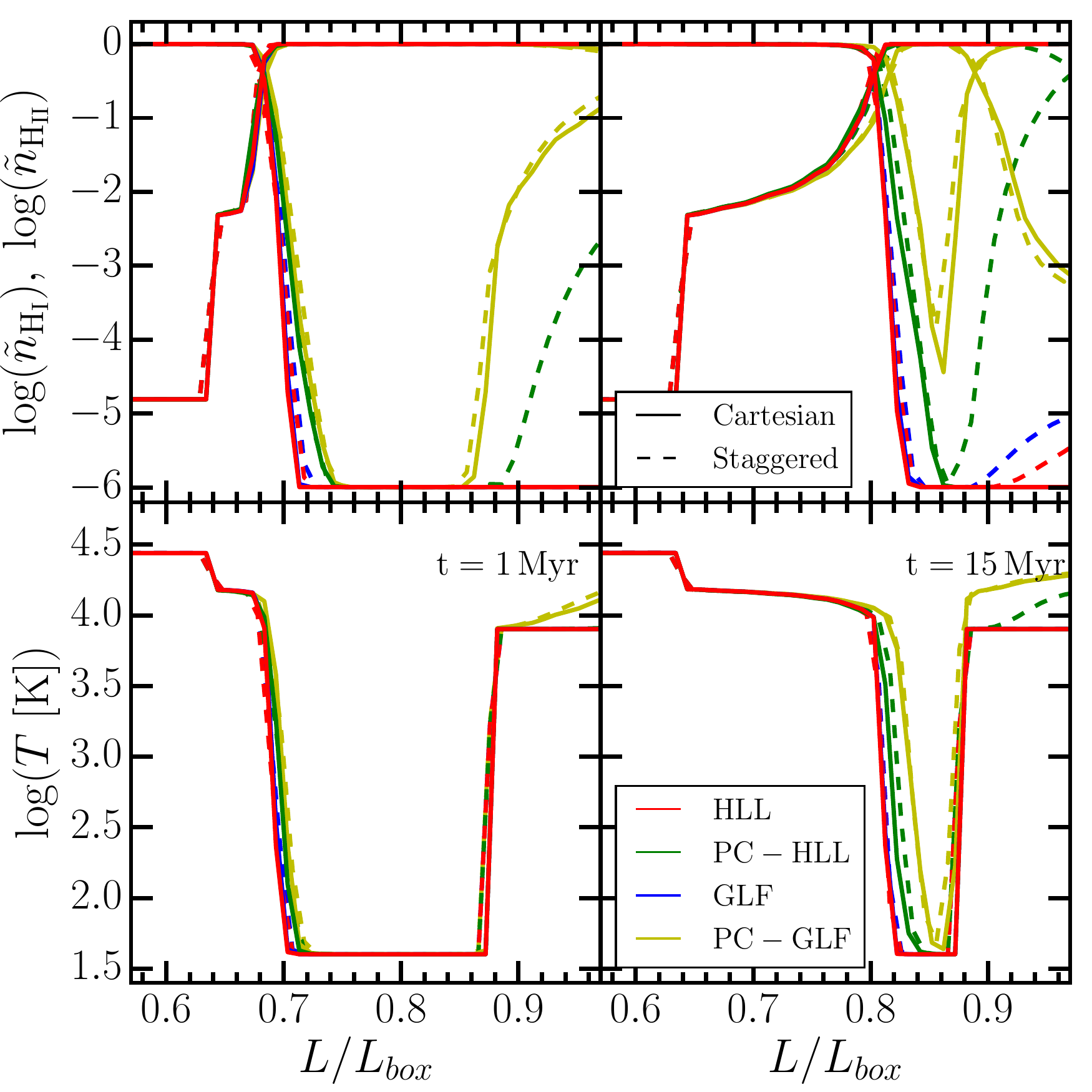}
\caption{{\bf I-front trapping in a dense clump and the formation of a shadow}: The ionization fraction (top panels) and temperature (bottom panels) profiles calculated in a thin cylindrical shell around the axis of symmetry after ${1\,\text{Myr}}$ (left panels) and ${15\,\text{Myr}}$ (right panels) of evolution. The solid curves depict the results for a simulation with an underlying Cartesian mesh while the dashed curves are for a staggered grid. The red, green, blue and yellow curves show the result for HLL, PC-HLL, GLF and PC-GLF schemes respectively. Numerical photon diffusion increases the temperature and \ion{H}{II} fraction in the shadow region if the PC schemes are used.}
\label{fig:htemp}
\end{figure}

The I-front travels fast through the diffuse medium outside the cloud, but
moves much more slowly inside of it, and a shadow is cast behind it. As the UV
radiation slowly ionizes and heats the cloud, the shadow very slowly diminishes
in width because some photons manage to cross through the edges of the cloud.
Fig.~\ref{fig:h11} presents the \ion{H}{I} fraction maps in the simulations using HLL
(first column), PC-HLL (second column), GLF (third column) and PC-GLF (fourth
column) schemes after ${1\,\text{Myr}}$ of evolution.  The top panel denotes the
results where the underlying mesh is Cartesian. Fig.~\ref{fig:h115} shows the same after
${15\,\text{Myr}}$ of evolution. These maps reveal interesting differences
between the numerical schemes. On a regular Cartesian mesh, the HLL, PC-HLL and
GLF schemes are able to maintain the directionality of the photons, even after
${15\,\text{Myr}}$ of evolution resulting in accurate shadows,
which is not true for the PC-GLF scheme.  This verifies the results presented
in (\citealt{Rosdahl2013}; hereafter referred to as R13). Based on this test
R13 and \citet{Lupi2017} argued that switching to a HLL flux function will
reduce the diffusion of the scheme and maintains the directionality of the
photons.  The directionality of the photons is only exactly maintained when the
photon flux is either completely parallel or perpendicular to the cell
interface. The eigenvalues of the system are exactly $\pm c$ and $0$ when the
angle between the photon flux and the interface is $90$ and $0$ degrees
respectively, and the value of the reduced flux ${f=1}$ (see Fig.1 of
\citealt{Gonzalez2007}). In this test, the photon flux is exactly perpendicular
to the cell interfaces along the photon propagation direction and parallel to
the other faces, in a Cartesian mesh. This necessitates that the perpendicular
numerical diffusion is exactly $0$ by construction. 

\begin{figure}
\includegraphics[width=\columnwidth]{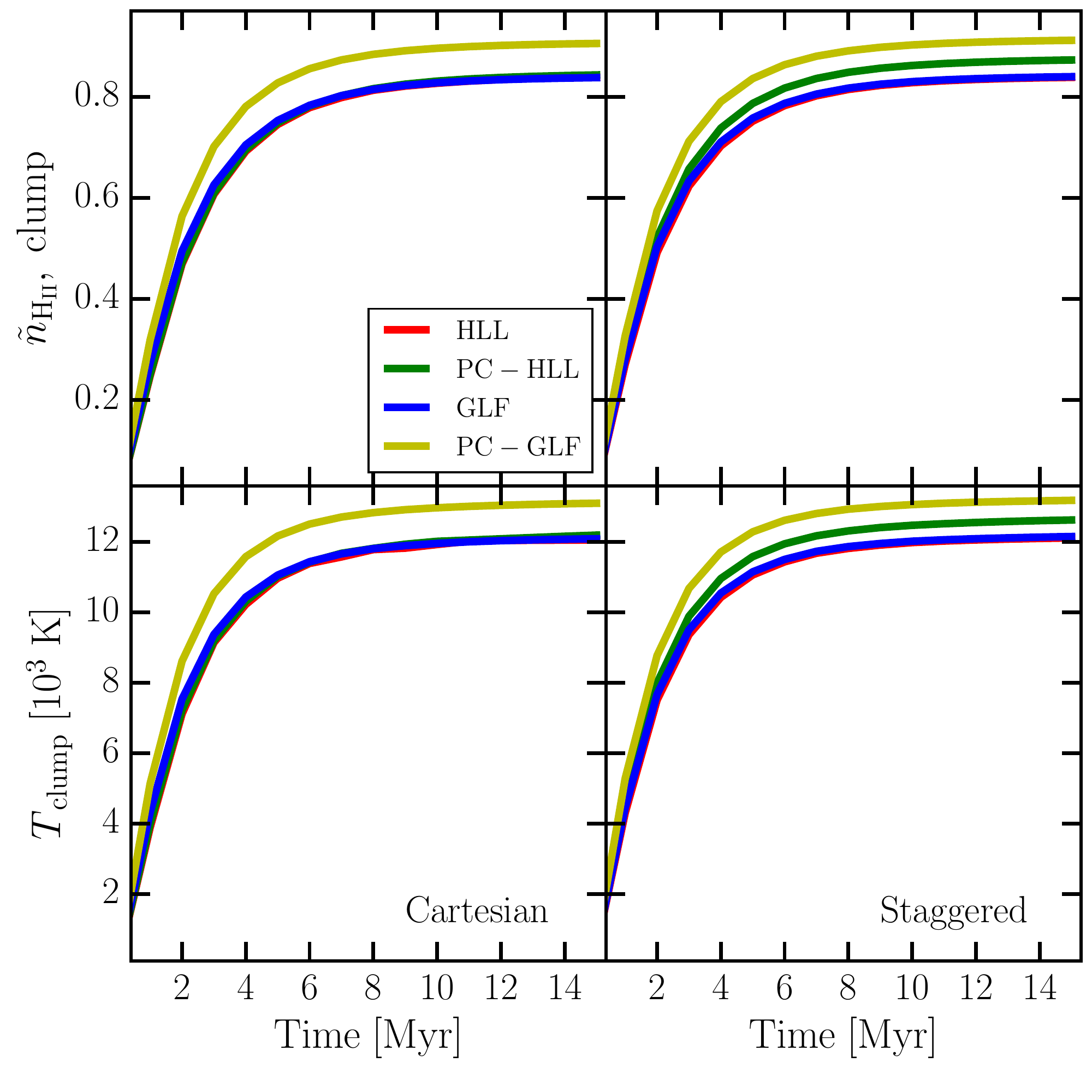}
\caption{{\bf I-front trapping in a dense clump and the formation of a shadow}: The time evolution of the average ionization fraction (top panels) and temperature (bottom panels) of the gas clump for simulations using a HLL (red curve), PC-HLL (green curves), GLF (blue curves) and PC-GLF (yellow curves) schemes with an underlying Cartesian (left panels) and staggered (right panels) grids. The more diffusive PC schemes tends to increase the temperature and ionization fraction of the high density clump.}
\label{fig:shadow}
\end{figure}

However, this geometry can only be achieved in test problems where the
radiation propagation direction is known beforehand and will almost never occur
in realistic simulations. To elucidate this point, we perform the simulations
also on a regular staggered mesh (bottom panels of Figs.~\ref{fig:h11} $\&$
\ref{fig:h115}) with ${2\times96^3}$ resolution elements. The photon flux is no
longer parallel/perpendicular to the cell interfaces therefore the PC-HLL
scheme no longer produces accurate and sharp shadows, and this is particularly
evident after ${15\,\text{Myr}}$. It still performs better than the
PC-GLF scheme, but performs unfavorably compared to fiducial schemes, which use
gradient extrapolations, regardless of the flux function used.  These results
reflect the findings of Section~\ref{sec:adv} for a more realistic setup.
Since, the temperature of the gas has a sharp dependence on the photon density
of the cell, the temperature maps (Figs.~\ref{fig:t1}~and~\ref{fig:t15}) show
similar trends.

Fig.~\ref{fig:htemp} quantifies the difference between the various schemes by
plotting the hydrogen ionization fraction (top panels) and temperature (bottom
panels) profiles calculated in a thin cylindrical shell around the axis of
symmetry after ${1\,\text{Myr}}$ (left panels) and ${15\,\text{Myr}}$ (right
panels) of evolution.  The temperature of the surrounding gas increases
slightly from ${8000\,\text{K}}$ to about ${3\times10^4\,\text{K}}$ very quickly.
Inside the clump, however, the I-front is trapped and its velocity slows down
considerably.  In the pre-ionization zone ahead of the main I-front all schemes
agree quite well at all times. Slight differences emerge in the position of
the I-front (defined as the point of 50 per cent ionized fraction) at
${15\,\text{Myr}}$, with the less diffusive fiducial schemes propagating to a
slightly smaller distance into the cloud.

The main differences arise in the shadow region behind the clump. The
temperature and the ionization fraction, in particular, of this low density gas
is extremely sensitive to any photons leaking into this region due to numerical
diffusion. On a Cartesian mesh (solid curves) the HLL (red curves), GLF (blue
curves) and the PC-HLL (green curves) schemes all perform quite well in the
shadow region.  The ionized hydrogen fraction stays at $10^{-6}$ indicating
very little numerical diffusion perpendicular to the direction photon flow. The
PC-GLF (yellow curves) scheme, however, has difficulty maintaining the
directionality of the photons and this drastically increases the ionization
fraction and temperature in the shadow region. In fact the gas in the shadow
region becomes fully ionized behind the clump at ${15\,\text{Myr}}$. The ability
of the PC-HLL scheme to produce accurate shadows is diminished as soon as we
move to a staggered mesh (dashed curves), which mimics cross mesh transport of
photons.  The ionization fraction in the shadow region increases almost as
dramatically as the PC-GLF scheme implying that there is no longer a sharp
shadow. Using a HLL flux function only slightly improves the numerical accuracy over
using a GLF flux function if a PC approximation is used. Only our fiducial
gradient extrapolated schemes are able to maintain low numerical diffusivity
even when the photon propagation direction is across the mesh interfaces.

Finally, Fig.~\ref{fig:shadow} shows the time evolution of the average
ionization fraction and temperature of the clump. The evolution
matches in general the range of results seen in tests performed using various RT
schemes presented in \citet{Iliev2006}. This confirms that the position and
velocity of the I-front match the expected solution very well. The
evolution in the average quantities only slightly differs between the schemes,
with the more diffusive PC-GLF (on both meshes) and PC-HLL (only on a staggered
mesh) schemes showing slightly higher ionization fractions and temperatures as
expected.

We conclude that this test demonstrates the importance of implementing higher
order, low diffusion schemes for radiative transfer in order to produce
accurate results that are independent of the geometry of the problem and that
of the underlying mesh.

\begin{figure*}
\begin{center}
\includegraphics[scale=0.46]{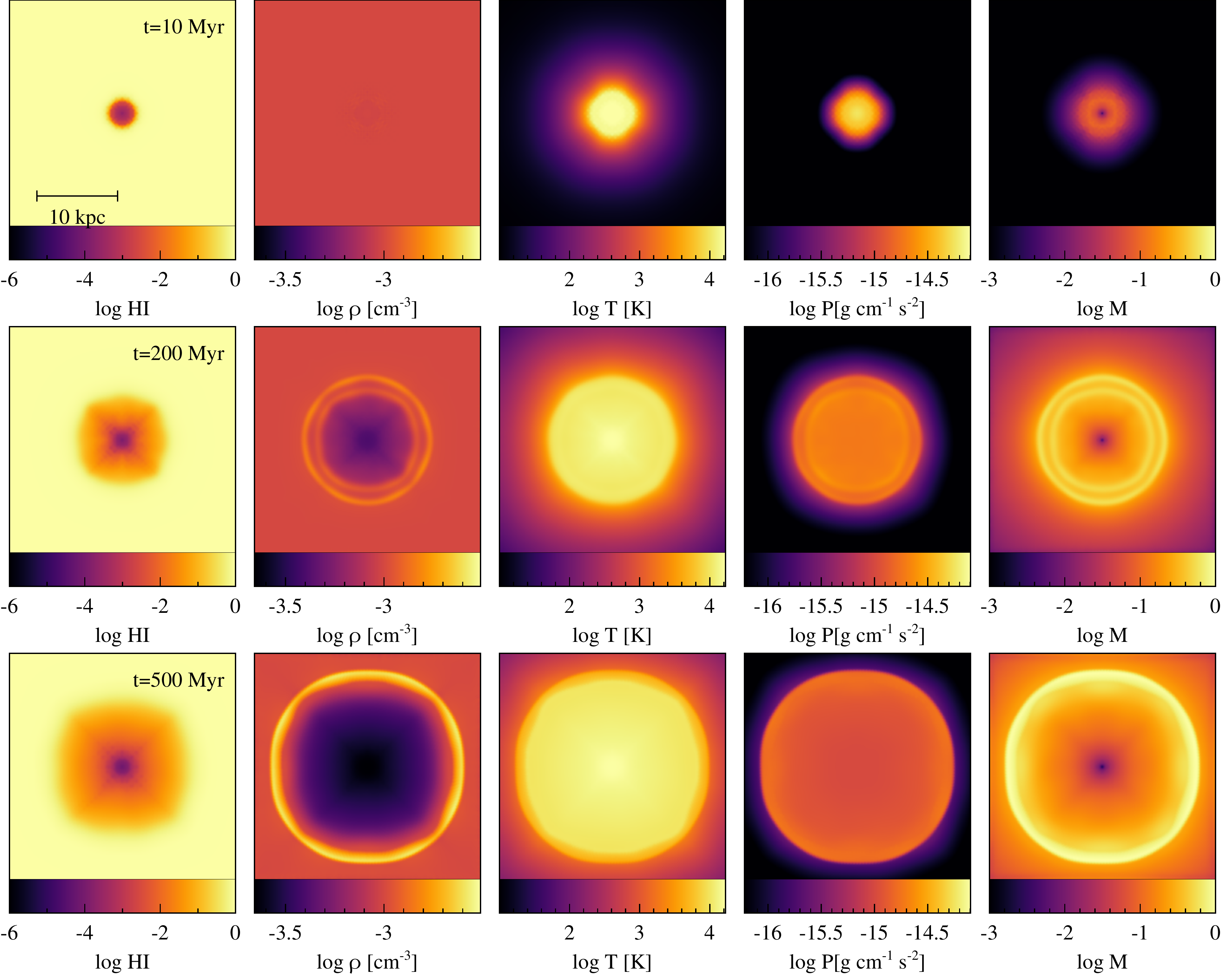}
\caption{{\bf Expansion of a $\ion{H}{\sc II}$ region}: Maps showing slices of the domain at ${z=15 \ \text{kpc}}$ for the simulation of the expansion of a \ion{H}{II} region in a constant density medium. The \ion{H}{I} fraction (first column), density (second column), temperature (third column), pressure (fourth column) and Mach number (fifth column) are plotted  at $10$ Myr (top panels), $200$ Myr (middle panels) and $500$ Myr (bottom panels).}
\label{fig:h21map}
\end{center}
\end{figure*}

\begin{figure*}
\includegraphics[scale=0.41]{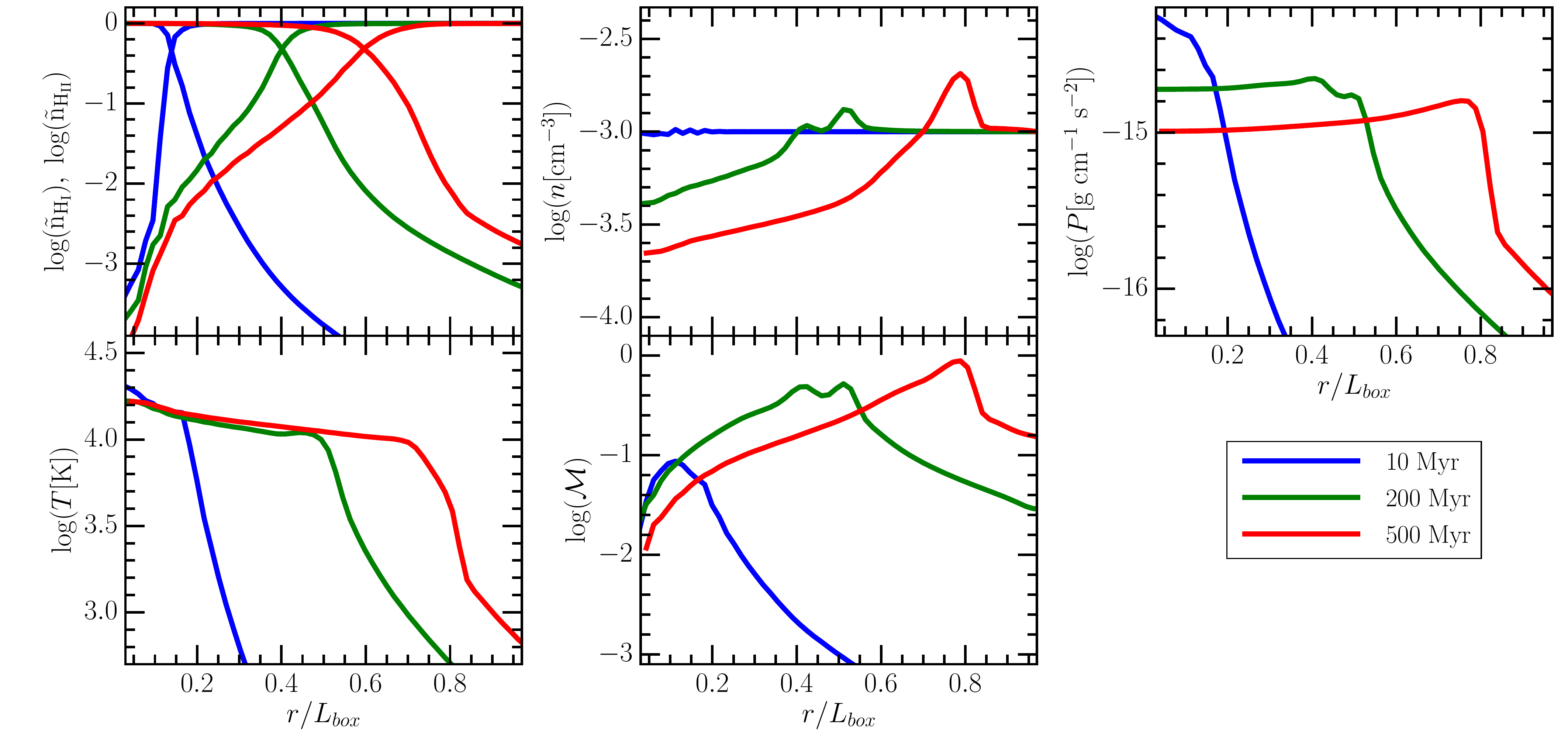}
\caption{{\bf Expansion of a $\ion{H}{\sc II}$ region}: The ionization (top left panel), density (top middle panel), pressure (top right panel), temperature (bottom left panel) and Mach number (bottom middle panel) profiles at $10$ Myr (blue curves), $200$ Myr (green curves) and $500$ Myr (red curves) in the simulation of the expansion of an \ion{H}{II} region in a constant density medium.  The profiles generally match the results from previous simulations of the same test reported in \citet[Test $5$]{Iliev2009}.}
\label{fig:h2const}
\end{figure*}

\subsection{Expansion of a $\ion{H}{\sc II}$ region}
\label{sec:h2exp}

In this test we explore the problem of the expansion of an I-front
due to photoheating from a point source (Test 5 $\&$ 6 of \citealt{Iliev2009}).
The temperature of the gas is allowed to vary and the hydrodynamics is switched
on. The photons heat the region around the source through photoheating
(Eqs.~\ref{eq:energy} and \ref{eq:UVheating}), producing an over-pressurised
region which drives the gas out with a certain velocity.   The I-fronts are
generally classified by comparing their speed, to two critical speed of the
gas : R- critical, defined as ${v_\ts{R} = 2c_\ts{s,I,2}}$,  and D-critical,
given by ${v_\ts{D} \sim c_\ts{s,I,1}^2/(2c_\ts{s,I,2})}$, where ${c_\ts{s,I,1} =
(p_1/\rho_1)^{1/2}}$ and ${c_\ts{s,I,2} = (p_2/\rho_2)^{1/2}}$ are the isothermal
sound speeds in the gas ahead of and behind the I-front, respectively.
Typically, the I-front is initially R-type (${V_\ts{I} \geq V_\ts{R}}$), where it
expands supersonically with respect to the neutral gas ahead, which means RT
post-processing is a fairly good approximation. The I-front then begins to slow
down once it approaches the Str\"omgren radius. When ${v_\ts{D} < v_\ts{I} <
v_\ts{R}}$ (sometimes referred to as an M-type I-front), the I-front is
necessarily led by a shock which compresses the gas entering the I-front
sufficiently to slow it down and guarantees that it is converted to a D-type
front (${v_\ts{I} \leq v_\ts{D}}$).

We test the performance of our implementation under two different physical conditions,
first, we simulate the expansion of the I-front in a uniform density medium
and then move on to a more realistic situation where a source is at the center
of a spherically symmetric, steeply decreasing power-law density profile with a
small flat central core. For the first simulation we initialise a box of size
${2L\ts{box}=30\,\text{kpc}}$ on a side, which is resolved initially with
${2\times80^3}$ resolution elements placed in a regular staggered grid. As the
gas starts to move, the  mesh is allowed to move and distort according to the
local fluid motion. The box is initialised with a pure Hydrogen gas of density
and temperature ${n_\ts{H}=10^{-3}\,\text{cm}^{-3}}$ and ${T=100\,\text{K}}$
respectively. A constant luminosity source is placed at the center of the
domain that emits a black body spectrum with ${T_\ts{eff} = 10^5\,\text{K}}$ at
the rate of ${\dot{N}_\gamma=5\times10^{48}\,\text{photons s}^{-1}}$.  We use a
reduced speed of light with $f_r=0.01$ and the run the simulation for
${500\,\text{Myr}}$.  A multi-frequency radiative transfer scheme is employed, where,
the emitted photons are grouped into three separate bins with frequency ranges
[${13.6\,{\rm eV}, \, 24.6\,{\rm eV}}$),  [${24.6\,{\rm eV}, \, 54.4\,{\rm eV}}$)
and  [${54.4\,{\rm eV}, \, 100.0\,{\rm eV}}$).

\begin{figure}
\includegraphics[width=\columnwidth]{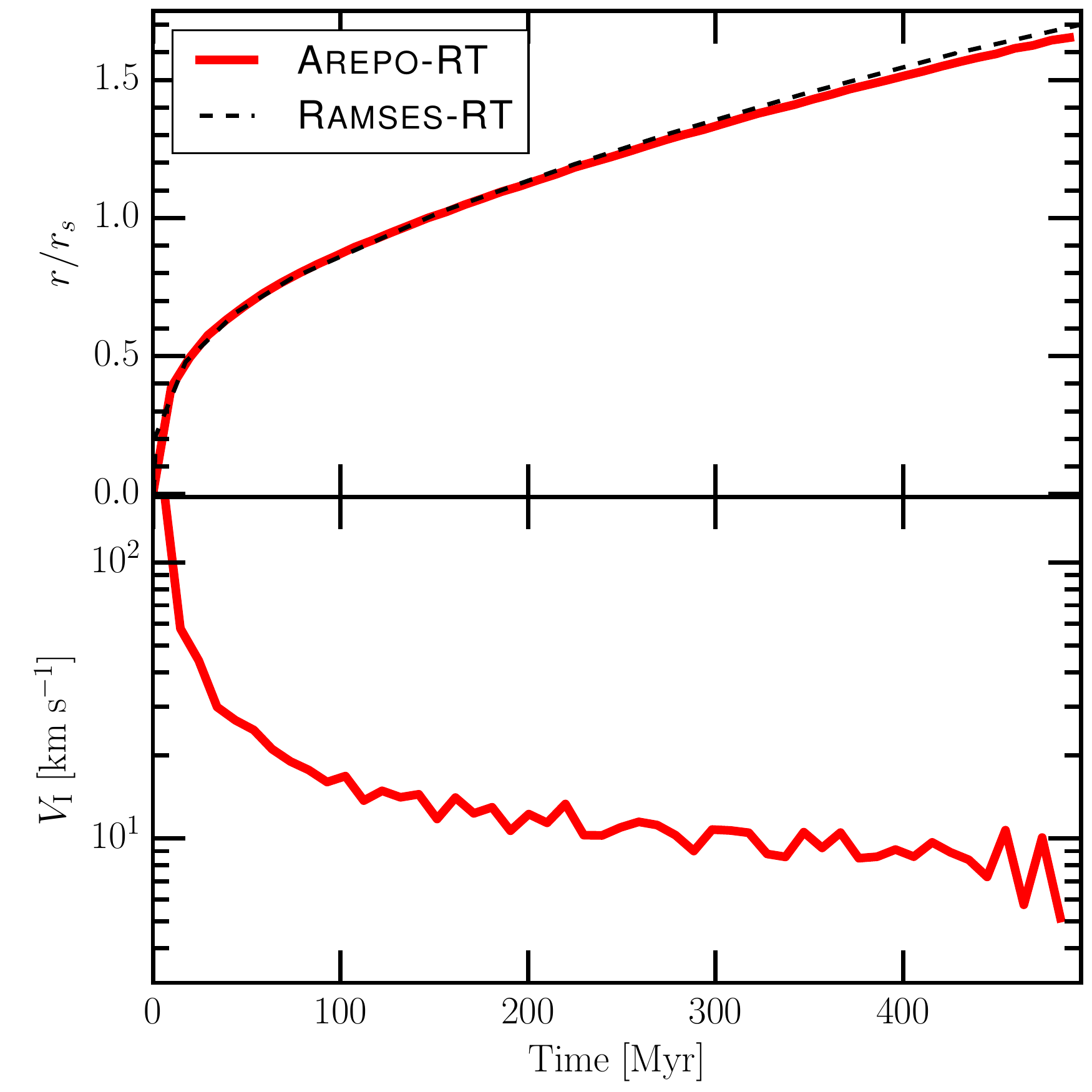}
\caption{{\bf Expansion of a $\ion{H}{\sc II}$ region}: The radius (top panel) and velocity (bottom panel) of the ionization front as a function of time in the expansion of a \ion{H}{II} region in a constant density medium. The evolution of the ionization front matches well with results obtained for the same test with {\sc ramses-rt} (dashed black curve) as reported in R13.}
\label{fig:h2t_const}
\end{figure}

\begin{figure*}
\begin{center}
\includegraphics[scale=0.46]{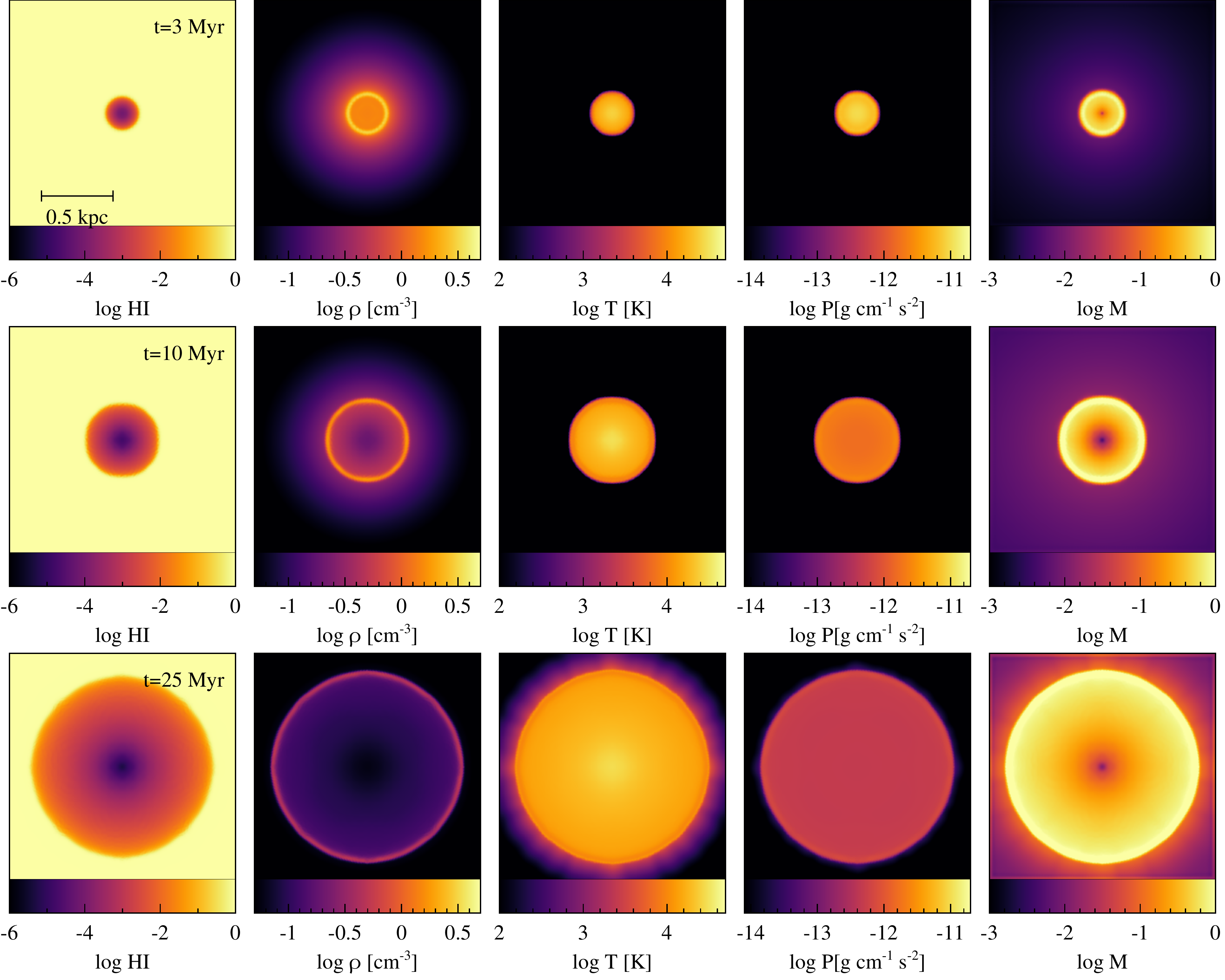}
\caption{{\bf Expansion of a $\ion{H}{\sc II}$ region}: Maps showing slices of the domain at $z=0.8 \ \text{kpc}$ for the simulation of the expansion of a \ion{H}{II} in a $r^{-2}$ initial density profile. The \ion{H}{I} fraction (first column), density (second column), temperature (third column), pressure (fourth column) and Mach number (fifth column) are plotted  at $3$ Myr (top panels), $10$ Myr (middle panels) and $25$ Myr (bottom panels).}
\label{fig:h22map}
\end{center}
\end{figure*}

\begin{figure*}
\includegraphics[scale=0.4]{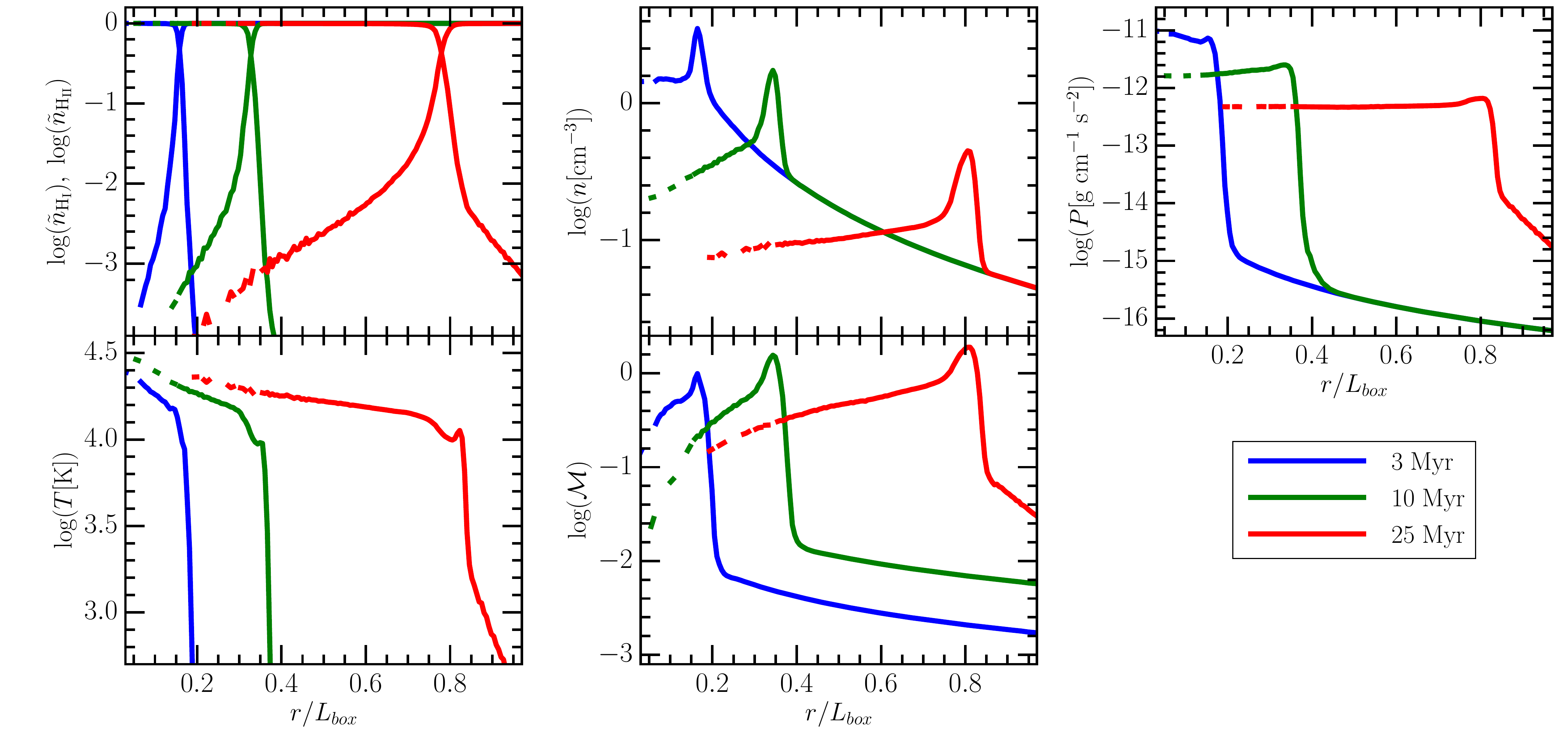}
\caption{{\bf Expansion of a $\ion{H}{\sc II}$ region}: The ionization (top left panel), density (top middle panel), pressure (top right panel), temperature (bottom left panel) and Mach number (bottom middle panel) profiles at ${3\,\text{Myr}}$ (blue curves), ${10\,\text{Myr}}$ (green curves) and ${25\,\text{Myr}}$ (red curves) in the simulation of the expansion of an \ion{H}{II} region in a $r^{-2}$ density profile. The profiles generally match the results from previous simulations of the same test reported in \citet[Test $6$]{Iliev2009}.}
\label{fig:h2var}
\end{figure*}

Fig.~\ref{fig:h21map} presents cross section (at ${z=15 \ \text{kpc}}$) maps
of the \ion{H}{I} fraction (first column), density (second column), temperature
(third column), pressure (fourth column) and Mach number (${\mathcal{M} =
v/c_s}$; fifth column) at $10$ Myr (top panels), $200$ Myr (middle panels) and
$500$ Myr (bottom panels). Although the radiation is able to ionize and heat
the gas in the central ${\sim 2 \ \text{kpc}}$ within $10$ Myr, there is very
little change in the density of the gas. The gas velocity is still very low
implying that it has not yet started evacuating the central regions. By $200$
Myr, the over pressurised region manages to push enough gas out of the central
regions and a high density expanding gas shell is formed around a low density
region. Interestingly, a second transient density peak forms beyond the
I-front, which is replicated in the pressure and Mach number maps. The I-front
expands almost to $L_\text{box}$ by $500$ Myr. We note that the maps show some
asymmetric artifacts that can be attributed to the geometry of the underlying
mesh.  In our experiments, these kind of features emerge only when there is a large scale
coherence in the geometry of the underlying mesh, and disappear as the mesh
becomes more unstructured as we will demonstrate later in this section.
 
In Fig.~\ref{fig:h2const} we plot the ionization fraction (${{\tilde n}_\ion{H}{I},
{\tilde n}_\ion{H}{II}}$), density ($\rho$), pressure ($P$), temperature ($T$) and
Mach number (${{\mathcal M}=v/c_s}$) as a function of radius ($r$) at
${t=10\,\text{Myr}}$ (blue curves), ${t=200\,\text{Myr}}$ (green curves) and
${t=500\,\text{Myr}}$ (red curves). By 10 Myr, the temperature increases
behind the I-front due to photoheating which in turn increases the pressure.
The density has not changed by much because the gas has not had time to react
to these changes. This is a classic R-type front which moves supersonically to
about the Str\"omgren radius (${\sim 5.4\,\text{kpc}}$), within a single
recombination timescale (${t_\ts{rec}\simeq125\,\text{Myr}}$). By ${200\, \text{Myr}}$, the
expansion of the I-front has progressed beyond what is expected from a pure
R-type expansion. The I-front is now D-type meaning that the front moves along
with the gas. The density behind the front is reduced, as the gas reacts to the
pressure jump inside the front causing it to flow radially outwards. The gas
piles up at the position of the I-front inducing a small density peak. It is
important to note, however,  that there is a second density peak beyond the
I-front, which is reproduced in the temperature, pressure and Mach number
profiles. This is the temporary effect of photoheating by high energy photons.
The front then expands slowly outwards till a pressure equilibrium is reached.
The final radius of the \ion{H}{II} region ($r_f$) is given by,
\begin{equation}
r_f\simeq\left(\frac{2T}{T_e}\right)^{2/3}r_{s} \sim 220\,\text{kpc}\, ,
\end{equation}
where $T$ is the temperature inside the  \ion{H}{II} region, $T_e$ is the
background temperature and $r_s$ is the Str\"omgren radius
(Eq.~\ref{eq:str_rad}). This indicates that the domain needs to be much larger
in order to simulate the equilibrium state of the solution. Therefore, we stop
the simulation at ${500\,\text{Myr}}$, a time at which the I-front is still
within the domain and plot the profiles as shown. We note that the positions of
the fronts and profiles match previous results in \citet{Iliev2009} and R13 very well.

Fig.~\ref{fig:h2t_const} shows the position and velocity of the I-front
(defined as where the radial average of ${\tilde n}_\ion{H}{II}$ is equal to
$0.5$).  For comparison we also overplot the position of the front obtained
using the {\sc ramses-rt} code presented in R13. The curves for the two codes
are virtually identical, with a very slight difference in the speed at late
times, which can be attributed to improper boundary conditions as the size of
the I-front becomes comparable to the box-size. 

\begin{figure}
\includegraphics[width=\columnwidth]{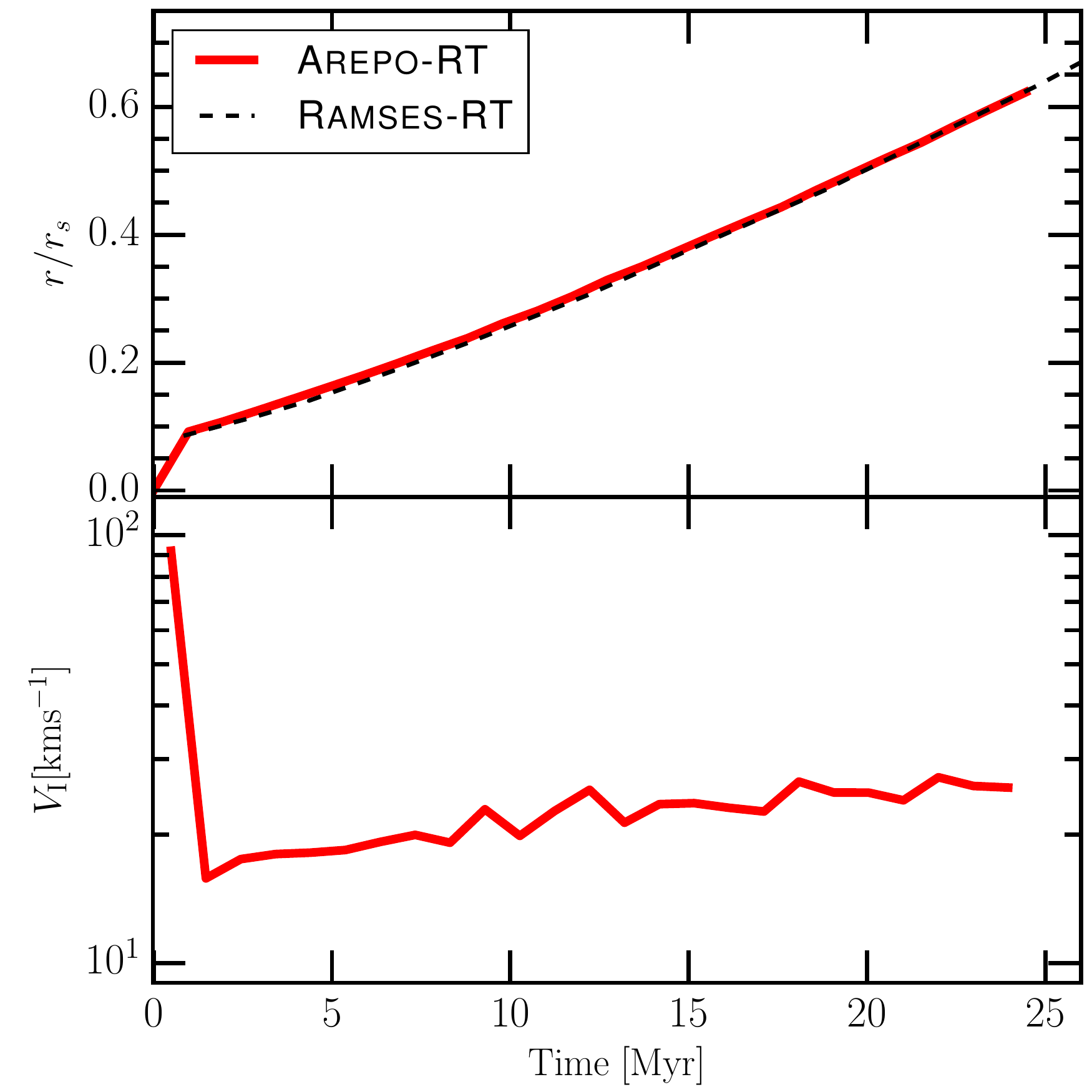}
\caption{{\bf Expansion of a $\ion{H}{\sc II}$ region}: The radius (top panel) and velocity (bottom panel) of the ionization front as a function of time in the of the expansion of a \ion{H}{II} region in a $r^{-2}$ density profile. The evolution of the ionization front matches well with results obtained for the same test with {\sc ramses-rt} (dashed black curve) as reported in R13.}
\label{fig:h2t_var}
\end{figure}

We now turn our attention to an I-front expansion created by a point source at
the centre of a spherically symmetric, steeply decreasing power-law density
profile with a small flat central core of density $n_0$ and radius $r_0$:
\begin{equation}
n_\ts{H}(r) =  
\begin{cases}
 n_0 &\quad\text{if} \ r<r_0 \\
 n_0 (r_0/r)^2 &\quad\text{if} \ r \geq r_0 \, .
\end{cases}
\end{equation}
Within the central core, the I-front propagates similarly as in the previous
simulation. The propagation of an I-front in $r^{-2}$ density profiles with
full gas dynamics does not have an exact analytical solution, but some insights
can be gained by comparing the core radius ($r_0$) to the Str\"omgren radius
($r_s$). For ${r_s<r_0}$ the I-front stalls within the core, converts to D-type,
but starts to re-accelerate upon entering the steep density gradient.
Alternatively, when ${r_s \geq r_0}$, the source flash-ionizes the cloud on
time-scales shorter than the dynamical time of the gas. We simulate the first
case, which is more interesting and is a better test of the coupling between
the radiation field and the hydrodynamics. 

For this test, the domain of side length ${2L_\ts{box}=1.6\, \text{kpc}}$ is resolved with 
${2\times80^3}$ resolution elements placed on a regular staggered grid. As
before, the mesh is allowed to move and distort according to local fluid flow.
The central core has a density of ${n_0=3.2\,\text{cm}^{-3}}$ and a radius of
${r_0=91.5\,\text{pc}}$. The central source is a black body spectrum with
${T_\ts{eff}=10^5\, \text{K}}$ and emits at a rate of ${10^{50}\,\text{photons
s}^{-1}}$. The initial temperature of the gas is ${100\,\text{K}}$. For these
parameters, ${r_s\simeq 70 \, \text{pc}}$, meaning that the I-front changes from
R-type to D-type within the core. The simulation is run for ${25\,\text{Myr}}$,
which is much larger than the recombination timescale within the core, which is
about ${t_\ts{rec}\sim0.04\,\text{Myr}}$. 

Fig.~\ref{fig:h22map} shows the cross section (at ${z=0.8 \ \text{kpc}}$) maps of
the \ion{H}{I} fraction (first column), density (second column), temperature
(third column), pressure (fourth column) and the Mach number (fifth column) at
$3$ Myr (top panels), $10$ Myr (middle panels) and $25$ Myr (bottom panels).
The photo-heated gas pressure is able to evacuate most of the gas from the
central regions, completely changing the background density profile. We see
that the I-front is more spherical than in the previous simulation because the
gas velocities are higher which distorts the mesh and reduces the large scale
coherence in its geometry. 

Fig.~\ref{fig:h2var} plots the ionization fraction (${\tilde n}_\ion{H}{I},
{\tilde n}_\ion{H}{II}$), density ($\rho$), pressure ($P$), temperature ($T$) and
Mach number (${{\mathcal M}=v/c_s}$) as a function of radius ($r$) at
${t=3\,\text{Myr}}$ (blue curves), ${t=10\,\text{Myr}}$ (green curves) and
${t=25\,\text{Myr}}$ (red curves). By ${3\,\text{Myr}}$ the I-front has moved out
of the central core and is expanding rapidly as a D-type front due to the steep
density gradient outside the core. We obtain very sharp I-fronts and the profiles of the various hydrodynamic quantities
match quite well with same test performed with an array of numerical schemes
presented in \citet{Iliev2009}.

Finally, Fig.~\ref{fig:h2t_var} shows the position and velocity of the I-front
(defined as where the radial average of ${\tilde n}_\ion{H}{II}$ is equal to
$0.5$).  For comparison, we also overplot the position of the front obtained
using the {\sc ramses-rt} code presented in R13.  

These tests verify the accuracy and reliability of our RT scheme on randomly
oriented, moving meshes. They also validate the accuracy of the coupling
between the hydrodynamics and the radiation field in our implementation.

\subsection{Radiation pressure driven outflows}
\label{sec:radpress}

Next we assess the accuracy of momentum injection into the gas due to
photon absorption, i.e., radiation pressure (Eq.~\ref{eq:uvmomentum}). During 
photon-matter interactions, both energy and momentum are conserved.
Photoheating is the product of thermalisation of the left-over energy above
the ionization threshold of the ionic species. The photon's momentum is also
transferred to the ion, which receives a kick (${\Delta v = E/cm}$) in the
direction of the absorbed photon. An estimate of the change in momentum of a
optically thick shell of gas under spherically symmetric geometry can be
written as ${\dot{P}=L/c}$, where $L$ is the luminosity of the source.  If
$L$ is invariant then the change in velocity of the shell in time ${\Delta t}$ is 
\begin{equation}
\Delta v_\ts{shell} = \frac{L \Delta t}{m_\ts{shell}c}\, ,
\label{eq:rad1}
\end{equation}
where $m_\ts{shell}$ is the mass of the shell. In case of a shell expanding
into a uniform background of density $\rho_0$, the velocity of the shell
$v_\ts{shell}$ at any time $t$ is given by \citep{Wise2012}
\begin{equation}
v_\ts{shell} = tA(r_i^4+2A\,t^2)^{-3/4}\, ,
\label{eq:rad2}
\end{equation}
where ${A = 3L/4\pi\rho_0c}$ and $r_i$ is the starting position of the shell. The
shell will first form once ionization balances recombinations at the
Str\"omgren radius, therefore ${r_i = r_s}$.

\begin{figure}
\includegraphics[width=\columnwidth]{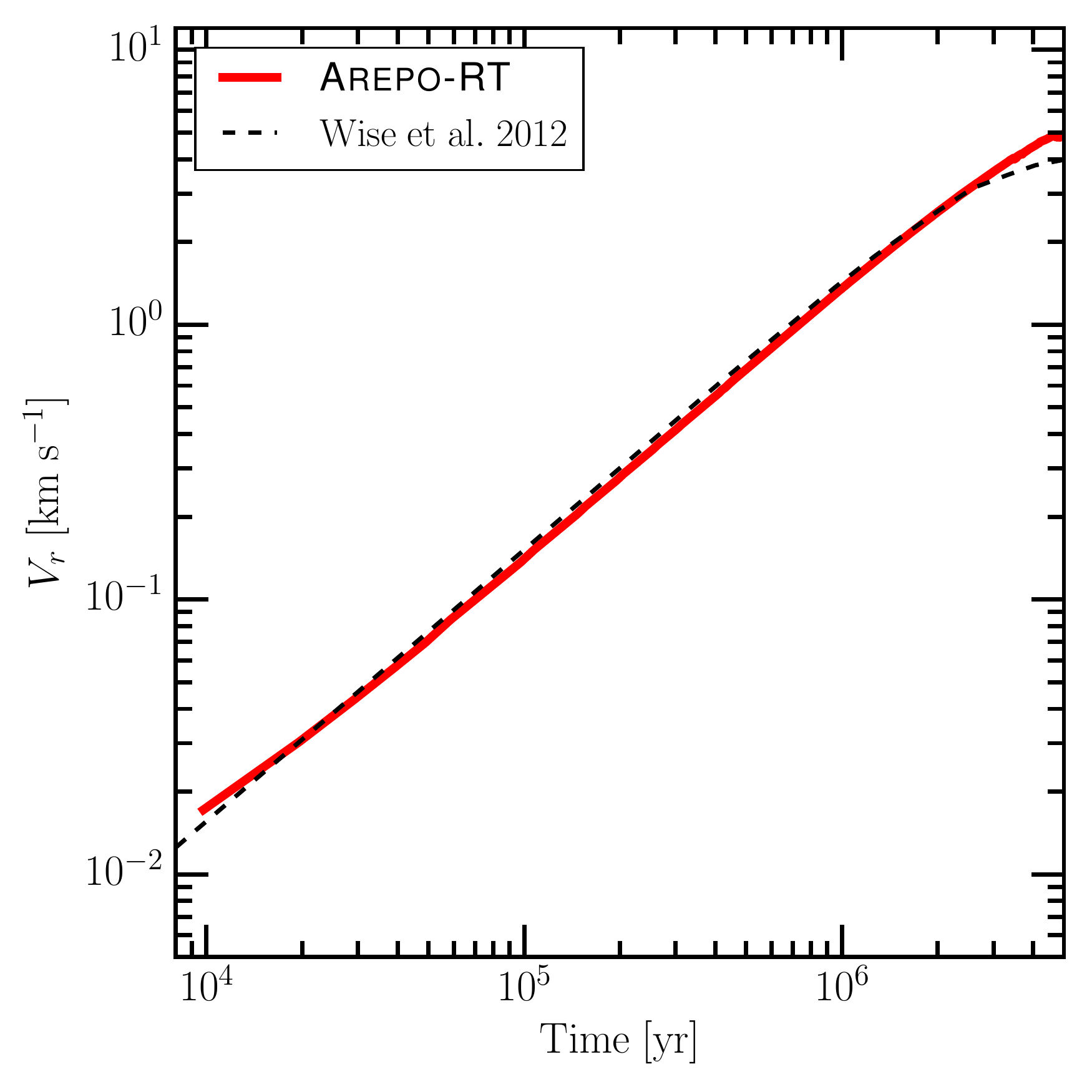}
\caption{{\bf Radiation pressure driven outflows}: Velocity as an function of time for gas with a initial density $n_{\rm H} = 1 \ \text{cm}^{-3}$ at a fixed temperature of  $T = 100 \ \text{K}$. The simulation (red curve) agrees well with analytical estimations from Eq.~\ref{eq:rad2} (dashed black curve).}
\label{fig:radpress}
\end{figure}

Here we follow the test setup presented in \citet{Sales2014}. Specifically, we
look at the outflow velocities generated when a constant
monochromatic (${E=13.6\,\text{eV}}$) source is placed in the center of a uniform
medium. The luminosity of the source is ${L=10^6L_\odot}$ which translates to a
photon injection rate of ${\dot{N}_\gamma = 1.8\times10^{50}\,\text{photons
s}^{-1}}$.  The initial density and temperature of the gas are ${n_{\rm H} =
1\,\text{cm}^{-3}}$ and ${T=100\,\text{K}}$ respectively, with the gas composed of
only neutral Hydrogen atoms. A monochromatic source with the energy equal to
the ionization potential of the Hydrogen atom implies that there is no photoheating
 and therefore the internal energy per unit
mass of the cells, $u$, experiences no change due to the presence of a luminous
source. This translates into an approximately constant temperature in the
cells, except the change in temperature due to the change in the mean molecular
weight of the gas due to ionization. For these parameters,
 ${r_s=51.7\,\text{pc}}$ and the corresponding recombination time is
${t_\ts{rec}=2.43\times 10^{3}\,\text{yr}}$.  The simulation domain is
${L_\text{box}=200\,\text{pc}}$ on a side, initialised with ${2\times80^3}$ resolution
elements placed in a regular staggered grid. As with previous RHD tests, the
mesh is allowed to move and distort according to local fluid flow. The
simulation is run for ${5 \,\text{Myr}}$. 

\begin{figure}
\includegraphics[width=\columnwidth]{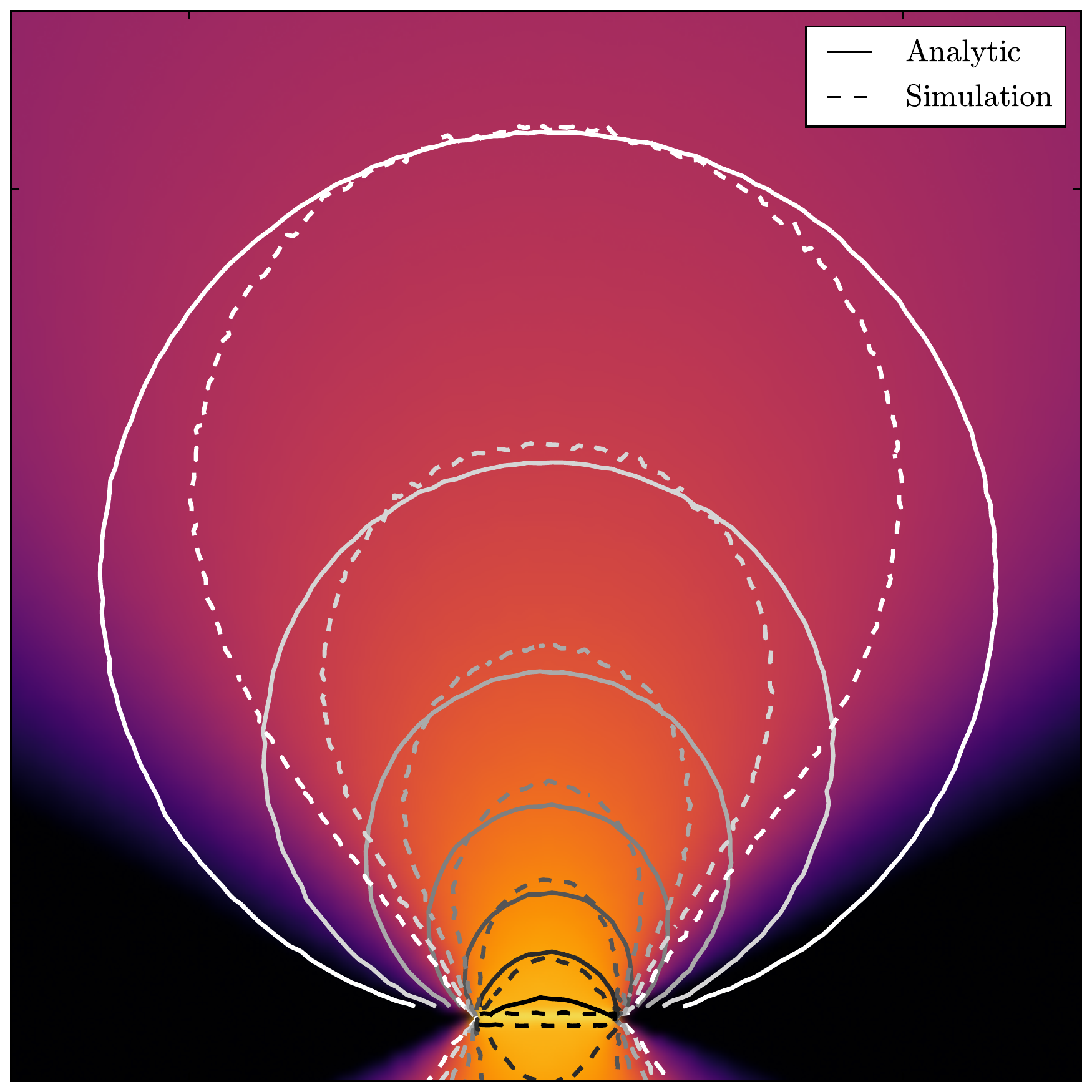}
\caption{{\bf Free-streaming radiation from a thin disc}: The radiation field morphology around a thin disc surrounded by an optically thick torus. The solid lines depict the analytic radiation field contours that are expected from this setup while the dashed lines plot the simulation results. The simulation slightly overshoots the analytic solution in the $y-$direction and undershoots it in the $x-$direction.}
\label{fig:disc}
\end{figure}

\begin{figure*}
\begin{center}
\includegraphics[scale=0.6]{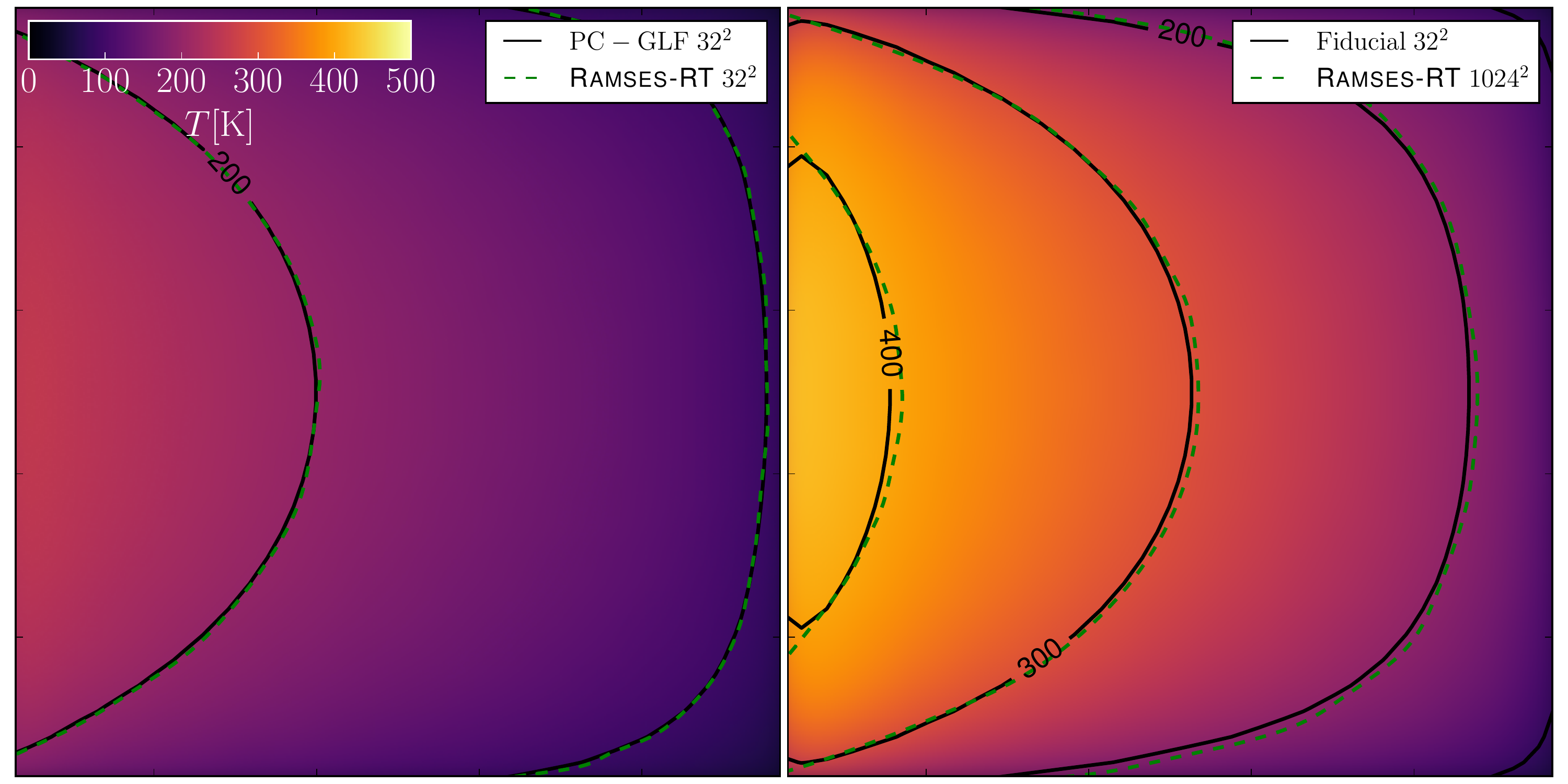}
\caption{{\bf Dust absorption in an optically thick medium}: The maps showing the stationary results from 2-D runs with a constant flux of photons into the box from the left. The color represents the radiation temperature, $T_r$, as indicated by the color bar and contours indicated by the solid black lines. The left panel shows the result using a PC-GLF scheme, which matches very well with the solution obtained in the $32^2$ simulation of R15 (dashed green curves). The plot on the right hand side shows the results from the same simulation but now run with our fiducial scheme. The radiation temperatures are about two times higher and is able to match the $1024^2$ simulation of R15. }
\label{fig:trap}
\end{center}
\end{figure*}

Fig.~\ref{fig:radpress} shows the velocity (red curve) of the ionized shell
defined as mass weighted mean velocity calculated within a radius at which
almost all the gas is fully ionized. For comparison we also plot (black dashed
curve) the expected analytic result from \citet[Eq.~\ref{eq:rad2}]{Wise2012},
which is in very good agreement with the simulated gas velocity. Initially, the
I-front has a larger velocity compared to the analytic result as the light
travels to the Str\"omgren radius before the gas has time to react to the
changes.  At later times (${t>1~\text{Myr}}$), mass entrainment will slow down the gas, an effect
which can be seen in the \citet{Wise2012} formulation as well. However, the
simulation curves turn over at larger velocities compared to the analytic estimate which is also seen in
\citet{Sales2014}. This test confirms the accuracy of our radiation pressure
implementation.

\subsection{Free-streaming radiation from a thin disc}
\label{sec:disc}
We have demonstrated in previous tests that the M1 closure relation performs
remarkably well for a wide range of problems. However, the previous tests only
simulate the propagation of radiation from a single source or a plane parallel
wave of radiation. A well known disadvantage of the M1 closure is the inability to
accurately model convergent rays (\citealt{Gonzalez2007}; R13). The closure
relation tends to produce spurious perpendicular flux when two rays are
converging to a point, instead of passing through each other. Therefore, it is
important to understand the limitations of this approximation and to test it
under realistic conditions. 

We therefore perform the test described in R15 to check the validity of our scheme in a multiple source
geometry. Specifically, we test the distribution of the radiation field
intensity, when a luminous thin disk is surrounded by an optically thick torus.
The rest of the domain is assumed to contain a tenuous optically thin gas that
allows for free transport of radiation. The 2D box of size ${L_{\rm box} = 1}$ on a
side is resolved by $128^2$ resolution elements.  A thin luminous disc of
height $1/128$ (corresponding to one cell width) and width ${L=0.125}$ is placed
parallel to the $x$-axis and is centered at ${(x,y)=(0.5,0.1)}$.  Surrounding this
disk is a one cell high torus that is is optically thick and for the purposes
of this simulation acts as a radiation sink.  The disc has a constant photon
energy density ($E_0$), which is imposed at every timestep. The disc and the torus
are made up of regular Cartesian cells while the rest of the domain is made up
of an irregular mesh obtained by randomly deviating the cell centers of a
Cartesian mesh by a ${0.2\Delta x}$ where ${\Delta x}$ is the cell width, mimicking
a typical deviation between mesh-generating points and cell centers in real
problems \citep{Vogelsberger2012}.

The field morphology can be obtained analytically for this setup as shown in
R15
\begin{equation}
E(x,y) = \frac{E_0}{2\pi} \left[  \arctan \left(\frac{L/2-x}{y}\right)  + \arctan \left(\frac{L/2+x}{y}\right)\right]\, ,
\end{equation}
in a coordinate system, whose center is defined at the center of the disc,
i.e., at $(x,y)=(0.5,0.1)\,\text{cm}$.

In Fig.~\ref{fig:disc} we present the histogram of the radiation field intensity in the
box, with the dashed contours plotting the simulation results and the solid
lines representing the analytic solution. The small irregularity in the
contours arises because of the irregularity of the underlying mesh. We see that
the simulation matches the analytic result at least qualitatively. However,
there are a few important differences which highlight the shortcomings of the M1
closure approximation. Namely, the radiation field intensity contours overshoot
the analytic solution in the $y-$direction by a small amount and undershoot
the solution in the $x-$direction. This is because the photons from the right
side of the disc are not able to propagate to the left side of the domain, because
they interact with the photons that emanate from the left side of the disc
going in the opposite direction. This causes spurious perpendicular fluxes,
thus overshooting the solution in the $y-$direction and undershooting the
solution in the $x-$direction. We note that R15 finds similar
results with their M1 closure scheme and that we have a qualitative agreement
with the analytic estimate for the radiation field morphology.

\subsection{Dust absorption in an optically thick medium}
\label{sec:dustabs}

In this section, we test the coupling between dust, gas and the IR radiation
field. Specifically we examine how well the IR dust-gas coupling performs in
the case of absorption in an optically thick regime. It is well know that in
highly optically thick media the photons propagate in a random walk reducing
the radiation transport equations to an isotropic diffusion equation.

Previous works \citep{Liu1987, Bouchut2004} have shown that if the numerical
diffusion of the scheme becomes larger than the true radiation diffusion then
the operator split approach to solve the RT equations is not valid anymore as
the source terms become extremely stiff compared to the hyperbolic transport
terms. Therefore, the ability of a numerical scheme to model radiation
transport in an optically thick regime is very sensitive to the inherent
numerical diffusivity of the scheme. There are a couple of ways to overcome
this problem. \citet{Berthon2007} proposed to modify the Riemann solution such
that it explicitly takes care of the source terms. This solution to the Riemann
problem becomes much more complicated but it does recover the right asymptotic
limit in the optically thick regime. However, it is unclear how radiation
pressure can be accounted for in such a scheme. 
 
R15 on the other hand propose an alternative method based on the Isotropic
Diffusion Source Approximation (ISDA) methodology \citep{Lieb2009}. The IR
photon group is split into two components, a trapped component ($E_t$) and a
free streaming component ($E_s$). The trapped radiation energy is assumed to be
strictly isotropic in angular space, corresponding to the asymptotic limit of
vanishingly small mean free path. The amount of trapped photons within a cell
can be obtained by comparing the numerical diffusivity of the scheme with the
expected analytic diffusivity in the diffusion limit. Specifically the free
streaming flux in the diffusion limit is 
 \begin{equation}
  {\bf F}_s \simeq  \frac{{\tilde c}}{3\kappa_\ts{R} \, \rho}{\bf \nabla}E_t\, ,
 \end{equation}
and the numerical diffusion for a PC-GLF scheme is
 \begin{equation}
  {\bf F}_s \simeq  \frac{{\tilde c \, \Delta x}}{3}{\bf \nabla}E_s\,.
  \label{eq:numdiff}
 \end{equation}
Equating these two equations gives ${E_t = 3\tau_c \, E_s/2}$, where $\tau_c$ is
the optical depth of the cell. The trapped photons are advected along with the
gas and the radiation pressure from trapped photons is accounted for by adding
an additional non-thermal pressure component from the radiation field, ${P =
P_{\text{therm}} + P_{\text{rad}}}$, where ${P_{\text{rad}} = {\tilde c}
E_t/(3c)}$, to the momentum conservation equation (Eq.~\ref{eq:momentum}). This
method produces accurate results in the diffusion limit. 
 
There are, however, a few drawbacks with this scheme. First, we note that
Eq.~\ref{eq:numdiff} is only valid if the Riemann problem at the interface is
solved using a GLF flux function and more importantly the left and right state
inputs to the Riemann solver must be the cell centered values and no
longer works if the the left and right states of the interface are the gradient
extrapolated values. This forces the underlying numerical scheme to follow a
piece-wise constant approach, which, as seen before, is very diffusive and
has suboptimal convergence properties (see Section~\ref{sec:adv} for more
details). Secondly, even at relatively low optical depths (${\sim 1}$), $60\%$ of
the total radiation flux is deposited into the trapped component. This
component is isotropic in angular space and hence the diffusion is isotropic.
Therefore, the M1 scheme reverts back to a flux limited diffusion (FLD;
\citealt{Lucy1977, Krumholz2007}) scheme, thereby, erasing the directionality
of the initial photon field. By the same rationale, the radiation pressure will
be isotropic and will generate isotropic velocities even when the underlying
radiation field has an inherent directionality.
 
For these reasons, we chose not to implement sub-grid schemes to model the
diffusion limit. Our fiducial scheme is extremely accurate and has
very low diffusivity. Moreover, the convergence order of our scheme is $\sim 2.0$. This implies that a small improvement in the resolution
will decidedly improve the accuracy of the solution. Since, {\sc Arepo} is
a moving mesh code, which automatically refines the high density regions (and
correspondingly high opacity regions), we are able to obtain accurate
solutions in many realistic problems as we will show in the present and
forthcoming tests. 
 
To elucidate the points made in this section we perform a quantitative test, proposed in R15, as
a simple demonstration of the accuracy of our scheme. We initialise a 2D domain
of size ${1\,{\text{pc}}}$ on a side. The opacity is ${\kappa_\ts{R} \rho = 6.48
\times 10^{-17}\,{\text{cm}^{-1}}}$, which sets ${\tau_\text{box} = 200}$. The
underlying mesh is a regular Cartesian grid with 32 resolution elements on a
side, translating to an optical depth in the cell of ${\tau_c = 6.25}$. We chose
a Cartesian grid so as to compare our scheme to that of R15. The left boundary
emits a constant IR flux of ${5.44\times 10^4\,\text{erg s}^{-1}\text{cm}^{-1}}$.
The rest of the boundaries act as sinks for radiation. The hydrodynamics is
turned off and the only source of cooling and heating of gas is through the
gas-dust IR coupling. For this test we use the full speed of light. The initial
temperature of the gas is $10$ K. The simulation is run till the result has
converged to a steady state solution.
 
We perform two different simulations, one with a piecewise constant
approximation and a GLF flux (PC-GLF) function and another with our fiducial
scheme. The PC-GLF run mimics the scheme outlined in R15. The left panel of
Fig.~\ref{fig:trap} shows the temperature map in the PC-GLF run. The solid
black lines show the contours of the temperature map for the simulation and the
dashed green lines are the contours from the same test performed by R15. The
PC-GLF run matches quite nicely with the $32^2$ run of R15, confirming that our
PC approximation performs equally well with their scheme. The maximum
temperature reached is about ${\sim 260\,\text{K}}$. 

As the optical depth of the cell is quite high, the numerical diffusion of the
PC-GLF scheme exceeds that of the analytic value and hence the IR heating of
the gas is massively underestimated. We rerun the same low resolution
$32^2$  simulation with our fiducial scheme. The right panel shows
the temperature map in our fiducial run with black solid contours indicating
the temperature obtained in our simulation and the dashed green line now
instead shows the results from R15's high resolution $1024^2$ simulation. We
see that our low resolution $32^2$ simulation using the fiducial scheme accurately reproduces R15's
$1024^2$ simulation. This is because our scheme is able to reduce the numerical
diffusivity by a large amount, thereby accurately capturing the diffusion limit
without the need for sub-grid diffusion models.

\subsection{Diffusion of constant luminosity source}
\label{sec:diff}
In this sub-section we redo the test proposed in R15 (Section 3.6), with the aim to quantify the numerical diffusivity of our scheme
in the radiation diffusion regime. A 3D domain of ${L_\text{box}=500\,\text{pc}}$
on a side is initialised with $32^3$ resolution elements arranged in a
Cartesian mesh. A source with a constant luminosity  of ${L =
10^{50}\,\text{photons s}^{-1}}$ is placed in the center of the domain. The gas
is assumed to have an opacity of ${\kappa_\ts{R} = 10\,\text{cm}^2\,\text{g}^{-1}}$.
The hydrodynamics is turned off, but the radiation is allowed to propagate
radially outward and the simulation is stopped when a steady state solution is reached. The density of the gas is varied from
${n_\ion{H}{}=5-10^4\,\text{cm}^{-3}}$ corresponding to cell optical depths of
${\tau_c = 0.004-8}$. 

The diffusion equation in a uniform optically thick medium is given as
\begin{equation}
 \frac{\partial N}{\partial t} - \frac{\tilde c}{3\kappa_\ts{R} \, \rho}{\bf \nabla}^2 N + \mathcal{L} = 0 \,,
\end{equation}
where $\rho$ is the density of the gas, $\mathcal{L}$ is the luminosity per unit volume and $N$ is the number density of photons. The steady state solution of this equation is then
\begin{equation}
 {\tilde c}N(r) = \frac{3\rho \, \kappa_\ts{R} \, L}{4\pi \, r}\, ,
 \label{eq:diffconv}
\end{equation}
where $r$ is the distance from the source. So, ideally, we should expect the photon number density to diminish as $1/r$. 

The analytic expectation is derived assuming an infinite homogeneous medium and
the diffusion equation only formally achieves a steady state solution at
${t=\infty}$. Both these approximations are broken in our simulation setup. In
order, to approximate the infinite spatial medium, the values of the radiation
variables are set such that they roughly match the expected slope of
Eq.~\ref{eq:diffconv}, i.e.,
\begin{equation}
 \boldsymbol{\mathcal{U}}_s = \boldsymbol{\mathcal{U}}_f \left(1-\frac{\Delta x}{L_\text{box}}\right)\, ,
\end{equation}
where ${\boldsymbol{\mathcal{U}} = (N,{\bf F})}$, ${\Delta x}$ is the distance
between the cell centers at the boundary and the subscripts $s$ and $f$ refer
to the solid layer cell and the fluid layer cell at the boundary, respectively.
The steady state solution is assumed to have been reached once the change in
$\boldsymbol{\mathcal{U}}$ is less than $1\%$ throughout the domain. 

\begin{figure}
\includegraphics[width=\columnwidth]{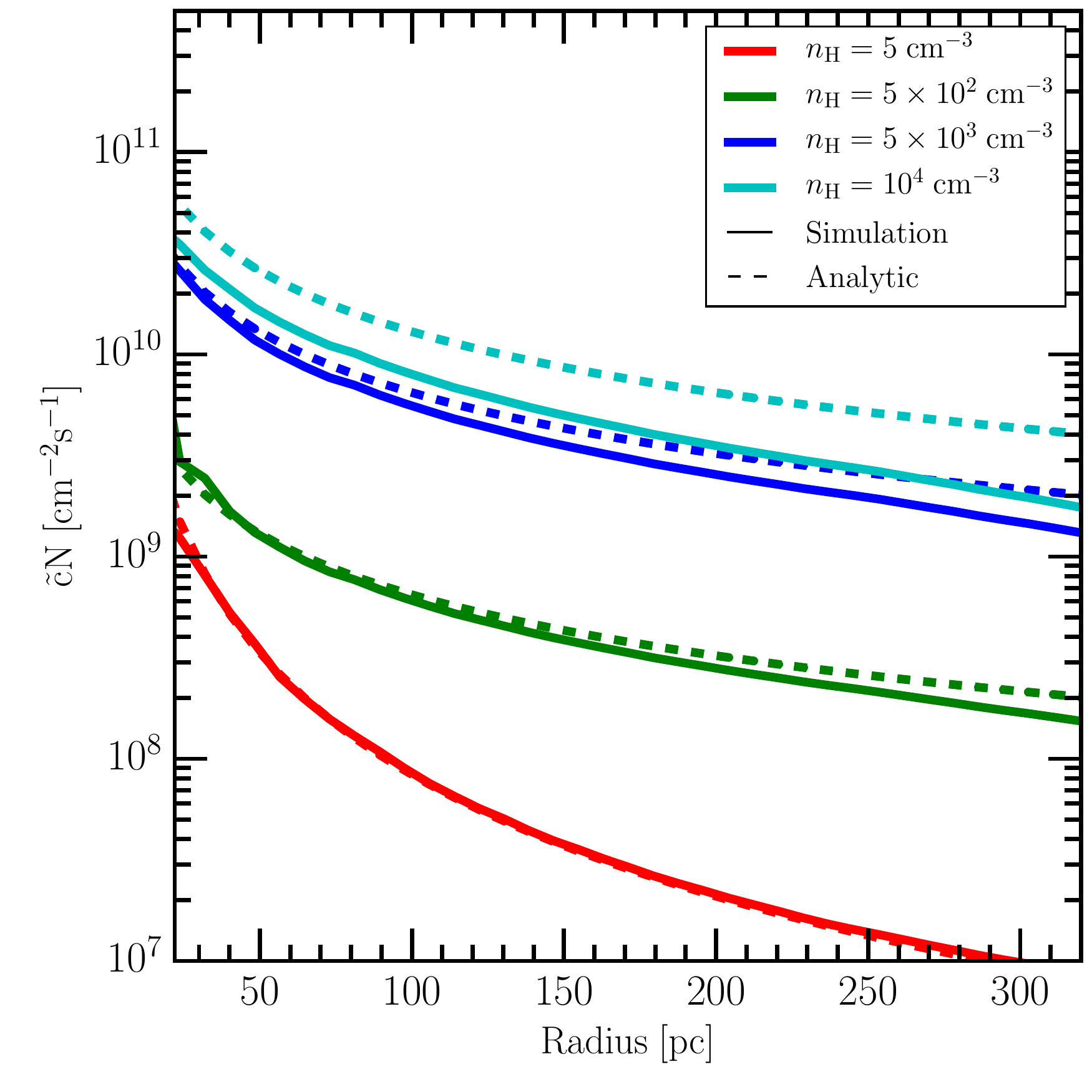}
\caption{{\bf Diffusion of constant luminosity source}: The plots show time-converged radiation profiles from the source at the center of the box in radiation tests with optical depths of $0.004$ (red curves), $0.4$ (green curves), $4$ (blue curves) and $8$ (cyan curves). The simulation results (solid curves) match the analytic solution (dashed curves) up to a cell optical depth of about ${\tau_c \sim 4}$ above which the numerical diffusion starts to dominate.}
\label{fig:diffprof}
\end{figure}

Fig.~\ref{fig:diffprof} shows the simulated (solid curves) and analytic (dashed
curves) photon number density profiles for runs with gas densities,
${n_\ion{H}{}=5\,\text{cm}^{-3}}$ (red curves),
${n_\ion{H}{}=5\times10^2\,\text{cm}^{-3}}$ (green curves),
${n_\ion{H}{}=5\times10^3\,\text{cm}^{-3}}$ (blue curves) and
${n_\ion{H}{}=10^4\,\text{cm}^{-3}}$ (cyan curves). The simulation with the
lowest density (and hence the lowest optical depth) gas is optically thin to
the radiation and hence the photons stream out of the box without any
hindrance. This leads to a ${N\propto1/r^2}$ dependence which is reproduced by
the simulation as expected. 

As the density increases the gas gradually becomes optically thick and the
radiation field starts to diffuse through the medium rather than streaming out
and the photon density profiles gradually starts to follow the $1/r$ relation
given by Eq.~\ref{eq:diffconv}. In the simulations with
${n_\ion{H}{}=5\times10^2\,\text{cm}^{-3}}$ the cells have optical depths of
${\tau_c=0.4}$, which can just be resolved by the R15 scheme. We do recover the
results with great accuracy in our scheme as well. However, this is the limit
to which the R15 scheme can resolve the photon diffusion and it has troubles capturing
the correct solution for higher optical depths as evidenced by very little
change in the simulation result for higher densities (left two panels of Fig.8
in R15). This is because, beyond this opacity the numerical diffusion of their
scheme becomes larger than the expected analytic diffusion. This is corrected
by using a sub-grid diffusion model which has its own drawbacks as discussed in
the previous section (\ref{sec:dustabs}). However, in our simulation with a gas
density of ${n_\ion{H}{}=5\times10^2\,\text{cm}^{-3}}$ having a optical depth of
${\tau_c\sim 4}$, our scheme is able to reproduce the analytic solution very 
well. This implies that the numerical diffusion of the fiducial scheme is so
low that it captures the right solution even when the photon mean free path is
under resolved. This is also seen in the previous section, where the scheme was
able to obtain the right results even when the cell optical depth was
${\tau_c\sim6.25}$. There is a slight discrepancy between the analytic and
simulation results at large radii which is attributed to the fact that the
boundary conditions are imperfect. 

Obviously, the numerical diffusion of our scheme cannot be identically zero.
The results from the highest density
(${n_\ion{H}{}=5\times10^2\,\text{cm}^{-3}}$) show the limitation of the scheme.
The cell optical depth is about ${\tau_c\sim 8}$, and the simulation undershoots
the photon number density at all radii in this regime, indicating that the
numerical diffusion is larger at such high optical depths.  

To conclude, our numerical algorithm performs well up to about a cell
optical depth of ${\tau_c \sim 6}$. Above this threshold, the numerical diffusion
dominates, degrading the accuracy of the solution. We note that this threshold
is about an order of magnitude larger than the one for the scheme presented in
R15. This allows us to capture the accurate solutions even at high optical
depths without resorting to sub-grid diffusion models. Additionally, our scheme
converges at a much higher rate ${\sim 2.0}$  and can therefore efficiently achieve higher accuracy by improving the spatial resolution.

\subsection{Levitation of optically thick gas}
\label{sec:lev}

Radiation pressure, both from direct UV and multiscattered IR, has been
hypothesised to drive significant galactic-scale outflows ($\sim 100\,\text{km
s}^{-1}$; \citealt{Hopkins2011, Agertz2013, Hopkins2014}). While the momentum
injection rate of the single scattered UV photons is just ${\dot{P}_{UV} = L/c}$,
the reprocessed IR radiation field can be trapped in a high optical depth medium
boosting the momentum injection, i.e., ${\dot{P}_{IR} = \tau_{IR}L/c}$. The
efficiency with which the gas can trap the IR radiation field is unknown and
different RT schemes seem to produce different results \citep{Krumholz2012,
Davis2014, Rosdahl2015, Zhang2017}.

In this section, we perform the experiment first outlined in
\citet{Krumholz2012}, which tests the radiation pressure, radiation-temperature
coupling and the multiscattering of IR radiation. A thin layer of gas is placed
in an external gravitational potential, and a certain amount of IR flux is injected in
the direction opposite to gravity. This setup mimics the physical conditions
found in stellar nurseries or in the central plane of an optically thick
galactic disk. It allows us to study how gravitationally bound gas responds to
multiscattering IR radiation. \citet{Krumholz2012} argue that as the gas lifts,
it becomes Rayleigh-Taylor unstable, which leads to a significant reduction in
the coupling between the IR radiation and the gas. The Rayleigh-Taylor instability
(RTI) creates chimneys through which the radiation escapes rather than
coherently lifting the gas. In their work, the radiation field has been modelled using the flux
limited diffusion (FLD) approach, which assumes that the radiation flux always
points in the direction of gradient of the photon energy density. As discussed
in Section~\ref{sec:RT}, the directionality of the underlying photon field is
washed out by using this scheme. It is therefore possible that the radiation field
diffuses out through the path of least resistance. 

\citet{Davis2014} performed the same experiment using a more accurate VET closure
scheme, which constructs the radiation flux vector for every volume element by
sweeping the whole domain with short characteristics rays, effectively
incorporating the contribution from all sources and sinks. Their results show
that the VET closure relation coherently lifts the gas even in the presence of
RTI. The velocity of the gas also significantly increases with the VET scheme.
This is despite the fact that the average optical depths and the radiation
force on the gas is quite similar to the FLD runs. The difference stems from
the fact that VET scheme manages to maintain the average Eddington ratio to just above
one, while the FLD only achieves Eddington ratios which are below
unity. 

This discrepancy between the different numerical schemes can be explained by
their varying accuracy in estimating the direction of the
underlying photon field. The diffusion approximation only transports photons in
the direction of energy gradient, which is a very good approximation in highly
thick media, but fails in optically thin or even slightly optically thick
systems. R15 performed the same test but using the M1 closure relation and a
PC-GLF scheme (called M1-R15 scheme from now on) and found results which were
closer to the FLD results rather than the VET. They argue that although the M1
closure locally stores the bulk direction of the radiation field, its inability
to accurately capture the propagation direction in the presence of multiple
sources creates artificial diffusion. This causes the radiation to escape out
of the chimneys created  by the RTI.  As we also employ the M1 closure
relation, we expect to get similar results as R15. However, it is interesting
to see if our low diffusion, higher order scheme will perform better.
Moreover, the quantitative results using the FLD, VET and M1 closures in terms
of optical depths, Eddington ratios and gas velocities are close enough that
this experiment can act as a good test of our implementation.  

 The simulation setup consists of a 2D domain of boxsize ${L(x,y) = (L_{\rm
box}/2 , 2L_{\rm box}) =  (512,2048) h_*}$, where $h_*$ is the characteristic
scale height of the initial density profile. A layer of gas is placed at the
bottom of the box and given an exponential density profile ${\rho(y) = \rho_*
\exp(-y/h_*)}$, where ${h_* = 2 \times 10^{15} \ \text{cm}}$ and ${\rho_* = 7.1
\times 10^{-16} \ \text{g cm}^{-3}}$. The column density of the box is then
${\Sigma_* = 1.4 \ \text{g cm}^{-2}}$. On top of this initial density profile a
perturbation of the form
\begin{equation}
\frac{\partial \rho}{\rho} = 0.25 (1\pm \chi) \sin(2\pi x/L_{\rm box}) \, ,
\end{equation}
is added, where $\chi$ is random number uniformly distributed between $[-0.25,
0.25]$. This mimics the turbulent nature of the gas present in the star forming
birth cloud and in the ISM. The density profile has a floor at ${10^{-10}
\rho_*}$ and the initial temperature of the gas is uniformly set to
${T_*=82 \  \text{K}}$.  The gas is acted upon by a homogeneous gravitational
acceleration pointing downwards with a magnitude of ${g = 1.46 \times 10^{-6} \
\text{cm s}^{-2}}$.  We note  that under these conditions the system is unstable
as gas pressure cannot counteract gravity. 

The initial mesh consists of a high resolution Cartesian mesh at the bottom in
order to resolve the high density gas, with the volume of the resolution
element set to ${(0.5 h_*)^2}$ (similar to the resolution used in
\citealt{Davis2014} and  R15). The resolution is then degraded slowly till
a minimum resolution with a volume of $(8 h_*)^2$ is reached. The initial
configuration of the mesh is not that important, as the mesh moves and distorts
according to local fluid flow. The mesh is regularised where needed and refined
and de-refined such that the minimum ($(0.5 h_*)^2$) and maximum ($(8 h_*)^2$)
cell sizes are approximately maintained throughout the simulation run.

\begin{figure*}
\includegraphics[scale=0.85]{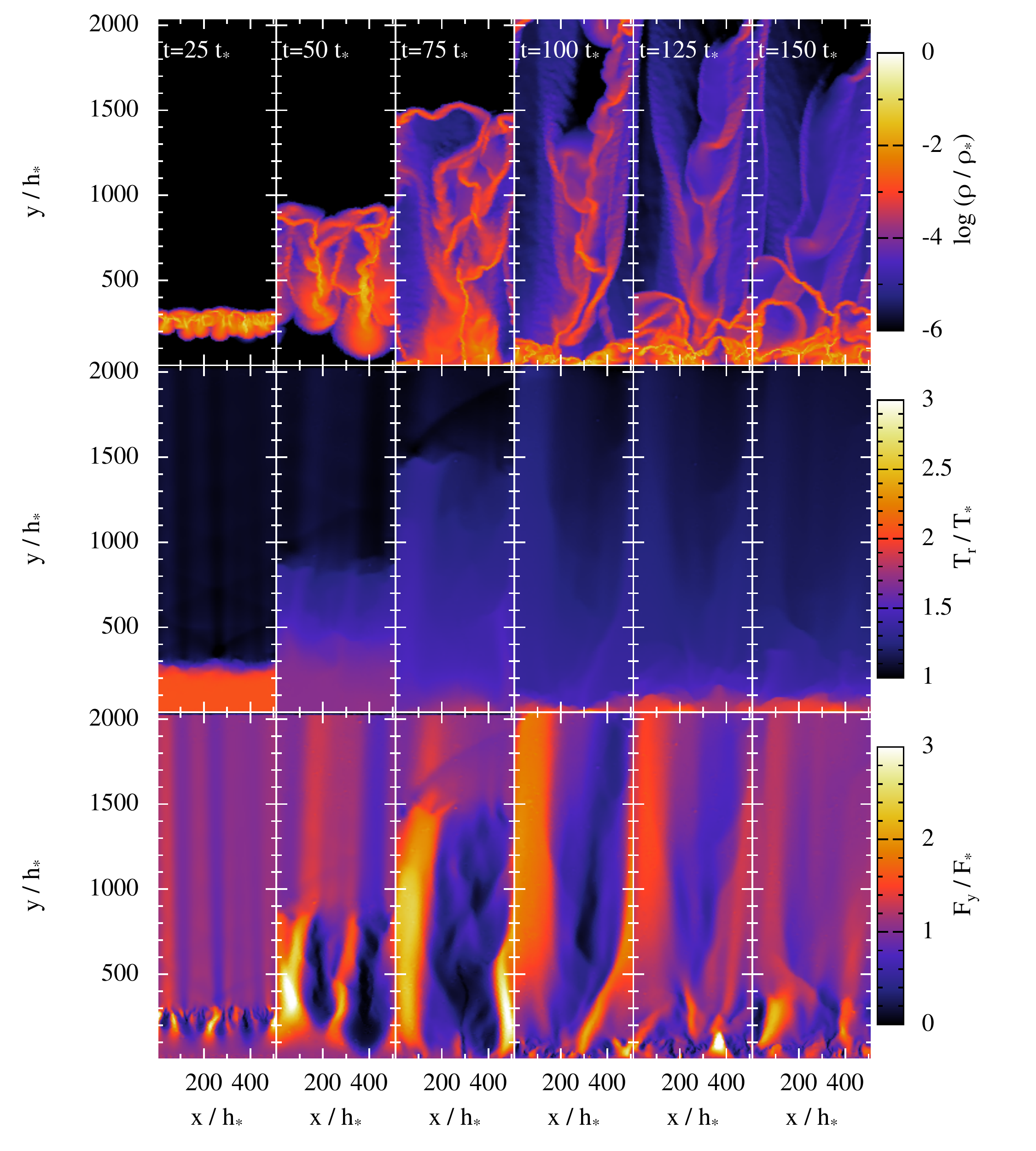}
\caption{ {\bf Levitation of optically thick gas}: Maps showing the normalised density (top panels),  radiation temperature (middle maps) and the radiation flux in the vertical direction ($F_y$; bottom panels) at ${t=25 t_*}$ (first column), ${t=50 t_*}$ (second column), ${t=75 t_*}$ (third column), ${t=100 t_*}$ (fourth column), ${t=125 t_*}$ (fifth column)  and ${t=150 t_*}$ (fourth column).}
\label{fig:optlevmaps}
\end{figure*}

\begin{figure}
\includegraphics[width=\columnwidth]{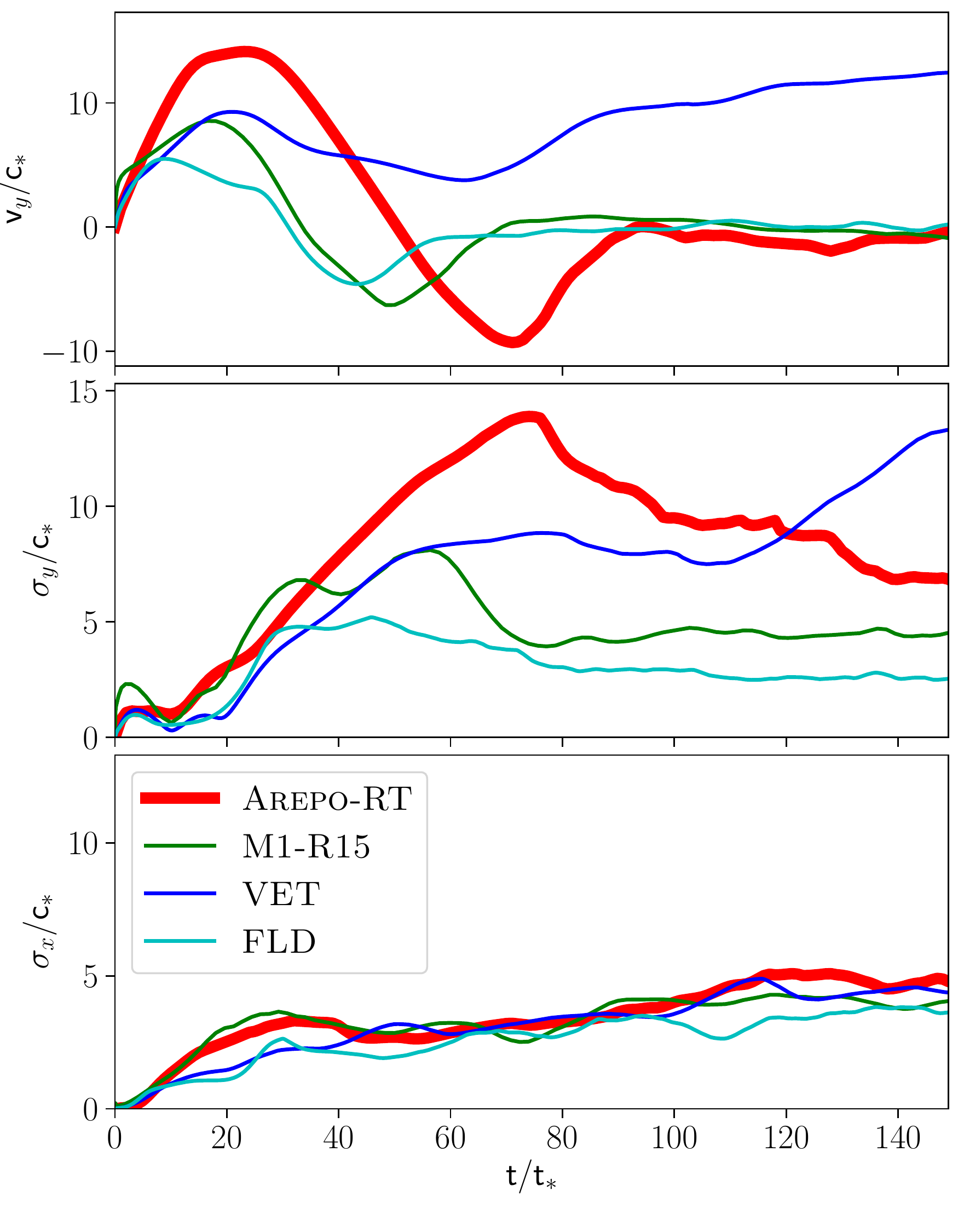}
\caption{ {\bf Levitation of optically thick gas}: Plot showing the mass weighted $y$-velocity (top panel), the vertical velocity dispersion (middle panel) and  the horizontal velocity dispersion (bottom panel) as a function of characteristic time $t_*$ in our simulation ({\sc Arepo-RT}; red curves) compared to M1-R15 (green curves), VET (blue curves) and FLD (cyan curves) schemes.  Initially the mean $y-$velocity increases to about ${15 c_*}$ due to the efficient trapping of photons. However, RTIs form low density chimneys through which the radiation escapes and the gas falls back to the bottom of the box and settles into a turbulent state.}
\label{fig:vsigma}
\end{figure}

The bottom boundary of the domain emits a constant IR radiation flux of ${F_* =
2.54 \times 10^{13} \ \text{L}_\odot \ \text{kpc}^{-2} (1.03 \times 10^4 \
\text{erg cm}^{-2} \ \text{s}^{-1})}$ and the box is initialised to contain an
upwards radiation flux of the same magnitude, with ${\tilde{c}E = F_y = F_*}$ and
${F_x=0}$. This sets the radiation temperature to
\begin{equation}
T_{r*} = \left(\frac{F_*}{ca}\right)^{1/4} = T_* \, ,
\end{equation}
where $a$ is the radiation constant.  The radiation is coupled to the gas-dust
fluid using the equations described in Section~\ref{sec:IR}. The radiation
energy density and the radiation flux are coupled to the gas using the Planck
($\kappa_\ts{P}$) and Rosseland ($\kappa_\ts{R}$) mean opacities, given by
\begin{equation}
\kappa_\ts{P} = 0.1 \left(\frac{T}{10 \ \text{K}}\right)^2 \ \text{cm}^2 \ \text{g}^{-1} \, ,
\end{equation}
\begin{equation}
\kappa_\ts{R} = 0.0316 \left(\frac{T}{10 \ \text{K}}\right)^2 \ \text{cm}^2 \ \text{g}^{-1} \, ,
\end{equation}
which sets the initial value of the Rosseland mean opacity to ${\kappa_{\ts{R}*} =
2.13  \ \text{cm}^2 \ \text{g}^{-1}}$.  These opacity functions are the same as
the ones used in previous works \citep{Krumholz2012, Davis2014, Rosdahl2015}
and mimic the observed dust opacity functions in cold (${T\leq 150 \ \text{K}}$)
gas obtained by \citet{Semenov2003}.  We note that the only non-adiabatic
source of heating-cooling for the gas is through IR radiation-dust-gas coupling
\begin{equation}
\frac{\partial u}{\partial t} = - \frac{\partial E}{\partial t} = \kappa_\ts{P} \, \rho(\tilde{c} E  - c \, a \, T^4 ) \, .
\end{equation} 

The gravitational force and the radiation pressure compete with each other and
the eventual motion of the gas will be determined by the Eddington ratio
\begin{equation}
f_E = \frac{f_{y,\text{rad}}}{g\rho} \, ,
\end{equation}
where $f_{y,rad}$ is the radiation force in the vertical direction
\begin{equation}
f_{y,\text{rad}} = \frac{\kappa_\ts{R} \, \rho \, F_y}{c} \, .
\end{equation}
Therefore, the Eddington ratio at the start of the simulation is ${f_{E*} =
0.5}$,  implying that the radiation is initially unable to overcome gravity and
lift the gas. However, the gas is optically thick to the IR radiation with an
initial optical depth of
\begin{equation}
\tau_* = \kappa_{\ts{R}*} \, \Sigma_* = 3 \, ,
\end{equation}
and the radiation can get trapped by the gas, increasing the Eddington ratio
and driving the gas upwards.

The simulation domain is periodic in the $x-$direction for both gas and the
radiation field. The top layer boundary cells have fixed values of temperature
and density of ${\rho = 10^{-13} \rho_*}$ and ${T = 10^{-3} T_*}$ respectively. The
velocity of the gas is set to zero at this boundary and the energy density and
flux of the radiation field is also set to zero. This allows for free flow of
gas and radiation field out of the top boundary. For the gas, the bottom
boundary is reflective and allows for no escape or entry of gas. The bottom
boundary should also emit radiation vertically at a the rate of $F_*$. This is
accomplished by setting the radiation energy density and flux at the bottom
solid layer to $E_s$ and ${{\bf F}_s=(0,F_*)}$, where
\begin{equation}
\tilde{c}E_s = F_* - F_{y,f} + \tilde{c}\, E_f \, ,
\end{equation}
where the subscripts `s' and `'f' refer to the solid layer and fluid layer
cells respectively.  We use our fiducial scheme to solve the transport
equations. The simulation is run for ${t=150t_*}$, where ${t_* = h_*/c_*}$ is the
characteristic sound crossing time and ${c_* = \sqrt{k_B T/(\mu m_p)} = 0.54 \
\text{km s}^{-1}}$ is the sound speed. The mean molecular weight is set to
${\mu=2.33}$.  We run the simulation with a reduced speed of light fraction set
to ${f_r=0.01}$. 
 
\begin{figure} 
\includegraphics[width=\columnwidth]{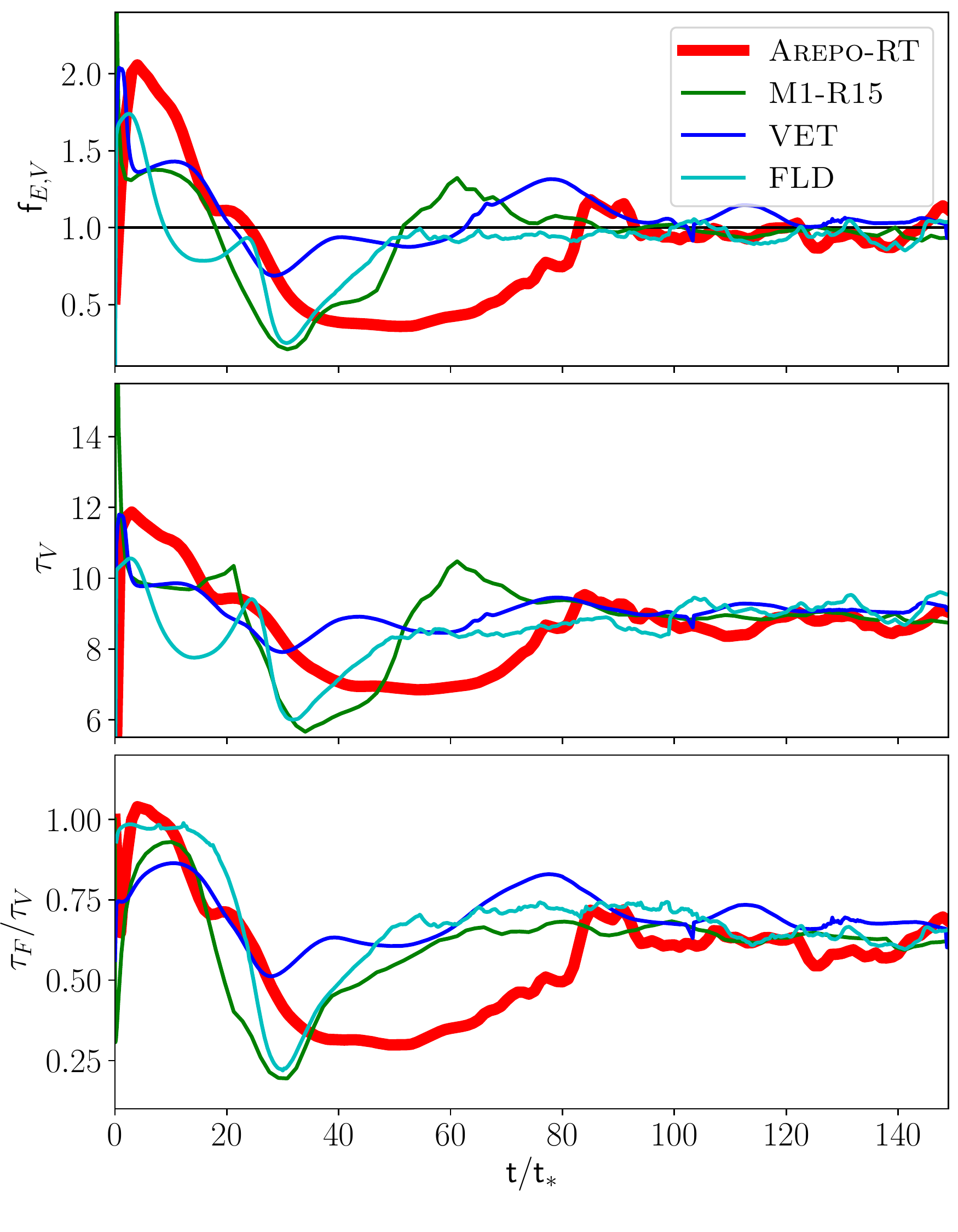}
\caption{{\bf Levitation of optically thick gas}: Plot showing the volume weighted Eddington ratio, optical depth and the ratio between the flux weighted mean optical depth to the volume weighted optical depth as a function of the characteristic time in our simulation ({\sc Arepo-RT}; red curves) compared to M1-R15 (green curves), VET (blue curves) and FLD (cyan curves) schemes. All three quantities rise sharply in the beginning and these high values are maintained for about ${15 t_*}$, after which all three quantities decline sharply as the RTIs build up. It is only after most of the gas falls back to the bottom (${\sim 50 t_*}$) that the optical depth starts to build up again and pushes the Eddington ratio closer to unity.}
\label{fig:fe}
\end{figure}

Fig.~\ref{fig:optlevmaps} shows the maps of normalised density (top panels),
radiation temperature (middle maps) and the radiation flux in the vertical
direction ($F_y$; bottom panels) at ${t=25 t_*}$ (first column), ${t=50 t_*}$
(second column), $t{=75 t_*}$ (third column), $t{=100 t_*}$ (fourth column), $t{=100 t_*}$ (fisth column) and ${t=150 t_*}$ (sixth column).
Initially, the trapped radiation field greatly increases the radiation
temperature ($T_*$), which in turn increases the gas temperature due to the
close coupling between the dust, gas and the radiation field.  The increased
gas temperature leads to an increase in $\kappa_\ts{R}$, resulting in increasing the
radiation force and pushing ${f_E>1}$. The gas becomes super-Eddington and it can
be driven upwards in a thin shell. By ${t=25t_*}$, the gas becomes
Rayleigh-Taylor unstable, creating dense filamentary structures punctuated with
diffuse chimneys through which the radiation escapes (as seen in the bottom
left panel). However, the gas continues to lift and a significant fraction ($\sim 15 \%$) of the gas lifts beyond $1024 h_*$ by $75 t_*$.  Some of the most dense
filaments stall and fall back, but the radiation is still trapped within
turbulent medium as evidenced by the temperature map at this time.  However, as the chimneys
widen the radiation eventually escapes and the gas falls back down to the
bottom (${\lesssim 500 h_*}$) where it is kept turbulent by the competition
between the radiation pressure and gravity.  
 
 Fig.~\ref{fig:vsigma} shows the mass-weighted mean vertical velocity (top
panel), the mass-weighted vertical (middle panel) and horizontal (bottom panel) velocity dispersions as a function of the simulation time in units
of the characteristic timescale ($t_*$).  The velocity and velocity dispersions
are plotted in units of the characteristic sound speed $c_*$. The initial gas
acceleration obtained using our simulation setup (red curves) is much higher than either of the VET (blue curves), FLD (cyan curves), or the M1-R15 (green curves) schemes.
The maximum average velocity reached by the gas in our simulation is about
${\left<v_y\right>_\text{max} \sim 15 c_*}$ which is about $50\%$ larger than the
velocities achieved by the other schemes. This velocity is maintained for a
brief amount of time (${\sim 25 t_*}$) and then the velocity starts to
decrease quickly as the gas starts to fall back to the bottom of the domain.
The chimneys created by the RTI also allow the radiation to escape, reducing
the radiation pressure on the gas. This causes strong deceleration and the minimum
velocity reaches ${\left<v_y\right>_\text{min} \sim -10 c_*}$. As the gas
falls back the radiation pressure starts to build up again and eventually
reaches a turbulent state, with velocity dispersions larger than the ones
obtained by the M1-R15 and FLD schemes. 

To understand this behaviour quantitatively, we plot the volume
averaged Eddington ratio
\begin{equation}
f_{E,V} = \frac{\left< f_{y,\text{rad}}\right>}{F_y} \, ,
\end{equation}
as a function of characteristic time in the top panel of Fig.~\ref{fig:fe}. As
mentioned earlier, this ratio reveals the competition between the radiation
pressure and gravity with ${f_{E,V} >1}$ corresponding to the case where the
radiation pressure wins and drives outflows, while ${f_{E,V}<1}$ indicates that
gravity wins and the gas just falls back to the bottom of the box. The
Eddington ratio quickly jumps to ${f_{E,V}\sim2}$ from an initial value of $0.5$
within the one $t_*$ and maintains that value for about ${15 t_*}$. This
behaviour is different from the VET and M1-R15 schemes, where the gas motions
cause their Eddington ration to drop below $1.5$ very quickly. The reason
for this discrepancy can understood by looking at the volume averaged optical
depth from top to bottom (middle panel of Fig.~\ref{fig:fe}),
\begin{equation}
\tau_V =  \ L_\text{box} \left< \kappa_\ts{R} \, \rho \, \right> \, .
\end{equation}
The optical depth rises from the initial value of ${\tau_V = 3}$ to ${\tau_V \sim
12}$ within one $t_*$ similar to the VET simulations.  However, while the
optical depth falls below $\sim 10$ quickly in their simulations, it is
maintained for about $15 t_*$ in our simulation. This increased optical depth
results in larger values of the Eddington ratio, which in turn drives the gas
to larger velocities. This indicates that our scheme is much more efficient in
trapping the photons initially compared to the other schemes used in
literature. The bottom panel of Fig.~\ref{fig:fe} shows the ratio between the
flux-weighted mean optical depth
\begin{equation}
\tau_F = L_\text{box} \frac{ \left< \kappa_\ts{R} \, \rho \, F_{ry} \right> }{ \left< F_{ry} \right> } \, ,
\end{equation} 
and $\tau_V$. $\tau_F$ gives the momentum per unit area transferred from the
radiation to the gas. Therefore, the ratio gives us an estimate of the fraction
of actual momentum in the radiation field transferred to the gas. This quantity
hovers around unity for the initial ${15 t_*}$ implying that there is almost
perfect coupling between radiation field and the gas at these early times. This
is again better than the VET and M1-R15 schemes which only manage to couple
about ${\sim 85\%}$ of the radiation  momentum into the gas.  After about $15
t_*$, all three quantities decline sharply, as the RTI instabilities build up.
 There is a slight rebound at about ${25 t_*}$ but declines again quite quickly. All
three quantities rebound slowly as the gas falls back to the bottom, the optical depth builds up and the gas reaches a
turbulent state with the Eddington ratio hovering around unity. 

This test has allowed us to gauge the advantages and limitations of our scheme.
The general results agree well with the FLD, VET and M1-R15 schemes used in
literature. Our scheme performs better than the other schemes at early times by
trapping the radiation more efficiently, increasing the optical depth and
powering a higher velocity outflow. The maximum average velocity is about
$50\%$ higher than the ones achieved by either of the VET or M1-R15 schemes.
However, the late time behaviour of the gas is closer to the FLD scheme
than the VET scheme. Most of the gas in our scheme falls back to the bottom of
the domain (like FLD and M1-R15 schemes), while the VET scheme still continues
to evacuate gas at a significant rate. The reason for this difference is the
approximation made in the M1 closure relation. Although the M1 closure locally
stores the bulk direction of the radiation field, it is unable to accurately
capture the propagation direction in the presence of multiple sources creating
artificial diffusion. This causes the radiation to escape out of the
chimneys created  by the RTI. Unfortunately, this is physical limitation of the
of the M1 closure approximation and not a limitation of the implemented
numerical scheme. It might turn out that the only way to accurately capture the
exact coupling between radiation and gas in this regime might be to run quite
expensive ray tracing short/long characteristic RT methods.  However, we note
that our scheme is better at trapping photons and driving outflows compared to
the M1-R15 scheme which has the same physical limitations as our method.

\section{Summary and Conclusions}
\label{sec:conclusions}
In this paper we have presented {\sc Arepo-RT}, a novel implementation of an
accurate and computationally efficient radiation hydrodynamics scheme on
unstructured, moving Voronoi meshes.  The scheme is based on a fluid
description of the radiation field obtained by taking the zeroth and first
order moments of the continuity equation of specific luminosity. These moment
equations are a pair of hyperbolic conservation laws for photon energy density
and photon flux. The system is then closed using the M1 closure relation that
equates the pressure tensor to the energy density using a specific form of the
Eddington tensor, which locally stores the the bulk direction of the radiation
field.  The ability of the M1 closure to obtain an estimate of the Eddington
tensor from just the local properties of the cell renders it very useful for
computationally challenging problems. 

We employ an operator split approach based on dividing the moment equations into equations
for pure radiation transport and equations for the source and sink terms.  We
achieve high order accuracy by replacing the piecewise constant approximation
of Godunov's scheme with a slope-limited piece-wise linear spatial
extrapolation and a first order time prediction to obtain the states
of the primitive variables on both sides of the cell interface. The spatial
extrapolation is carried out using a least-square-fit gradient estimator that
has been shown to work well in meshes where the center of mass of the cell can
be offset from the mesh generating point. Two different flux functions have
been implemented to solve the Riemann problem at the interface: a second order
Harten-Lax-van Leer flux function that uses the exact eigenvalues that
represent the wave speeds of the RT transport equation and a
Global-Lax-Friedrichs flux functions that sets the eigenvalues to the light
speed irrespective of the geometry of the problem.

A conservative time integration scheme is implemented using Heun's method,
which is a variant of the second order Runge-Kutta scheme. The fluxes are
computed as an average of fluxes at the beginning and end of the timestep. A
reduced speed of light approximation is implemented in order to overcome the
problem of small timesteps required due to the large speed of light.
Additionally, a conservative sub-cycling scheme is implemented that is fully
compatible with the individual time stepping scheme of {\sc Arepo}. 

The radiation field couples to the gas and dust via photoionization,
photoheating and momentum injection. A
multi-frequency approach is used to model UV, optical and IR radiation fields.
We implement atomic Hydrogen and Helium thermochemistry using a semi-implicit
approach that is quite stable and allows for reasonably large timesteps. The
local nature of the M1 closure relation allows the scheme to account for
radiation emitted from collisional recombinations and discard OTSA.  IR
radiation is accounted for by coupling it to the semi-empirical dust model of
\citet{McKinnon2016}  which treats dust as a passive scalar, whose motion is
closely coupled to the gas motions. This allows us to assume that the system is
close to local thermodynamic equilibrium, which is a good approximation for cold
high density regions of the ISM. The main advantage of this coupling is that
the dust opacities are self-consistently calculated from the properties of the
dust in the cell, thereby eliminating the need for ad hoc scaling relations
used in previous works \citep{Bieri2017, Costa2017}. 

We test our implementation on a variety of problems. The
implementation works well overall and reproduces analytic results in all the
tests performed in this work.  We first start with a test designed to gauge the
accuracy of our radiation transport scheme in vacuum by simulating the radial
advection of a thin Gaussian pulse. We find that using different flux functions (GLF or
HLL) to solve the Riemann problem makes little difference to the obtained
solution.  It is much more important to use gradient extrapolated values at the
face as inputs to the Riemann solver instead of the piecewise constant
approximation used in R13 and R15. Such higher order fiducial schemes have very low
numerical diffusion and the convergence order is ${\sim 2.0}$ compared to ${\sim
0.5}$ for the PC schemes. In fact, the L$^1$ error in the simulation with
${2\times64^2}$ resolution elements run with the fiducial scheme is much lower
than the error for a simulation with ${2\times256^2}$ resolution elements using a
PC approximation. 

The veracity of the multi-frequency scheme coupled to the H-He photochemistry
is verified by simulating Str\"omgren spheres around an ionizing source in a
constant density medium.  The ionization structure and the time evolution of
the Str\"omgren radius match very well with the analytic expectations. We
also simulated the case of an I-front trapping inside a dense clump of gas. The
I-front is trapped inside the clump forming a shadow behind the clump.  The sharpness of the
shadow correlates with the diffusivity of the scheme. We show that the PC
approximation is unable to produce accurate shadows if the radiation transport is even slightly non-parallel to the
mesh geometry.  On the other hand our fiducial schemes which use gradient
extrapolations are able to produce sharp shadows irrespective of the flux
function or the mesh geometry used. 

The coupling between radiative transfer and hydrodynamics is tested by
simulating the expansion of a \ion{H}{II} region in both a constant density
medium and in a medium with a steep power law density slope which mimics more
realistic situations. The evolution of the ionization structure, density, 
temperature, pressure and the Mach number of the gas matches very well with the
results obtained by previous works. Additionally, we also address the accuracy
of the momentum injection into the gas due to photon absorption by simulating
radiation pressure driven outflows. We show that after an initial period of
supersonic expansion, a linear relation between the gas velocity and time is
reproduced which is directly proportional to the luminosity of the source.
These tests together validate the accuracy of the coupling between
hydrodynamics and the radiation field. 

One of the main shortcomings of the M1 closure relation is its inability to
accurately determine the direction of the underlying photon field in a multiple
source geometry. We quantify this deficiency by simulating the topology of the
radiation field emanating from a thin disc surrounded by an optically thick
torus. Although we achieve qualitative agreement with the analytic expectation,
the simulation slightly overshoots the field geometry perpendicular to the disc
and undershoots it in the transverse direction. This is because the rays from
one side of the disc intersects the rays emanating from the opposite side causing spurious
perpendicular flux. We stress that this is a fundamental  limitation of the M1
closure approximation. In fact, R15 find similar results with their M1 closure
scheme.

Next, we test the implementation of the multi-scattering IR-dust gas coupling.
Especially in optically thick regimes multiple scatterings lead to an
isotropisation of the radiation flux and the radiation tends to diffuse rather than
advect through the medium. It is difficult to capture this transition
because if the numerical diffusion of the scheme becomes larger than the true
radiation diffusion then the operator split approach to solve the RT equations
is not valid anymore. Therefore, the ability of a numerical scheme to
model the radiation transport in an optically thick regime is extremely
sensitive to the inherent numerical diffusivity of the scheme. Our simulations of dust coupling in a
optically thick media show that our fiducial scheme manages to reproduce accurate results for
${\tau_c\lesssim 6}$, while that limit without using a sub-grid diffusion model for
R15 is  ${\tau_c\sim 0.6}$. We conclude that our scheme is able to attain the right solution even when the optical depth is moderately under-resolved. 

As a final test of our scheme, we explore the ability of a trapped IR radiation
field to accelerate a layer of gas in the presence of an external gravitational
field that points in the opposite direction to the radiation pressure.
Simulations using a FLD scheme argue that as the gas lifts it becomes
Rayleigh-Taylor unstable creating  chimneys through which the radiation escapes
rather than coherently lifting the gas. Simulations performed with a more
accurate VET closure scheme, however, coherently lift the gas even in the
presence of RTI. The difference stems from the fact that VET manages to
accurately estimate the direction of the underlying photon field while the FLD
approximation only transports photons in the direction of energy gradient. Our scheme performs
better than the other schemes at early times by trapping the radiation more
efficiently, increasing the optical depth and powering a higher velocity
outflow. However, the late time behaviour of the gas is closer to the FLD
scheme than the VET scheme. Most of the gas in our scheme falls back to the
bottom of the domain, while the VET scheme still continues to evacuate gas at a
significant rate.  This is because,  although the M1 closure locally stores the
bulk direction of the radiation field,  it is unable to accurately capture the
propagation direction in the presence of multiple sources, creating artificial
diffusion. Unfortunately, this is a physical limitation of the M1 closure
approximation. It might turn out that the only way to accurately capture the
exact coupling between radiation and gas in this regime might be to run quite
expensive ray tracing short/long characteristic RT methods.  However, we note
that our scheme is much better at trapping photons and driving outflows
compared to the R15 scheme which have the same physical limitations as our
scheme. 

We conclude that we have implemented an efficient, robust and accurate
radiation hydrodynamics solver in the moving-mesh code {\sc Arepo}. In
forthcoming work, we plan to use this implementation to study timely problems
in astrophysics related to radiative transfer, such as the role of radiative
stellar feedback in driving galactic scale outflows, radiation pressure from
quasars and its role in quenching high redshift galaxies and modeling the
re-ionization history of the Universe.

\section*{Acknowledgements}
 We would like to thank the referee for giving constructive comments which substantially improved the quality of the paper. We thank Shane Davis and Joakim Rosdahl for kindly sharing their data. RK  acknowledges support from NASA through Einstein Postdoctoral Fellowship
grant number PF7-180163 awarded by the {\it Chandra} X-ray Center, which is
operated by the Smithsonian Astrophysical Observatory for NASA under contract
NAS8-03060. MV acknowledges support through an MIT RSC award, the Alfred P.
Sloan Foundation, NASA ATP grant NNX17AG29G, and a Kavli Research Investment
Fund. RM acknowledges support from the DOE CSGF under grant number
DE-FG0297ER25308. VS acknowledges support through subproject EXAMAG of the
Priority Program 1648 SPPEXA of the German Science Foundation. VS and RP are
also supported by the European Research Council through ERC-StG grant
EXAGAL-308037. The simulations were performed on the joint MIT-Harvard
computing cluster supported by MKI and FAS. The figures in this work were produced
by using the
{\sc matplotlib}
graphics environment \citep{Hunter2007} and the interactive visualization tool {\sc splash} \citep{Price2007}.

\bibliographystyle{mnras}
\bibliography{paper}

\appendix

\section{Sub-Cycling of the RT step}
\label{sec:subcycle}

In this section we describe a scheme designed to subcycle multiple RT steps for each hydro step that is compatible with the local timestepping scheme of {\sc Arepo}. Implementing such a scheme relies on how well the artificial domain boundaries created by inactive cells are  handled. It was shown that the imposition of Dirichlet boundary conditions \citep{Commercon2014} can lead to a violation of energy conservation which can get quite acute in the presence of large energy gradients. We instead chose to follow the method outlined in \citep{Pakmor2016}. The number of RT sub-cycles to every hydro step ($N_\ts{sub}$) is an input variable. The hydro timestep of every cell in the simulation is then set as
\begin{equation}
\Delta t_\ts{hydro} = \text{min}\left(N_\ts{sub}\Delta t_\text{RT}, \Delta t_\text{hydro}\right) \, .
\label{eq:substep}
 \end{equation}
In order to maximise the impact of subcycling, the number of subcycles should be set to ${N_\ts{sub} = 2^n}$, since the local timestepping scheme of {\sc Arepo} uses a discretization of the allowed time-step sizes into a power-of-two hierarchy \citep{Springel2010}.

For every hydro step the following RT loop is executed ${N_{sub}}$ times:
\begin{itemize}
\item The timestep of each subcycle step is set to ${\Delta t = \Delta t_\ts{hydro}/N_\ts{sub}}$. 
\item The thermo-chemistry and momentum injection steps are executed for all the active cells. 
 \item A list of all active interfaces, i.e. interfaces
with at least one adjacent active cell is made.
\item All cells that share at least one corner
with  an  active  interface are collected.  This  includes  a  layer  of  inactive
cells around the active cells.
\item The fluxes are exchanged over the active interfaces (as described in Section~\ref{sec:transport}), with the timestep set to the minimum of the timesteps of the
two adjacent cells divided by ${N_\ts{sub}}$.
\item The primitive variables of the active cells are then updated.
\end{itemize}
This form of subcyling is fully conservative and only requires to solve
the RT equations  on  the  active  cells  plus  a  one-cell
boundary layer.

\begin{figure}
`\includegraphics[width=\columnwidth]{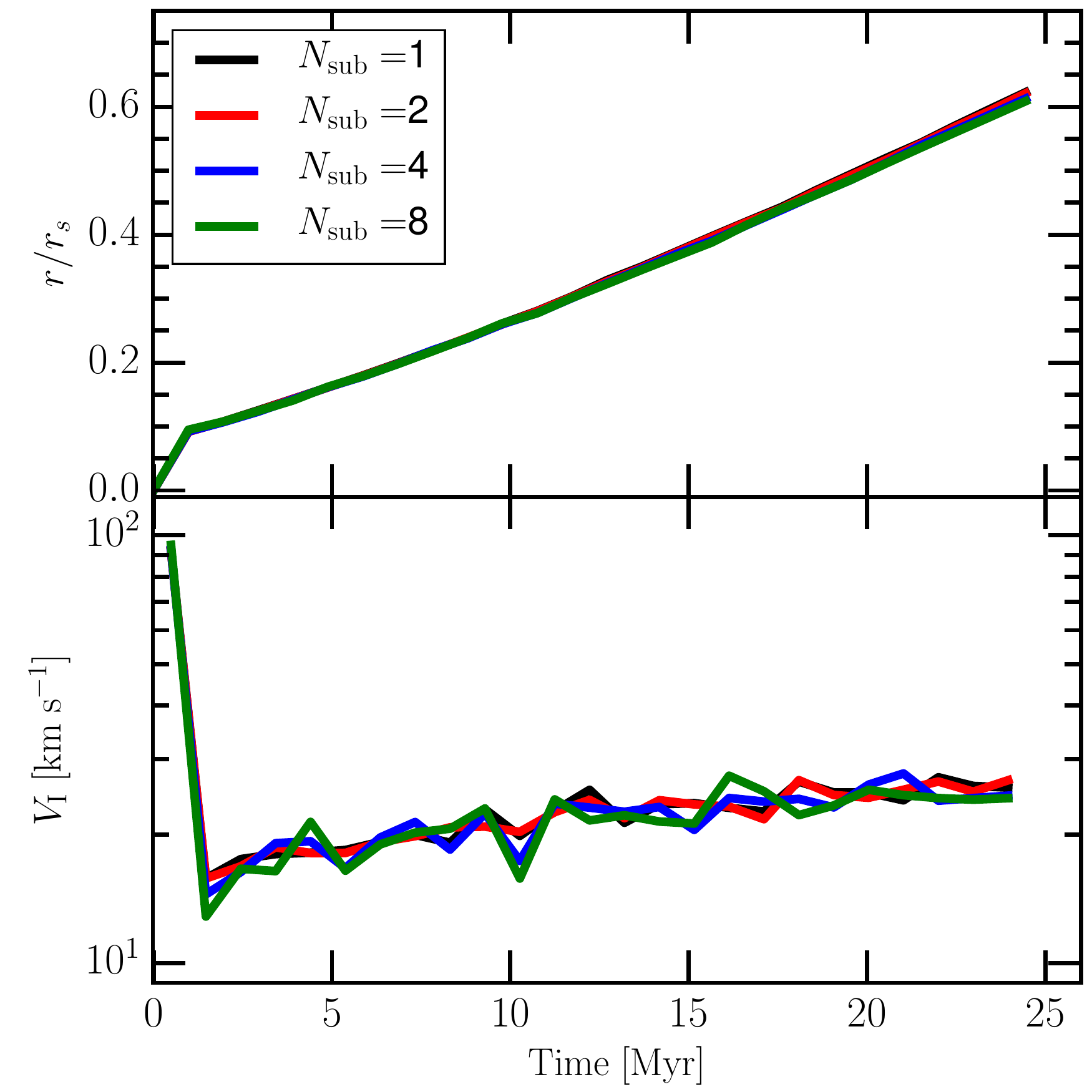}
\caption{{\bf Sub-cycling of the RT step}: The radius (top panel) and velocity (bottom panel) of the ionization front as a function of time for the expansion of an \ion{H}{II} region in a ${r^{-2}}$ density profile, in a simulation with $N{_\ts{sub}=1}$ (black curves),  $N{_\ts{sub}=2}$ (red curves),  $N{_\ts{sub}=4}$ (blue curves) and  $N{_\ts{sub}=8}$ (green curves). The simulation results are converged with respect to the number of subcycles used. }
\label{fig:subcycle}
\end{figure}

As a test of this scheme we re-simulate the expansion of a \ion{H}{II} region in a varying density field as described in Section~\ref{sec:h2exp}, with a varying number of RT subcycles.  Briefly, a constant luminosity source with a black body spectrum ($T{_\ts{eff}=10^5\,\text{K}}$) emitting at a rate of ${10^{50}\,\text{photons s}^{-1}}$ is placed at the
center of a spherically symmetric, steeply decreasing power-law ($-2$) density profile with a small flat central core of gas. 
A domain of side length ${2L_\ts{box}=1.6\, \text{kpc}}$ is resolved by ${2\times80^3}$ resolution elements placed on a regular grid. The central core has a density of ${n_0=3.2\,\text{cm}^{-3}}$ and a radius of ${r_0=91.5\,\text{pc}}$. The simulation is run for ${25\,\text{Myr}}$. For a fiducial run with ${N_\ts{sub}=1}$, the position and velocity of the ionization front are plotted in Fig.~\ref{fig:h2t_var} and the corresponding quantities are shown in Fig.~\ref{fig:subcycle} for ${N_\ts{sub}=1}$ (black curves), ${N_\ts{sub}=2}$ (red curves), ${N_\ts{sub}=4}$ (blue curves) and ${N_\ts{sub}=8}$ (green curves). All the runs reproduce the same result irrespective of the number of subcycles used, proving the validity and accuracy of our scheme. We also note that during the simulation the time bin hierarchy reaches up to $4$ bins deep, proving that the subcyling scheme is compatible with {\sc Arepo}'s local time-stepping scheme.

\begin{figure}
\includegraphics[width=\columnwidth]{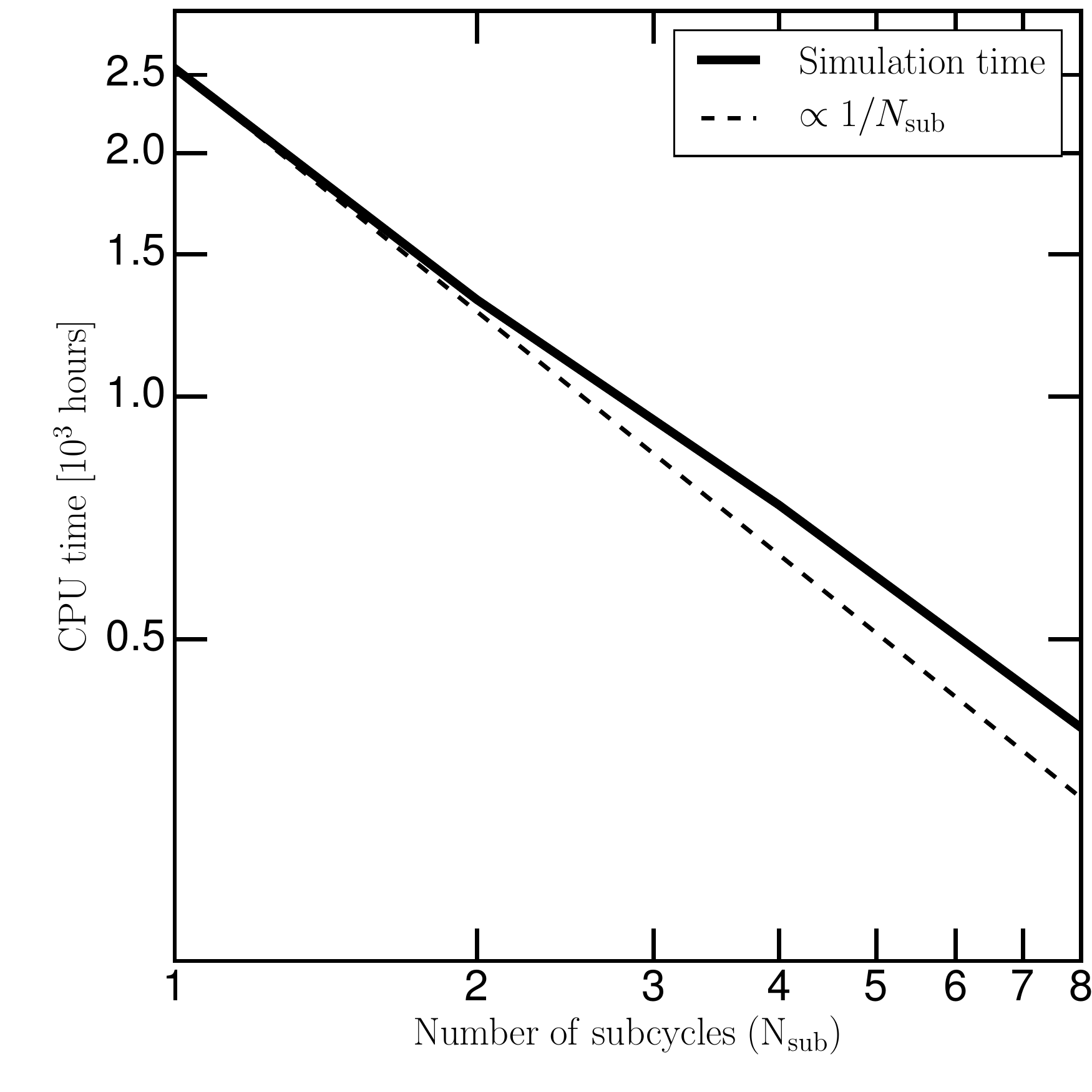}
\caption{{\bf Sub-cycling of the RT step}: The total run time (solid black line) of the simulation (in CPU hours) as a function of the number of RT subcycles used (${N_\ts{sub}}$) compared to the ideal ${\propto 1/N_\ts{sub}}$ scaling (dashed line). }
\label{fig:tottime}
\end{figure}

\begin{figure*}
\includegraphics[scale=0.55]{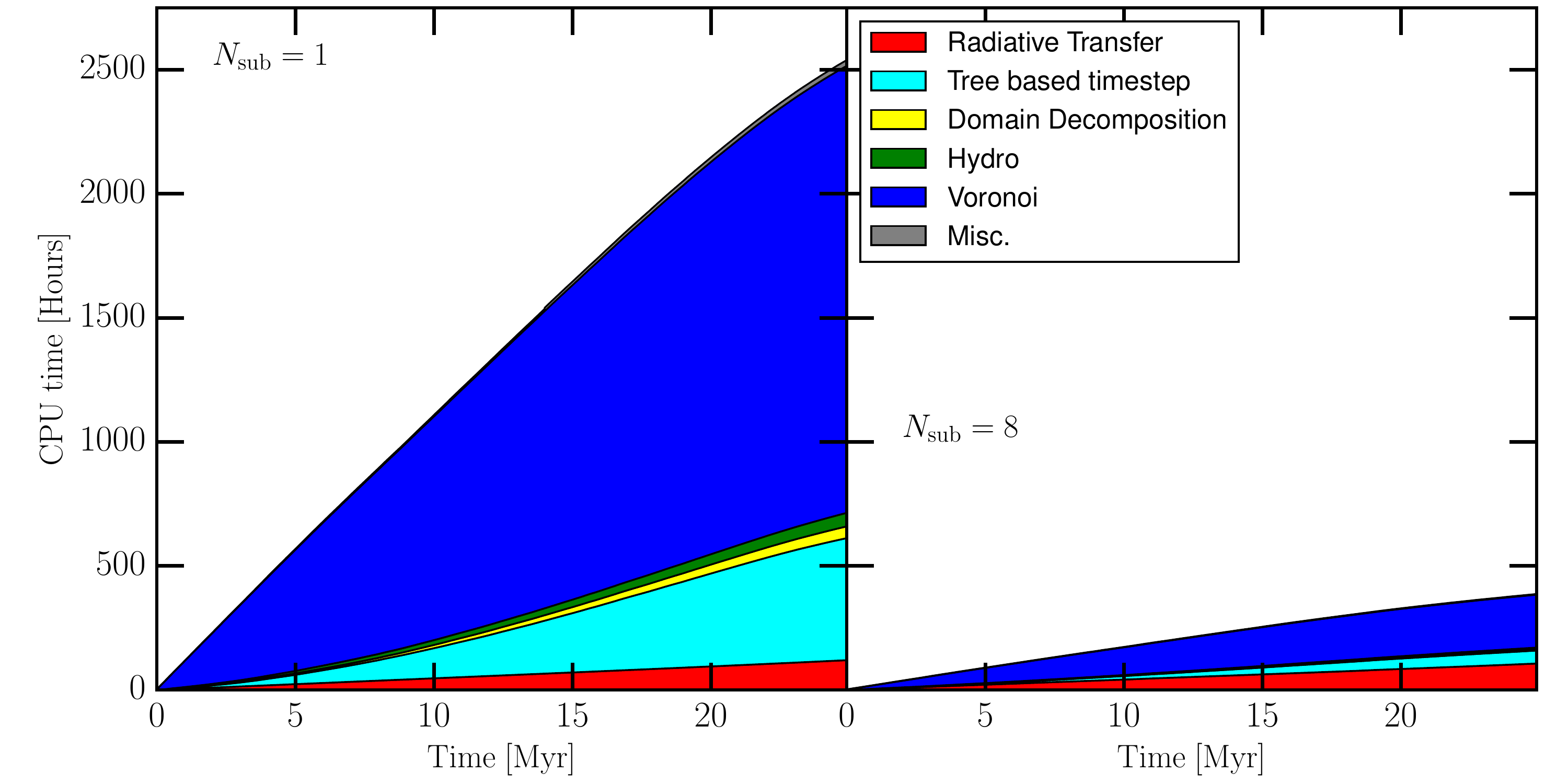}
\caption{{\bf Sub-cycling of the RT step}: The total amount of computing time taken by radiative transfer (red region), tree based timestep estimate (cyan region), domain decomposition (yellow region), hydrodynamic flux calculation (green region) and Voronoi mesh construction (blue region) routines in simulations with N${_\ts{sub}=1}$ (left panel) and N${_\ts{sub}=8}$ (right panel). Increasing the number of sub-cycles reduces the frequency with which the computationally expensive routines (such as the Voronoi mesh construction) are called, thereby reducing the total computing time. }
\label{fig:tsplit}
\end{figure*}

Fig.~\ref{fig:tottime} shows the total amount of time taken to run the simulation (in CPU hours) as a function of ${N_{sub}}$ (solid black curve). Satisfyingly, the run time of the simulation reduces by the expected factor of ${\sim 1/N_{sub}}$. This drop in the run time is attributed to the fact that the time consuming routines are called less frequently due to the larger hydro step (Eq.~\ref{eq:substep}). This can be seen more clearly in Fig.~\ref{fig:tsplit}, which compares the amount of time taken by individual sub-routines, such as, radiative transfer (red shaded region; including cooling and chemistry), tree based timestep calculation (cyan shaded region), domain decomposition (yellow shaded region), hydro (green shaded region) and Voronoi mesh construction (blue shaded region) for a simulation with ${N_\ts{sub}=1}$ (left panel) and ${N_\ts{sub}=8}$ (right panel). The time taken by the RT routine stays about constant because it still has to follow the timestep imposed by the speed of light. The increased hydro step reduces the frequency with which the other routines are called thereby massively reducing the time spent on them.

It is important to note that although Voronoi mesh construction is a computationally  expensive process, it does not usually take up as much time as depicted in our simulation (${\sim 70\%}$). This only happens because we start out with a regular mesh and this introduces a lot of degeneracies in mesh geometry which require a large number of geometric predicates that need to be carried out with exact floating point arithmetic to lift the degeneracies, thereby,  artificially boosting the computing time. Starting from more realistic, and therefore less regular, particle geometries will drastically bring down the time taken by this routine. However, this high computing time for Voronoi mesh construction acts as a proxy for other routines not included in the simulation such as gravity and galaxy formation physics that will definitely be present in realistic simulations. 

Finally, the speed of {\sc Arepo-RT}, compared to the parent code {\sc Arepo} is problem dependent. Specifically, it depends on the value of the reduced speed of light used in the simulations. If the timestep due to the speed of light is comparable to the timestep obtained from hydrodynamic considerations, then the overhead due to the RT flux calculations is pretty small, of order $\sim 10\%$. The non-equilibrium chemistry and cooling can cause significant ($\sim 30\%$) overhead in simulations with high density gas due to extremely small cooling times. The weak scaling of {\sc Arepo-RT} is as good as {\sc Arepo} (which has shown excellent weak scaling upto $10$'s of thousands of cores) because it uses the same domain decomposition, parallelization and communication algorithms.

\section{H-He Thermochemistry}
\label{sec:uvchem}
In this section we describe in detail the semi-implicit scheme used to solve the thermochemical network of H and He (Eqs.~\ref{eq:uvchemN}-\ref{eq:uvchemend}). 
First let us define
\begin{equation}
\begin{split}
 A &= \Delta t \, \sigma_{e\hi} \, (n_e)^n, \\
 B &= \Delta t \, {\tilde c} \sum_i {\bar \sigma}_{i\hi} \left(N_\gamma^i\right)^n  \ \text{and}\\ 
 C &= \Delta t \, \alpha_\hii \left(n_e\right)^n,
 \end{split}
\end{equation}
where ${\Delta t}$ is the time interval over which we are integrating the equation and the superscript `n' denotes the values of the quantity at the present time.
Then Eq.~\ref{eq:uvchemnhii} then can be written in a semi-implicit form
\begin{equation}
\begin{split}
  \left({\tilde n}_\hii\right)^{n+1} &= \left({\tilde n}_\hii\right)^{n} + A (1 - \left({\tilde n}_\hii\right)^{n+1}) + B (1 - \left({\tilde n}_\hii\right)^{n+1}) \\ &- C \left({\tilde n}_\hii\right)^{n+1},
  \end{split}
\end{equation}
which gives
\begin{equation}
  \left({\tilde n}_\hii\right)^{n+1} = \frac{\left({\tilde n}_\hii\right)^{n} + A  + B } {1 + A + B +C }
\end{equation}
where ${\{{\tilde n}_\hi, {\tilde n}_\hii \} = \{{n}_\hi, {n}_\hii \}/n_\text{H}}$, therefore ${{\tilde n}_\hi + {\tilde n}_\hii= 1}$ which in turn sets the value of ${\left({\tilde n}_\hi\right)^{n+1}}$. 

For He chemistry we represent ${{\tilde n}_j = n_j/n_\text{He}}$ and define
\begin{equation}
\begin{split}
 D &= \Delta t \, \sigma_{e\heii} \, (n_e)^n, \\
 E &= \Delta t \, \alpha_\heiii \left(n_e\right)^n, \\
 F &= \Delta t \, \sigma_{e\hei} (n_e)^n, \\
 G &= \Delta t \, \alpha_\heii \left(n_e\right)^n, \\
 H &= \Delta t \, {\tilde c} \sum_i {\bar \sigma}_{i\hei} \left(N_\gamma^i\right)^n, \ \text{and} \\
 I &= \Delta t \, {\tilde c} \sum_i {\bar \sigma}_{i\heii} \left(N_\gamma^i\right)^n.
\end{split}
\end{equation}
Therefore the change in ${n_\heiii}$ (Eq.~\ref{eq:uvchemnheiii}) can be written as
\begin{equation}
 \begin{split}
  \left({\tilde n}_\heiii\right)^{n+1} &=  \left({\tilde n}_\heiii\right)^{n} + D \left({\tilde n}_\heii\right)^{n+1} - E   \left({\tilde n}_\heiii\right)^{n+1} \\ &+ I \left({\tilde n}_\heii\right)^{n+1}, \\
 \end{split}
\end{equation}
which gives
\begin{equation}
 \left({\tilde n}_\heiii\right)^{n+1} = \frac{\left({\tilde n}_\heiii\right)^{n} + (D+I)\left({\tilde n}_\heii\right)^{n+1}}{1+E}.
 \label{eq:nheiii}
\end{equation}
We note that ${\left({\tilde n}_\heii\right)^{n+1}}$ is still an unknown, which is given by
\begin{equation}
 \begin{split}
  \left({\tilde n}_\heii\right)^{n+1} &= \left({\tilde n}_\heii\right)^{n} -D \left({\tilde n}_\heii\right)^{n+1} + E   \left({\tilde
  n}_\heiii\right)^{n+1} \\ &- G\left({\tilde n}_\heii\right)^{n+1} - I \left({\tilde n}_\heii\right)^{n+1}\\ & + F\left(1 - \left({\tilde n}_\heii\right)^{n+1} - \left({\tilde n}_\heiii\right)^{n+1}\right) \\ &+ H \left(1 - \left({\tilde n}_\heii\right)^{n+1} - \left({\tilde n}_\heiii\right)^{n+1}\right).
 \end{split}
 \label{eq:nheii}
\end{equation}
Substituting the value of ${\left({\tilde n}_\heiii\right)^{n+1}}$ from Eq.~\ref{eq:nheiii} into Eq.~\ref{eq:nheii} we get
\begin{equation}
  \left({\tilde n}_\heii\right)^{n+1} = \frac{\left({\tilde n}_\heii\right)^{n}+F+H - 
 \displaystyle\frac{H+F-E}{1+E}\left({\tilde n}_\heiii\right)^{n}}{1+D+F+G+H+I+\displaystyle\frac{(H+F-E)(D+I)}{1+E}}.
\end{equation}
Then the value of $ {\left({\tilde n}_\heii\right)^{n+1}}$ is used in Eqs.~\ref{eq:nheiii} to obtain ${\left({\tilde n}_\heiii\right)^{n+1}}$. Finally ${\left({\tilde n}_\hei\right)^{n+1}}$ is given as
\begin{equation}
 \left({\tilde n}_\hei\right)^{n+1} = 1 - \left({\tilde n}_\heii\right)^{n+1} - \left({\tilde n}_\heiii\right)^{n+1},
\end{equation}
and the electron density is set to
\begin{equation}
 (n_e)^{n+1} =\left({ n}_\hii\right)^{n+1} +\left({ n}_\heii\right)^{n+1} + 2\left({ n}_\heiii\right)^{n+1}.
\end{equation}

Finally, the photon number density and flux are updated as 
\begin{equation}
\left(N_\gamma^i\right)^{n+1} = \frac{\left(N_\gamma^i\right)^{n} + \Delta t \sum_j s_{ij}}{1+\Delta t {\tilde c} \left( \sum_j \left(n_j\right)^n {\bar \sigma}_{ij} + \kappa_i \, \rho \right)}  \, ,
\end{equation}
and
\begin{equation}
\left({\bf F}_\gamma^i\right)^{n+1} = \frac{\left({\bf F}_\gamma^i\right)^{n}}{1+\Delta t {\tilde c} \left( \sum_j \left(n_j\right)^n {\bar \sigma}_{ij} + \kappa_i \, \rho \right)} \, .
\end{equation}

\section{Dust Opacities}
\label{sec:dust}

The coupling between dust and radiation is a function of both dust grain size
and wavelength (or frequency) of incident radiation.  A grain of size $a$ and
geometric cross-section ${\pi a^2}$ subject to incident radiation at wavelength
$\lambda$ has radiation pressure cross-section ${Q(a, \lambda) \pi a^2}$, where
$Q$ is the dimensionless radiation pressure coefficient ${Q(a, \lambda) =
Q_\text{abs}(a, \lambda) + (1 - g(a, \lambda)) \times Q_\text{sca}(a,
\lambda)}$.  Here, $Q_\text{abs}$ is the contribution from absorption,
$Q_\text{sca}$ is the contribution from scattering, and $g = \langle \cos
\theta \rangle$ is the average cosine of the angle of scattered light.

We use tabulated values of $Q(a, \lambda)$ from \citet{Draine1984} and
\cite{Laor1993}, who present absorption and scattering data for $10^{-3} \leq
\lambda / \mu\text{m} \leq 10^3$ and for graphite and silicate grain
compositions.  Since we do not follow detailed grain chemistry, we average the
radiation pressure coefficients for graphite and silicate grains to calculate
effective $Q(a, \lambda)$ values.  We note that $Q(a, \lambda)$ is related to
grain opacity by $\kappa(a, \lambda) = 4 Q(a, \lambda) / 3 a \rho_\text{gr}$,
where $\rho_\text{gr} \approx 2.4 \, \text{g} \, \text{cm}^{-3}$ is the
internal density of a solid dust grain.

Astrophysical dust grains come in a range of different sizes, and the size
distribution is typically defined in terms of a function $\diff n / \diff a$,
where $\diff n / \diff a \times \diff a$ gives the number of grains with size
in the interval $[a, a + \diff a]$.  We assume that dust grains follow a
power-law size distribution $\diff n / \diff a \propto a^{-3.5}$
\citep{Mathis1977}, with minimum size $a_\text{min} = 0.001 \, \mu\text{m}$ and
maximum size $a_\text{max} = 1 \, \mu\text{m}$.

Using these radiation pressure coefficients, the Planck mean opacity
$\kappa_\ts{P}$ at gas temperature $T$ for the frequency interval
$[\nu_\text{min}, \nu_\text{max}]$ averaged over the grain size distribution is
given by
\begin{equation}
\kappa_\ts{P} = \frac{\displaystyle \int_{a_\text{min}}^{a_\text{max}} \int_{\nu_\text{min}}^{\nu_\text{max}} B_\nu(T) \kappa(a, c/\nu) \frac{\diff n}{\diff a} \diff \nu \diff a}{\displaystyle \int_{a_\text{min}}^{a_\text{max}} \frac{\diff n}{\diff a} \, \diff a  \int_{\nu_\text{min}}^{\nu_\text{max}} B_\nu(T) \diff\nu  }\, ,
\label{eqn:kappa_P_dust}
\end{equation}
and the Rosseland mean opacity $\kappa_\ts{R}$ is calculated from
\begin{equation}
\frac{1}{\kappa_\ts{R}} = \frac{\displaystyle \int_{a_\text{min}}^{a_\text{max}} \int_{\nu_\text{min}}^{\nu_\text{max}} \frac{\partial B_\nu(T)}{\partial T} \frac{1}{\kappa(a, c/\nu)} \frac{\diff n}{\diff a} \diff \nu \diff a}{\displaystyle \int_{a_\text{min}}^{a_\text{max}} \frac{\diff n}{\diff a} \, \diff a \int_{\nu_\text{min}}^{\nu_\text{max}} \frac{\partial B_\nu(T)}{\partial T} \diff\nu }\, ,
\label{eqn:kappa_R_dust}
\end{equation}
where $B_\nu(T)$ is Planck black body function.  By calculating mean opacities
over specific frequency ranges, these values can be used in the multi-frequency
approach detailed in Section~\ref{sec:chemistry}.

In practice, we create a lookup table by precomputing these dust opacities over
a range of gas temperatures and during simulations calculate the opacities of
individual gas cells by interpolating to the cells' temperatures.
Additionally, since the opacities from Eqs.~\ref{eqn:kappa_P_dust}
and~\ref{eqn:kappa_R_dust} are given in units of area per unit mass of dust, we
multiply by a gas cell's dust-to-gas ratio so that opacities are per unit mass
of gas.

\bsp
\label{lastpage}
\end{document}